\documentclass[12pt,english]{article}
\usepackage{amsmath}
\usepackage{amsthm}

\usepackage{newtxmath}

\usepackage[T1]{fontenc}
\usepackage[latin9]{inputenc}
\usepackage{color}
\usepackage{babel}
\usepackage{array}
\usepackage{float}
\usepackage{multirow}
\usepackage{graphicx}
\usepackage[a4paper]{geometry}
\usepackage{rotfloat}
\usepackage{setspace}
\usepackage[authoryear]{natbib}
\usepackage[pdfusetitle,
 bookmarks=true,bookmarksnumbered=false,bookmarksopen=false,
 breaklinks=true,pdfborder={0 0 0},pdfborderstyle={},backref=false,colorlinks=true]
 {hyperref}
\hypersetup{
 citecolor=DarkBlue,linkcolor=DarkBlue,urlcolor=DarkBlue}

\makeatletter

\providecommand{\tabularnewline}{\\}
\newcommand{\lyxdot}{.}

\theoremstyle{plain}
\newtheorem{lem}{\protect\lemmaname}
\theoremstyle{plain}
\newtheorem{cor}{\protect\corollaryname}
\theoremstyle{plain}
\newtheorem{prop}{\protect\propositionname}

\usepackage[svgnames]{xcolor}
\usepackage{bbm}

\ifdefined\showcaptionsetup
 \PassOptionsToPackage{caption=false}{subfig}
\fi
\usepackage{subfig}
\makeatother

\providecommand{\corollaryname}{Corollary}
\providecommand{\lemmaname}{Lemma}
\providecommand{\propositionname}{Proposition}

\begin{document}
\title{The Race between Technology and Woman:\\
{\Large Changes in Relative Wages and Labor Shares in OECD Countries}\thanks{We are grateful for comments and suggestions to Shoya Ishimaru, Dirk
Krueger, Christopher Taber, Takashi Yamashita, anonymous referees,
and numerous seminar participants at Hitotsubashi University, National
University of Singapore, Osaka University, Tohoku University, University
of Tokyo, the Japanese Economic Association Spring Meeting, Kansai
Labor Economics Workshop, and the Society of Labor Economists Annual
Meeting, and UCL Economics PhD Alumni Conference. Yamada acknowledges
support from JSPS KAKENHI grant numbers 17H04782 and 21H00724.}\medskip{}
}
\author{Hiroya Taniguchi\thanks{Okayama University. \texttt{taniguchi.hi@okayama-u.ac.jp}}
\and Ken Yamada\thanks{Kyoto University. \texttt{yamada@econ.kyoto-u.ac.jp}}\medskip{}
}
\date{October 2025}
\maketitle
\begin{abstract}
\begin{onehalfspace}
The era of technological change entails complex patterns of changes
in wages and employment. We develop a unified framework to evaluate
the effects of capital-embodied technological change on, as well as
the contributions of factor inputs to, the relative wages and income
shares of different types of labor. We obtain the aggregate elasticities
of substitution among different types of capital and labor by estimating
and aggregating sectoral production function parameters with cross-country
cross-industry panel data from OECD countries. We show that advances
in information, communication, and computation technologies contribute
significantly to narrowing the gender wage gap, widening the skill
wage gap, and declining labor shares.\bigskip{}
\end{onehalfspace}

\global\long\def\E{\mathbb{E}}%
\end{abstract}
\newpage{}

\section{Introduction}

Over the decades, there have been substantial changes in the structure
of wages and employment. The wage gap between male and female workers
has narrowed in many countries, while that between skilled and unskilled
workers has widened in some countries \citep*{Krueger_Perri_Pistaferri_Violante_RED10}.
At the same time, the share of labor in national income has declined
in many countries \citep*{Karabarbounis_Neiman_QJE14}. During the
period, there have been tremendous advances in information, communication,
and computation technologies in industrialized nations. Such technological
advances have been recognized as one of the causes for the widening
of the skill wage gap and the decline in the labor share \citep*{Hornstein_Krusell_Violante_HEG05,Acemoglu_Autor_HLE11,Grossman_Oberfield_ARE22,Karabarbounis_JEP24}.
Nevertheless, changes in relative wages and income shares vary in
direction, magnitude, or both across different types of labor classified
by gender and skill, as described later.

The direction and magnitude of changes in the relative wages and income
shares of different types of labor due to the expansion of technologically
advanced new equipment depend on the presence and degree of capital\textendash skill
or capital\textendash gender complementarity. Male and female workers
possess different sets of skills \citep*{Welch_AERPP00,Borghans_terWeel_Weinberg_ILRR14,Cortes_Jaimovich_Siu_JHR23,Rendall_wp24},
as skilled and unskilled workers do \citep*{Griliches_RESTAT69}.
Communication skills, in which women can possibly have a comparative
advantage, are more likely to be complementary to information and
communication technologies. Physical skills, in which men can possibly
have a comparative advantage, are more likely to be substitutable
with computation and automation technologies. The widespread use of
new technologies might raise the demand for brains relative to brawn,
which could result in not only a widening of the skill wage gap but
also a narrowing of the gender wage gap and a decline in the income
share of male unskilled labor.

This study develops a unified framework to evaluate the effects of
capital-embodied technological change on, as well as the contributions
of factor inputs to, the relative wages and income shares of different
types of labor. Our framework builds upon a multi-sector, multi-factor
extension of the \citet*{Greenwood_Hercowitz_Krusell_AER97} model
of investment-specific technological change.\footnote{The terms, capital-embodied technological change and investment-specific
technological change, are used interchangeably in this paper.} We incorporate technological change embodied in new capital into
an economy consisting of multiple sectors, where multiple consumption
and investment goods are produced from multiple types of capital and
labor. The presence and degrees of capital\textendash skill and capital\textendash gender
complementarities in the multi-sector economy are determined by the
aggregate elasticities of substitution among different types of capital
and labor. We show that the aggregate elasticities of substitution
are sufficient statistics for measuring the contributions of factor
inputs to the relative wages and income shares of different types
of labor. In this context, we generalize a model that explains changes
in the wage premium to skill in terms of the race between technology
and education \citep*{Tinbergen_K74,Goldin_Katz_bk10} and a model
that explains changes in the labor share in terms of capital deepening
\citep*{Hicks_bk32,Elsby_Hobijn_Sahin_BPEA13}. We derive analytical
expressions that characterize not only the contribution of each input,
holding the other inputs constant, but also the effects of capital-embodied
technological change, allowing all inputs to vary. We further show
that the aggregate elasticities of substitution are sufficient statistics
for quantifying the effects of capital-embodied technological change
on the relative wages and income shares of different types of labor.
In this context, we synthesize models that explain a rise in the skill
premium or a decline in the labor share in terms of a fall in the
relative price of capital equipment \citep*{Krusell_Ohanian_RiosRull_Violante_EM00,Karabarbounis_Neiman_QJE14}.

We apply this framework to advanced economies consisting of multiple
sectors, where the production of goods and services requires multiple
types of capital and labor. Given the fact that capital-embodied technological
progress was accelerated by information processing equipment and software
in the 1980s and 1990s \citep*{Cummins_Violante_RED02}, we distinguish
equipment related to information, communication, and computation technologies
from other types of capital. We refer to the former as ICCT capital
and the latter as non-ICCT capital. We classify labor by gender and
skill into four types: male skilled, female skilled, male unskilled,
and female unskilled labor. In doing so, we analyze how technological
change embodied in ICCT capital shapes the observed differences in
the direction and magnitude of changes in the gender wage gap by skill,
the skill wage gap by gender, and labor shares by gender and skill.

Although it is notoriously difficult to identify the elasticities
of substitution in the aggregate production function using time-series
data \citep*{Diamond_McFadden_Rodriguez_CEA78,Acemoglu_JEL02}, it
is possible to obtain the aggregate elasticities of substitution by
estimating and aggregating the disaggregated production function parameters
with cross-sectional variation in disaggregated data \citep{Sato_bk75,Oberfield_Raval_EM21}.
We estimate the sectoral production function parameters and obtain
the aggregate elasticities of substitution between ICCT capital and
the four types of labor using cross-country cross-industry data from
OECD countries. The advantages of cross-country cross-industry data
over firm-level data are that it enables us to measure equipment related
to information, communication, and computation technologies and to
classify labor by gender and skill. Our estimates of the aggregate
elasticities of substitution indicate that ICCT capital is more complementary
not only to skilled labor than unskilled labor but also to female
labor than male labor. This finding is robust to the choice of instrumental
variables and the presence of labor market institutions, factor-biased
technological change, and price markups.

We decompose changes in the relative wages and labor shares of the
four types of labor into components attributable to the factor complementarity
effect associated with progress in ICCT and the relative quantity
effect associated with advances in educational attainment and female
employment, and then further into components attributable to the two
types of capital and the four types of labor. Our results suggest
that the expansion of ICCT equipment contributes to the widening of
the skill wage gap, the narrowing of the gender wage gap, and a fall
in the income share of male unskilled labor that accounts for most
of the decline in the labor share. The observed difference in the
rate of decrease (increase) in the gender (skill) wage gap by skill
(gender) can be explained in terms of the race between progress in
ICCT and advances in female employment (educational attainment). From
the 1980s to the 2000s in OECD countries, the rate of decline in the
gender wage gap was greater among unskilled workers than among skilled
workers, while the rate of increase in the skill wage gap was greater
among male workers than among female workers. Our results suggest
that these differences do not arise from an insufficient increase
in the relative demand for female skilled labor, but rather from a
substantial increase in the relative supply of female skilled labor.
Furthermore, we quantify the general equilibrium effects of technological
change embodied in ICCT equipment on the gender wage gap by skill,
the skill wage gap by gender, and labor shares by gender and skill.
On average across OECD countries, if all else were held constant,
a fall in the relative price of ICCT equipment between the years 1980
and 2005 would have reduced the skilled (unskilled) gender wage gap
by 49 (10) percentage points and raised the male (female) skill wage
gap by 26 (77) percentage points, while it would have raised the income
share of male (female) skilled labor by 1.4 (2.0) percentage points
and reduced the income share of male (female) unskilled labor by 7.0
(0.7) percentage points.

The remainder of this paper is structured as follows. The next section
reviews the related literature. Section \ref{sec: theory} establishes
the theoretical framework to evaluate the effects of technological
change on, as well as the contributions of factor inputs to, the relative
wages and income shares of different types of labor. Section \ref{sec: data}
describes the data used in the analysis and shows changes in the relative
prices, relative quantities, and income shares of factors. Section
\ref{sec: estimation} discusses the specification, identification,
estimation, and aggregation of parameters in the sectoral production
function. Section \ref{sec: results} presents the results regarding
the sources of changes in the relative wages and income shares of
different types of labor. The final section provides conclusions.

\section{Related Literature}

This study adds to several strands of literature. First, it is related
to the literature on technological change and wage inequality. This
strand of literature highlights the importance of factor-biased technological
change in accounting for changes in wage inequality in the United
States and other OECD countries, typically by estimating the aggregate
production function. The widening of the skill wage gap is attributed
to skill-biased technological change \citep{Bound_Johnson_AER92,Katz_Murphy_QJE92,Goldin_Katz_bk10},
technological change embodied in capital equipment \citep*{Krusell_Ohanian_RiosRull_Violante_EM00,Maliar_Maliar_Tsener_EL22,Taniguchi_Yamada_LE22,Ohanian_Orak_Shen_RED23},
or sector-specific disembodied technological change \citep*{Buera_Kaboski_Rogerson_Vizcaino_RES22}.
The narrowing of the gender wage gap is also attributed to gender-biased
technological change \citep{Johnson_Keane_JoLE13}. Relative to these
studies, we obtain the aggregate elasticities of substitution by estimating
and aggregating sectoral production function parameters, and examine
the sources of changes in the four wage gaps: skilled gender wage
gap, unskilled gender wage gap, male skill wage gap, and female skill
wage gap.

Second, this study is related to the literature on the labor market
impact of new technologies. This strand of literature indicates advances
in information, communication, computation, and automation technologies
as causes of changes in the structure of wages and employment in the
United States and other OECD countries. The rate of increase in the
wage bill share of skilled or female labor tends to be higher in ICCT
or computer-intensive industries \citep*{Autor_Katz_Krueger_QJE98,Michaels_Natraj_VanReenen_RESTAT14,Raveh_EB15}.
The increased employment and wages of skilled or female labor are
associated with a rise in occupations requiring nonroutine analytical
and interactive tasks that are complementary to ICCT and a decline
in routine cognitive and manual tasks that are substitutable with
ICCT \citep*{Autor_Levy_Murnane_QJE03,Black_SpitzOener_RESTAT10}.
The widening of the skill wage gap is attributed to a relative decline
in industries requiring routine tasks that are substitutable with
automation technology \citep{Acemoglu_Restrepo_EM22}. The widening
of the skill wage gap is also attributed to the increased use of computers
and other capital equipment, whereas the narrowing of the gender wage
gap is not necessarily attributable to that increased use \citep*{Beaudry_Lewis_AEJ14,Burstein_Morales_Vogel_AEJ19,Caunedo_Jaume_Keller_AER23}.\footnote{\citet*{Beaudry_Lewis_AEJ14} consider an aggregate production function
with a unit elasticity of substitution between computer equipment
and skilled labor and an infinite elasticity of substitution between
computer equipment and unskilled labor. \citet*{Burstein_Morales_Vogel_AEJ19}
consider an aggregate production function, in which final output is
produced from occupational outputs, which are in turn produced by
production units, each using one type of capital (computing or other
equipment) and one type of labor (classified by gender, education,
and age). They specify an occupational production function with a
unit elasticity of capital\textendash labor substitution for all types
of capital and labor. \citet*{Caunedo_Jaume_Keller_AER23} consider
an aggregate production function, in which final output is produced
from occupational outputs, which are in turn produced from capital
and labor. They specify an occupational production function with an
common elasticity of capital\textendash labor substitution for all
types of labor and an infinite elasticity of labor\textendash labor
substitution for all types of labor.} Relative to these studies, we allow all four types of labor disaggregated
by gender and skill to be imperfect substitutes for one another and
substitutable to varying degrees with ICCT equipment within sectors,
thereby permitting the separate identification of capital\textendash skill
and capital\textendash gender complementarities. We impose neither
specific values on the elasticities of substitution among the four
types of labor nor on those between ICCT equipment and labor.

Third, this study is related to the literature on technological change
and the labor share of income. This strand of literature points to
the relevance of technological change to the decline in the labor
share in the United States and elsewhere. The decline in the labor
share is attributed to capital-embodied technological change \citep*{Karabarbounis_Neiman_QJE14,Eden_Gaggl_RED18},
factor-biased technological change \citep*{Oberfield_Raval_EM21},
or the adoption of robotic technology \citep{Acemoglu_Restrepo_JPE20}.
Relative to these studies, we examine the sources of changes in labor
shares by gender and skill. We decompose changes in the income share
of the four types of labor into components attributable to the relative
quantity effect, whose sign depends on whether the aggregate elasticities
of substitution exceed one, and to the factor complementarity effect,
whose sign is determined by the relative magnitude of the aggregate
elasticities of substitution.

Finally, from a methodological point of view, this study is related
to \citet{Baqaee_Farhi_JEEA19} and \citet{Oberfield_Raval_EM21},
who develop a framework to characterize and estimate the aggregate
elasticity of substitution in an economy consisting of multiple sectors.
Relative to these studies, we consider a dynamic economy in which
new capital accumulates more rapidly as technology progresses. We
derive analytical expressions that decomposes changes in relative
wages and income shares into components attributable to factor inputs
and quantifies the general equilibrium effects of capital-embodied
technological change on relative wages and income shares. We further
extend the framework to allow for endogenous labor supply and imperfect
labor mobility across sectors. As in the wage inequality literature,
we focus on the value-added output.

This study contributes to these strands of literature in three ways.
First, we provide the first estimates of the elasticities of substitution
between ICCT equipment and four types of labor disaggregated by gender
and skill. These estimates enable us to explain the observed differences
in the direction and magnitude of changes in the gender wage gap by
skill, the skill wage gap by gender, and labor shares by gender and
skill in terms of technological change embodied in ICCT equipment.
Second, we analytically derive and quantitatively evaluate the contributions
of specific factor inputs to changes in the relative wages and income
shares of different types of labor in OECD countries. Our results
suggest that technological change embodied in ICCT equipment underlies
gender-biased technological change, as well as skill-biased technological
change noted in the literature. Finally, we analytically derive and
quantitatively evaluate the general equilibrium effects of technological
change embodied in ICCT equipment on the relative wages and income
shares of different types of labor in OECD countries. Our results
capture the effects of capital-embodied technological change more
comprehensively because the elasticity of capital\textendash labor
substitution is allowed to vary within and across sectors.

\section{Theory\label{sec: theory}}

We develop a unified framework to evaluate the effects of capital-embodied
technological change on, as well as the contributions of factor inputs
to, the relative wages and income shares of different types of labor.
Our framework builds upon the seminal work of \citet*{Greenwood_Hercowitz_Krusell_AER97}
on investment-specific technological change. We consider a multi-sector
economy in which multiple consumption and investment goods are produced
in each sector from multiple types of capital and multiple types of
labor, whereas \citet*{Greenwood_Hercowitz_Krusell_AER97} consider
a two-sector economy in which consumption and investment goods are
separately produced in different sectors from two types of capital
and one type of labor. We first show how the relative wages and income
shares of different types of labor can change in response to each
input, holding all other inputs constant. We then show how they can
change in response to a fundamental determinant, technological change,
allowing all inputs to vary. Furthermore, we extend the framework
to allow for endogenous labor supply and imperfect labor mobility
across sectors in Appendix \ref{sec: extension}. Our theoretical
results do not require specifying any aggregate or sectoral production
function.

\subsection{Environment}

\subsubsection{Household\label{subsec: household}}

The representative household has preferences over sequences of consumption
$\{c_{ft}\}_{t=0}^{\infty}$ characterized by
\begin{equation}
\sum_{t=0}^{\infty}\beta^{t}\sum_{f=1}^{F_{\ell}}\mu_{f}\mathcal{U}\left(c_{ft}\right),\label{eq: utility}
\end{equation}
where $\beta\in(0,1)$ is a discount factor, $\mu_{f}$ is the share
of individuals of type $f\in\{1,\ldots,F_{\ell}\}$, and $\mathcal{U}(\cdot)$
is an instantaneous utility function. Aggregate consumption by an
individual of type $f$ in period $t$ is a constant elasticity of
substitution (CES) composite of goods produced in each sector:
\begin{equation}
c_{ft}=\left(\sum_{n=1}^{N}\theta_{n}c_{fnt}^{\eta}\right)^{\frac{1}{\eta}},\label{eq: aggregator_c}
\end{equation}
where $c_{fnt}$ is consumption goods produced in sector $n$. The
share parameters $\theta_{n}$ lie between zero and one, and the substitution
parameter $\eta$ is less than one.

The household is endowed with an initial capital stock $k_{f0}$.
The individual of type $f$ is endowed with time $l_{ft}$ in each
period. Capital accumulates according to the law of motion:
\begin{equation}
k_{f,t+1}=\left(1-\delta_{f}\right)k_{ft}+x_{ft},\label{eq: capital}
\end{equation}
where $x_{ft}$ is investment in capital $f\in\{1,\ldots,F_{k}\}$,
and $\delta_{f}\in(0,1)$ is a depreciation rate. Aggregate investment
in capital $f$ is also a CES composite of goods produced in each
sector:
\begin{equation}
x_{ft}=q_{ft}\left(\sum_{n=1}^{N}\theta_{n}x_{fnt}^{\eta}\right)^{\frac{1}{\eta}},\label{eq: aggregator_x}
\end{equation}
where $x_{fnt}$ is investment goods produced in sector $n$, and
$q_{ft}$ is investment-specific technology. In each period, the household
faces the budget constraint:
\begin{equation}
\sum_{n=1}^{N}p_{nt}\sum_{f=1}^{F_{\ell}}\mu_{f}c_{fnt}+\sum_{n=1}^{N}p_{nt}\sum_{f=1}^{F_{k}}x_{fnt}=\sum_{f=1}^{F_{\ell}}\mu_{f}w_{ft}l_{ft}+\sum_{f=1}^{F_{k}}r_{ft}k_{ft},\label{eq: budget}
\end{equation}
where $p_{nt}$ is the price of goods produced in sector $n$, $w_{ft}$
is the wages of labor $f$, and $r_{ft}$ is the rental price of capital
$f$. The supply of labor is assumed to be exogenously given, but
this assumption will be relaxed in Appendix \ref{sec: extension}.

The household maximizes its utility by equalizing consumption across
individuals: $c_{fnt}=c_{nt}$ and $c_{ft}=c_{t}$ for all $f$. The
expenditure share of final goods $n$ in period $t$ is then defined
as $\zeta_{nt}=p_{nt}y_{nt}/\sum_{n=1}^{N}p_{nt}y_{nt}=p_{nt}y_{nt}/p_{t}y_{t}$,
where the aggregate price of consumption goods is $p_{t}=(\sum_{n=1}^{N}\theta_{n}^{1/(1-\eta)}p_{nt}^{-\eta/(1-\eta)})^{-(1-\eta)/\eta}$,
and the aggregate output is $y_{t}=c_{t}+\sum_{f=1}^{F_{k}}(x_{ft}/q_{ft})$.
The expenditure share can be derived as
\begin{equation}
\zeta_{nt}=\text{\ensuremath{\theta_{n}^{\frac{1}{1-\eta}}\left(\frac{p_{nt}}{p_{t}}\right)^{-\frac{\eta}{1-\eta}}}}.\label{eq: D.PnYn/PY}
\end{equation}
As in \citet*{Greenwood_Hercowitz_Krusell_AER97}, investment-specific
technology can be measured by the ratio of the aggregate price of
investment goods to the aggregate price of consumption goods.
\begin{equation}
\frac{1}{q_{ft}}=\frac{p_{ft}}{p_{t}}.\label{eq: 1/Qf}
\end{equation}
where the aggregate price of investment goods is $p_{ft}=(1/q_{ft})(\sum_{n=1}^{N}\theta_{n}^{1/(1-\eta)}p_{nt}^{-\eta/(1-\eta)})^{-(1-\eta)/\eta}$.

\subsubsection{Firms\label{subsec: firm}}

The firm in sector $n$ in period $t$ maximizes its profits:
\begin{equation}
p_{nt}y_{nt}-\sum_{f=1}^{F_{\ell}}w_{ft}\ell_{fnt}-\sum_{f=1}^{F_{k}}r_{ft}k_{fnt}\label{eq: profit}
\end{equation}
subject to a constant returns to scale technology:
\begin{equation}
y_{nt}=A_{nt}\mathcal{F}_{n}\left(\ell_{1nt},\ldots,\ell_{F_{\ell}nt},k_{1nt},\ldots,k_{F_{k}nt}\right),\label{eq: technology}
\end{equation}
where $y_{nt}$ is the real output in sector $n$ in period $t$,
$\ell_{fnt}$ is the quantity of labor $f\in\{1,\ldots,F_{\ell}\}$,
$k_{fnt}$ is the quantity of capital $f\in\{1,\ldots,F_{k}\}$, $A_{nt}$
is sector-specific factor-neutral technology, and $\mathcal{F}_{n}(\cdot)$
is a sectoral production function.

Below, to avoid notational clutter, we let $\ell_{fnt}$ denote not
only labor for $f\in\{1,\ldots,F_{\ell}\}$ but also capital for $f\in\{F_{\ell}+1,\ldots,F\}$
and $w_{ft}$ denote not only wages for $f\in\{1,\ldots,F_{\ell}\}$
but also the rental price of capital for $f\in\{F_{\ell}+1,\ldots,F\}$.

The income share of factor $f$ in sector $n$ in period $t$ is defined
as $\lambda_{\ell_{f}nt}=w_{ft}\ell_{fnt}/p_{nt}y_{nt}$. When the
production function is homogeneous of degree one, the profits \eqref{eq: profit}
can be written as $p_{nt}y_{nt}-y_{nt}\mathcal{\widetilde{C}}_{n}(w_{1t},\ldots,w_{Ft})/A_{nt}$,
where $\mathcal{\mathcal{\widetilde{C}}}_{n}(\cdot)/A_{nt}$ is a
unit cost function in sector $n$ in period $t$, and the demand for
factor $f$ in sector $n$ in period $t$ can be written as $\ell_{fnt}=y_{nt}\widetilde{\mathcal{G}}_{fn}(w_{1t},\ldots,w_{Ft})/A_{nt}$,
where $\widetilde{\mathcal{G}}_{fn}(\cdot)/A_{nt}$ is the demand
for factor $f$ per unit output in sector $n$ in period $t$. By
Shephard\textquoteright s lemma, $\lambda_{\ell_{f}nt}=\partial\ln\mathcal{\mathcal{\widetilde{C}}}_{n}(w_{1t},\ldots,w_{Ft})/\partial\ln w_{ft}$
and $\partial\ln\lambda_{\ell_{f}nt}/\partial\ln w_{gt}=\partial\ln\widetilde{\mathcal{G}}_{fn}\left(w_{1t},\ldots,w_{Ft}\right)/\partial\ln w_{gt}-\lambda_{\ell_{g}nt}+I_{\left(f,g\right)},$where
$I_{(f,g)}$ is the $fg$-th element of the $F\times F$ identity
matrix. While $\lambda_{\ell_{f}nt}$ can be observed directly from
data, $\partial\ln\lambda_{\ell_{f}nt}/\partial\ln w_{g}$ can be
estimated after $\mathcal{F}_{n}(\cdot)$ is specified and $\widetilde{\mathcal{G}}_{fn}(\cdot)$
is derived.

\subsubsection{Equilibrium}

Given a set of technologies \{$A_{nt}$, $q_{ft}$\} and an initial
capital stock $k_{f0}$, a competitive equilibrium is a set of prices
\{$p_{nt}$, $w_{ft}$, $r_{ft}$\} and a set of quantities \{$c_{fnt}$,
$x_{fnt}$, $\ell_{fnt}$, $k_{fnt}$\} such that: (i) given prices,
the representative household chooses $c_{fnt}$, $x_{fnt}$, and $k_{f,t+1}$
to maximize the utility \eqref{eq: utility} subject to the budget
constraint \eqref{eq: budget}, the capital law of motion \eqref{eq: capital},
and the consumption and investment aggregators \eqref{eq: aggregator_c}
and \eqref{eq: aggregator_x}, (ii) given prices, each firm chooses
$\ell_{fnt}$ and $k_{fnt}$ to maximize its profits \eqref{eq: profit}
subject to the technology \eqref{eq: technology}, (iii) goods markets
clear:

\begin{equation}
c_{nt}+\sum_{f=1}^{F_{k}}x_{fnt}=y_{nt},\label{eq: goods_market}
\end{equation}
and (iv) factor markets clear:
\begin{equation}
\sum_{n=1}^{N}\ell_{fnt}=\mu_{f}l_{ft}\qquad\text{and}\qquad\sum_{n=1}^{N}k_{fnt}=k_{ft}.\label{eq: factor_market}
\end{equation}

We consider the competitive equilibrium that satisfies conditions
(ii) and (iii) throughout this paper. When we extend this model to
allow for endogenous labor supply or imperfect labor mobility, we
modify condition (i) or (iv) as described in Appendix \ref{sec: extension}.

\subsection{Aggregate elasticity of substitution}

The aggregate elasticity of substitution can be expressed in terms
of the aggregate production or cost function ($\mathcal{F}$ or $\mathcal{C}$).\footnote{\citet{Baqaee_Farhi_JEEA19} describe the aggregate production function
as a maximal attainable output subject to consumer's preferences,
producer's technology, and resource constraints and the aggregate
cost function as a minimal attainable cost of production subject to
consumer's preferences, producer's technology, and resource constraints
in a static economy, where all factors are exogenously given and perfectly
mobile across sectors.} The \citet*{Morishima_KH67} elasticities of substitution between
factors $f$ and $g$ are defined as
\[
\frac{1}{\epsilon_{\ell_{f}\ell_{g}}^{\mathcal{F}}}=\frac{\partial\ln\left(\left.\frac{\partial\mathcal{F}}{\partial\ell_{f}}\right/\frac{\partial\mathcal{F}}{\partial\ell_{g}}\right)}{\partial\ln\left(\left.\ell_{g}\right/\ell_{f}\right)},
\]
\[
\epsilon_{\ell_{f}\ell_{g}}^{\mathcal{C}}=\frac{\partial\ln\left(\left.\frac{\partial\mathcal{C}}{\partial w_{f}}\right/\frac{\partial\mathcal{C}}{\partial w_{g}}\right)}{\partial\ln\left(\left.w_{g}\right/w_{f}\right)},
\]
where $\ell_{f}=\sum_{n=1}^{N}\ell_{fn}$ for $f\in\{1,\ldots,F\}$.\footnote{The Morishima elasticity of substitution is a measure of curvature
of the isoquant in line with the original concept by \citet{Hicks_bk32}
and a sufficient statistic for measuring the contributions of factor
prices to relative factor shares, as noted by \citet{Blackorby_Russell_AER89}.} The partial derivative of the log of the production (cost) function
with respect to the log of the factor quantity (price) equals the
factor share of income in the aggregate economy, defined as $\Lambda_{\ell_{f}}=w_{f}\ell_{f}/py$.
The aggregate elasticities of substitution can be derived by looking
at the extent to which the factor shares of income vary according
to the quantity or price of factors. Throughout this and next two
subsections, we suppress the time subscript $t$ for notational simplicity.
The following lemma recasts the results of \citet*{Baqaee_Farhi_JEEA19}
for value-added output.\footnote{\citet*{Baqaee_Farhi_JEEA19} derive expressions for the elasticities
of substitution in the aggregate production and cost functions when
a micro-level production function takes a nested CES form, and firms
are connected through input\textendash output linkages (Propositions
3 and 8). In the appendix, they refine these expressions without imposing
any functional form on the micro-level production function. \citet*{Oberfield_Raval_EM21}
derive expressions for the elasticity of substitution in the aggregate
cost function both when the micro-level production function takes
a nested CES form (Proposition 2) and when no functional form is imposed
on the micro-level production function (Proposition 2').}
\begin{lem}
\label{prop: epsilon_f =000026 epsilon_c}The aggregate elasticities
of substitution between factors $f$ and $g$ are given by
\begin{align}
\frac{1}{\epsilon_{\ell_{f}\ell_{g}}^{\mathcal{F}}}= & \Psi_{\left(f,g\right)}^{\ell}-\Psi_{\left(g,g\right)}^{\ell}+1,\label{eq: morishima_y}\\
\epsilon_{\ell_{f}\ell_{g}}^{\mathcal{C}}= & \Psi_{\left(f,g\right)}^{w}-\Psi_{\left(g,g\right)}^{w}+1,\label{eq: morishima_c}
\end{align}
where $\Psi_{(f,g)}^{\ell}$ is the $fg$-th element of the $F\times F$
matrix $\Psi^{\ell}=-(I-\Psi^{w})^{-1}\Psi^{w}$, and $\Psi^{w}$
is the $F\times F$ matrix whose $fg$-th element is given by $\Psi_{(f,g)}^{w}=\sum_{n=1}^{N}(\zeta_{n}\lambda_{\ell_{f}n}/\Lambda_{\ell_{f}})[\partial\ln\lambda_{\ell_{f}n}/\partial\ln w_{g}+[1-1/(1-\eta)](\lambda_{\ell_{g}n}-\Lambda_{\ell_{g}})]$.
\end{lem}
Appendix \ref{sec: proofs} contains the proofs of this lemma and
all propositions. The lemma and propositions apply not only to sector-level
production functions but also to firm- and establishment-level production
functions.

The aggregate elasticities of substitution do not generally coincide
with the sectoral elasticities of substitution for two reasons. First,
the elasticities of substitution in one sector may differ from those
in another sector. Second, factor intensities in one sector may differ
from those in another sector. The aggregate elasticities of substitution
depend not only on substitution within sectors but also on reallocation
across sectors.
\begin{cor}
\label{cor: epsilon_c}The aggregate elasticity of substitution \eqref{eq: morishima_c}
can be expressed as a weighted average of the elasticities of substitution
in production and consumption:
\begin{equation}
\epsilon_{\ell_{f}\ell_{g}}^{\mathcal{C}}=\left[\sum_{n=1}^{N}\left(\frac{\ell_{fn}}{\ell_{f}}\right)\epsilon_{\ell_{f}\ell_{g}n}^{\mathcal{C}}+\sum_{n=1}^{N}\left(\frac{\ell_{gn}}{\ell_{g}}-\frac{\ell_{fn}}{\ell_{f}}\right)\sum_{f\neq g}^{F}\lambda_{\ell_{f}n}\epsilon_{\ell_{f}\ell_{g}n}^{\mathcal{C}}\right]+\left[\sum_{n=1}^{N}\left(\frac{\ell_{gn}}{\ell_{g}}-\frac{\ell_{fn}}{\ell_{f}}\right)\lambda_{\ell_{g}n}\frac{1}{1-\eta}\right],\label{eq: morishima_c'}
\end{equation}
where the elasticity of substitution in the sectoral cost function
is defined as $\epsilon_{\ell_{f}\ell_{g}n}^{\mathcal{C}}=\partial\ln[(\partial\mathcal{C}_{n}/\partial w_{f})$
$/(\partial\mathcal{C}_{n}/\partial w_{g})]/\partial\ln(w_{g}/w_{f})$.
\end{cor}
The first square bracket captures the extent to which the producer's
optimal mix of factors changes in response to relative factor prices.
If factor intensities are equal across sectors (i.e., $\ell_{fn}/\ell_{gn}=\ell_{fn^{\prime}}/\ell_{gn^{\prime}}=\ell_{f}/\ell_{g}$
for $n\neq n^{\prime}$), the aggregate elasticity \eqref{eq: morishima_c'}
is the weighted average of the sectoral elasticities; that is, $\epsilon_{\ell_{f}\ell_{g}}^{\mathcal{C}}=\sum_{n=1}^{N}(\ell_{fn}/\ell_{f})\epsilon_{\ell_{f}\ell_{g}n}^{\mathcal{C}}$,
where the weight is the quantity share of factor $f$ in each sector.
Furthermore, if the elasticities of substitution are equal across
sectors (i.e., $\epsilon_{\ell_{f}\ell_{g}n}^{\mathcal{C}}=\epsilon_{\ell_{f}\ell_{g}n^{\prime}}^{\mathcal{C}}$
for $n\neq n^{\prime}$), the aggregate elasticity coincides with
the sectoral elasticity. The second square bracket captures the extent
to which the consumer's optimal mix of goods (i.e., the size of each
sector) changes in response to relative factor prices. An increase
in the factor price raises the price of goods due to an increase in
the marginal cost of production, which in turn reduces consumption.
An increase in the price of factor $g$ reduces the equilibrium quantity
of labor $g$ relative to factor $f$ (or equivalently raises the
equilibrium quantity of factor $f$ relative to factor $g$) in sectors
more intensive in factor $g$ than in factor $f$ (i.e., $\ell_{fn}/\ell_{f}<\ell_{gn}/\ell_{g}$).\footnote{Our representation in equation \eqref{eq: morishima_c'} differs from
\citet{Baqaee_Farhi_JEEA19}, who consider a production function with
capital, labor, and intermediate inputs and underscore input\textendash output
linkages, and from \citet{Oberfield_Raval_EM21}, who consider a production
function with capital, labor, and material inputs and focus on the
elasticity of substitution between capital and labor.}

The aggregate elasticities \eqref{eq: morishima_y} and \eqref{eq: morishima_c}
can be expressed in a simple form if production requires only one
type of capital and one type of labor, as described in Corollary \ref{cor: Oberfield-Raval}
in Appendix \ref{sec: corolloaries}.

\subsection{Contributions of factor inputs\label{subsec: factor_inputs}}

\subsubsection{Labor shares}

The aggregate elasticities of substitution, $\epsilon_{\ell_{f}\ell_{g}}^{\mathcal{F}}$
($\epsilon_{\ell_{f}\ell_{g}}^{\mathcal{C}}$), are sufficient statistics
for measuring the contributions of factor quantities (prices) to the
factor shares of income if the factor shares of income are known.
We denote by $\Delta\ln z$ the log change of variable $z$ from one
period to another.
\begin{prop}
\label{prop: D.WfLf/PY-Lf}Changes in factor shares are given by
\begin{multline}
\Delta\ln\Lambda_{\ell_{f}}=-\sum_{g\neq f}^{F}\left(\frac{1-\epsilon_{\ell_{g}\ell_{f}}^{\mathcal{F}}}{\epsilon_{\ell_{g}\ell_{f}}^{\mathcal{F}}}\right)\Lambda_{\ell_{g}}\Delta\ln\ell_{f}+\sum_{g\neq f}^{F}\left(\frac{1-\epsilon_{\ell_{f}\ell_{g}}^{\mathcal{F}}}{\epsilon_{\ell_{f}\ell_{g}}^{\mathcal{F}}}\right)\Lambda_{\ell_{g}}\Delta\ln\ell_{g}\\
+\sum_{g\neq f}^{F}\sum_{h\neq f,g}^{F}\left(\frac{\epsilon_{\ell_{h}\ell_{g}}^{\mathcal{F}}-\epsilon_{\ell_{f}\ell_{g}}^{\mathcal{F}}}{\epsilon_{\ell_{f}\ell_{g}}^{\mathcal{F}}\epsilon_{\ell_{h}\ell_{g}}^{\mathcal{F}}}\right)\Lambda_{\ell_{h}}\Delta\ln\ell_{g}+\psi_{\left(f\right)}^{\ell}\label{eq: D.WfLf/PY-Lf}
\end{multline}
or
\begin{multline}
\Delta\ln\Lambda_{\ell_{f}}=\sum_{g\neq f}^{F}\left(1-\epsilon_{\ell_{g}\ell_{f}}^{\mathcal{C}}\right)\Lambda_{\ell_{g}}\Delta\ln\left(\frac{w_{f}}{p}\right)-\sum_{g\neq f}^{F}\left(1-\epsilon_{\ell_{f}\ell_{g}}^{\mathcal{C}}\right)\Lambda_{\ell_{g}}\Delta\ln\left(\frac{w_{g}}{p}\right)\\
-\sum_{g\neq f}^{F}\sum_{h\neq f,g}^{F}\left(\epsilon_{\ell_{h}\ell_{g}}^{\mathcal{C}}-\epsilon_{\ell_{f}\ell_{g}}^{\mathcal{C}}\right)\Lambda_{\ell_{h}}\Delta\ln\left(\frac{w_{g}}{p}\right)+\psi_{\left(f\right)}^{w},\label{eq: D.WfLf/PY-Wf}
\end{multline}
where $\psi_{(f)}^{\ell}$ is the $f$-th element of the $F\times1$
vector $\psi^{\ell}=(I-\Psi^{w})^{-1}\psi^{w}$\textup{,} and $\psi^{w}$
is the $F\times1$ vector whose $f$-th element is given by $\psi_{(f)}^{w}=-[1-1/(1-\eta)]\sum_{n=1}^{N}(\zeta_{n}\lambda_{\ell_{f}n}/\Lambda_{\ell_{f}})(\Delta\ln A_{n}-\sum_{m=1}^{N}\zeta_{m}\Delta\ln A_{m})$.
\end{prop}
We refer to the first two terms in equation \eqref{eq: D.WfLf/PY-Lf}
as the relative quantity effect, the first two terms in equation \eqref{eq: D.WfLf/PY-Wf}
as the relative price effect, and the third term in equations \eqref{eq: D.WfLf/PY-Lf}
and \eqref{eq: D.WfLf/PY-Wf} as the factor complementarity effect.
The direction of the relative price and quantity effects on the income
share of factor $f$ depends on whether the elasticity of substitution
between factors $f$ and $g$ exceeds one, whereas the direction of
the factor complementarity effect on the income share of factor $f$
depends on whether factor $g$ is more complementary to factor $f$
than to factor $h$.

If production requires only one type of capital and one type of labor
(or equivalently, if all types of labor are perfectly substitutable,
and all types of capital are perfectly substitutable), the factor
complementarity effect may disappear. In this case, the direction
and magnitude of changes in the labor share are determined by the
elasticity of substitution between capital and labor and the degree
of capital deepening, as described in Corollary \ref{cor: Hicks}
in Appendix \ref{sec: corolloaries}.

However, when multiple types of capital and labor are used to produce
goods and services, the labor share can change not only due to the
relative price or quantity effect but also due to the factor complementarity
effect. Equations \eqref{eq: D.WfLf/PY-Lf} and \eqref{eq: D.WfLf/PY-Wf}
show how the income share of factor $f$ can change in response to
its own price or quantity change and other price or quantity change
in such a general case. On the one hand, a rise in the price or a
fall in the quantity of factor $f$ causes a rise in the relative
price of factor $f$ and a fall in the relative quantity of factor
$f$. A rise in the price (a fall in the quantity) of factor $f$
can result in a decline in the income share of factor $f$ if the
latter exceeds the former, in which case $\epsilon_{\ell_{g}\ell_{f}}^{\mathcal{C}}>1$
($\epsilon_{\ell_{g}\ell_{f}}^{\mathcal{F}}>1$). On the other hand,
a fall in the price or a rise in the quantity of factor $g$ causes
a rise in the relative price of factor $f$ and a fall in the relative
quantity of factor $f$. A fall in the price (a rise in the quantity)
of factor $g$ can result in a decline in the income share of factor
$f$ if the latter exceeds the former, in which case $\epsilon_{\ell_{f}\ell_{g}}^{\mathcal{C}}>1$
($\epsilon_{\ell_{f}\ell_{g}}^{\mathcal{F}}>1$). A fall in the price
(a rise in the quantity) of factor $g$ can further reduce the income
share of factor $f$ if factor $g$ is more substitutable with factor
$f$ than with factor $h$, in which case $\epsilon_{\ell_{f}\ell_{g}}^{\mathcal{C}}>\epsilon_{\ell_{h}\ell_{g}}^{\mathcal{C}}$
($\epsilon_{\ell_{f}\ell_{g}}^{\mathcal{F}}>\epsilon_{\ell_{h}\ell_{g}}^{\mathcal{F}}$).
The last term arises from sector-specific disembodied technological
change.

When production requires multiple types of labor, the rate of change
in the labor share can be expressed as the weighted average of the
rates of changes in the income shares of different types of labor:
\begin{equation}
\Delta\ln\Lambda_{\ell}=\sum_{f=1}^{F_{\ell}}\left(\frac{w_{f}\ell_{f}}{\sum_{f=1}^{F_{\ell}}w_{f}\ell_{f}}\right)\Delta\ln\Lambda_{\ell_{f}},\label{eq: D.WfLf/PY}
\end{equation}
where $\Delta\ln\Lambda_{\ell_{f}}$ is given by equation \eqref{eq: D.WfLf/PY-Lf},
and the weight is the wage bill share of each type of labor. Given
the fact that the majority of labor is male unskilled, the largest
weight is assigned to the rate of change in the income share of male
unskilled labor.

\subsubsection{Relative wages}

The aggregate elasticities of substitution, $\epsilon_{\ell_{f}\ell_{g}}^{\mathcal{F}}$
($\epsilon_{\ell_{f}\ell_{g}}^{\mathcal{C}}$), are sufficient statistics
for measuring the contributions of factor quantities (prices) to the
relative wages of (relative demand for) different types of labor.
\begin{prop}
\label{prop: D.Wf/Wg-Lf}Changes in relative factor prices are given
by
\begin{equation}
\Delta\ln\left(\frac{w_{f}}{w_{g}}\right)=-\sum_{h\neq f,g}\left(\frac{\epsilon_{\ell_{f}\ell_{h}}^{\mathcal{F}}-\epsilon_{\ell_{g}\ell_{h}}^{\mathcal{F}}}{\epsilon_{\ell_{f}\ell_{h}}^{\mathcal{F}}\epsilon_{\ell_{g}\ell_{h}}^{\mathcal{F}}}\right)\Delta\ln\ell_{h}-\frac{1}{\epsilon_{\ell_{g}\ell_{f}}^{\mathcal{F}}}\Delta\ln\ell_{f}+\frac{1}{\epsilon_{\ell_{f}\ell_{g}}^{\mathcal{F}}}\Delta\ln\ell_{g}+\left(\psi_{\left(f\right)}^{\ell}-\psi_{\left(g\right)}^{\ell}\right).\label{eq: D.Wf/Wg-Lf}
\end{equation}
Changes in relative factor demand are given by
\begin{equation}
\Delta\ln\left(\frac{\ell_{f}}{\ell_{g}}\right)=\sum_{h\neq f,g}\left(\epsilon_{\ell_{f}\ell_{h}}^{\mathcal{C}}-\epsilon_{\ell_{g}\ell_{h}}^{\mathcal{C}}\right)\Delta\ln\left(\frac{w_{h}}{p}\right)-\epsilon_{\ell_{g}\ell_{f}}^{\mathcal{C}}\Delta\ln\left(\frac{w_{f}}{p}\right)+\epsilon_{\ell_{f}\ell_{g}}^{\mathcal{C}}\Delta\ln\left(\frac{w_{g}}{p}\right)+\left(\psi_{\left(f\right)}^{w}-\psi_{\left(g\right)}^{w}\right).\label{eq: D.Lf/Lg-Wf}
\end{equation}
\end{prop}
We refer to the first term as the factor complementarity effect and
the second and third terms in equation \eqref{eq: D.Wf/Wg-Lf} as
the relative quantity effect again. The direction of the factor complementarity
effect on the wages of factor $f$ relative to factor $g$ depends
on whether factor $h$ is more complementary to factor $f$ than to
factor $g$, whereas the direction of the relative quantity effect
depends on whether each factor increases. Equation \eqref{eq: D.Wf/Wg-Lf}
is a multi-sector, multi-factor general equilibrium version of the
model of the race between technology and skills, which nests the standard
models of skill differentials, as described in Corollaries \ref{cor: Katz_Murphy}
and \ref{cor: Krusell_etal} in Appendix \ref{sec: corolloaries}.

Equation \eqref{eq: D.Wf/Wg-Lf} shows how gender and skill premia
can vary according to the shift in demand due to progress in ICCT
and the shift in supply due to advances in educational attainment
and female employment. The first term captures the shift in demand.
The wages of factor $f$ relative to factor $g$ can increase with
a rise in factor $h$ if factor $h$ is more complementary to factor
$f$ than to factor $g$ (i.e., $\epsilon_{\ell_{g}\ell_{h}}^{\mathcal{F}}>\epsilon_{\ell_{f}\ell_{h}}^{\mathcal{F}}$).
In particular, if ICCT capital is more complementary to skilled (female)
labor than to unskilled (male) labor, the skill (gender) wage gap
increases (decreases) with a rise in ICCT capital. The magnitude of
this factor complementarity effect is directly proportional to the
difference between $\epsilon_{\ell_{g}k_{i}}^{\mathcal{F}}$ and $\epsilon_{\ell_{f}k_{i}}^{\mathcal{F}}$,
as well as the rate of increase in $k_{i}$. The second and third
terms capture the shift in supply. The wages of factor $f$ relative
to factor $g$ can decrease with a rise (fall) in the quantity of
factor $f$ ($g$). The magnitude of this relative quantity effect
is inversely proportional to $\epsilon_{\ell_{f}\ell_{g}}^{\mathcal{F}}$
and $\epsilon_{\ell_{g}\ell_{f}}^{\mathcal{F}}$ and directly proportional
to the rate of increase in $\ell_{f}$ relative to $\ell_{g}$. Given
the fact that the rate of increase in skilled (female) labor is greater
than that in unskilled (male) labor, the relative quantity effect
should be negative (positive) for the skilled\textendash unskilled
(male\textendash female) wage gap. Consequently, the factor complementarity
effect and the relative quantity effect would work in opposite directions.
Whether the skill (gender) wage gap eventually widens (narrows) depends
on the relative magnitude of these two effects.

When production requires four or more types of labor, it may be of
interest to examine the sources of change in the average wage for
one group relative to that for another group. The average wage of
labor $h$ in group $H$ is defined as $\overline{w}_{H}=\sum_{h\in H}(\ell_{h}/\sum_{h\in H}\ell_{h})w_{h}$.
The rate of change in the average wage for group $H$ relative to
that for group $U$ can then be expressed as
\begin{multline}
\Delta\ln\left(\frac{\overline{w}_{H}}{\overline{w}_{U}}\right)=\sum_{h\in H}\left(\frac{w_{h}\ell_{h}}{\sum_{h\in H}w_{h}\ell_{h}}\Delta\ln\Lambda_{\ell_{h}}-\frac{\ell_{h}}{\sum_{h\in H}\ell_{h}}\Delta\ln\ell_{h}\right)\\
-\sum_{u\in U}\left(\frac{w_{u}\ell_{u}}{\sum_{u\in U}w_{u}\ell_{u}}\Delta\ln\Lambda_{\ell_{u}}-\frac{\ell_{u}}{\sum_{u\in U}\ell_{u}}\Delta\ln\ell_{u}\right),\label{eq: D.Wh/Wu}
\end{multline}
where $\Delta\ln\Lambda_{\ell_{f}}$ is given by equation \eqref{eq: D.WfLf/PY-Lf}.
The rate of change in the ratio of average wages between two groups
equals the difference between the weighted average of the rates of
change in the income shares of different types of labor, with each
type weighted by its wage bill share, and the weighted average of
the rates of change in the quantities of different types of labor,
with each type weighted by its quantity share.

\subsection{Effects of technological change\label{subsec: technological_change}}

The analysis thus far is useful for isolating the sources of changes
in the relative wages and income shares of different types of labor.
We can quantify the extent to which each factor contributes to changes
in relative wages and labor shares using equations \eqref{eq: D.WfLf/PY-Lf}
and \eqref{eq: D.Wf/Wg-Lf} for given values of the aggregate elasticities
of substitution. However, if the equilibrium prices and quantities
of factors can change in response to technological change, those equations
are insufficient to fully capture the effects of technological change.
Here, we derive the analytical expressions for changes in the equilibrium
relative wages and income shares of different types of labor due to
capital-embodied technological change and sector-specific disembodied
technological change. Throughout this and next subsections, we focus
on steady-state equilibria in which all variables are constant over
time and examine changes before and after information revolution.
We denote by $d\ln z$ the log change of variable $z$ from one steady
state to another.
\begin{prop}
\label{prop: D.Wf/Wg-Qf}Changes in relative wages due to technological
change are given by
\begin{equation}
d\ln\left(\frac{w_{f}}{w_{g}}\right)=\sum_{h=1}^{F_{k}}\left(\Upsilon_{\left(f,h\right)}^{q}-\Upsilon_{\left(g,h\right)}^{q}\right)d\ln q_{h}+\sum_{n=1}^{N}\left(\Upsilon_{\left(f,n\right)}^{A}-\Upsilon_{\left(g,n\right)}^{A}\right)d\ln A_{n}\quad\text{for }f,g=1,\ldots,F_{\ell},\label{eq: D.Wf/Wg-Qf}
\end{equation}
where $\Upsilon_{\left(f,g\right)}^{q}$ is the $fg$-th element of
the $F_{\ell}\times F_{k}$ matrix $\Upsilon^{q}$, and $\Upsilon_{\left(f,n\right)}^{A}$
is the $fn$-th element of the $F_{\ell}\times N$ matrix $\Upsilon^{A}$.
The matrices $\Upsilon^{q}$ and $\Upsilon^{A}$ are described in
Appendix \ref{sec: proofs}. Changes in labor shares due to technological
change are given by
\begin{multline}
d\ln\Lambda_{\ell_{f}}=\sum_{h=1}^{F_{k}}\left\{ \sum_{g\neq f}^{F}\left(1-\epsilon_{\ell_{g}\ell_{f}}^{\mathcal{C}}\right)\Lambda_{\ell_{g}}\Upsilon_{\left(f,h\right)}^{q}+\sum_{g\neq f}^{F_{\ell}}\left[-\left(1-\epsilon_{\ell_{f}\ell_{g}}^{\mathcal{C}}\right)\Lambda_{\ell_{g}}+\sum_{s\neq f,g}^{F}\left(\epsilon_{\ell_{f}\ell_{g}}^{\mathcal{C}}-\epsilon_{\ell_{s}\ell_{g}}^{\mathcal{C}}\right)\Lambda_{\ell_{s}}\right]\Upsilon_{\left(g,h\right)}^{q}\right.\\
\left.-\left[-\left(1-\epsilon_{\ell_{f}k_{h}}^{\mathcal{C}}\right)\Lambda_{k_{h}}+\sum_{s\neq f,h}^{F}\left(\epsilon_{\ell_{f}k_{h}}^{\mathcal{C}}-\epsilon_{\ell_{s}k_{h}}^{\mathcal{C}}\right)\Lambda_{\ell_{s}}\right]\right\} d\ln q_{h}+\sum_{n=1}^{N}\left\{ \Phi_{\left(f,n\right)}^{w}+\sum_{g\neq f}^{F}\left(1-\epsilon_{\ell_{g}\ell_{f}}^{\mathcal{C}}\right)\Lambda_{\ell_{g}}\Upsilon_{\left(f,n\right)}^{A}\right.\\
\left.+\sum_{g\neq f}^{F_{\ell}}\left[-\left(1-\epsilon_{\ell_{f}\ell_{g}}^{\mathcal{C}}\right)\Lambda_{\ell_{g}}+\sum_{s\neq f,g}^{F}\left(\epsilon_{\ell_{f}\ell_{g}}^{\mathcal{C}}-\epsilon_{\ell_{s}\ell_{g}}^{\mathcal{C}}\right)\Lambda_{\ell_{s}}\right]\Upsilon_{\left(g,n\right)}^{A}\right\} d\ln A_{n}\quad\text{for }f=1,\ldots,F_{\ell},\label{eq: D.WfLf/PY-Qf}
\end{multline}
where $\Phi^{w}$ is the $F\times N$ matrix whose $fn$-th element
is given by $\Phi_{\left(f,n\right)}^{w}=-[1-1/(1-\eta)]\zeta_{n}(\lambda_{\ell_{f}n}-\Lambda_{\ell_{f}})/\Lambda_{\ell_{f}}$.
\end{prop}
Equations \eqref{eq: D.Wf/Wg-Qf} and \eqref{eq: D.WfLf/PY-Qf} express
how the relative wages and income shares of different types of labor
can change in response to capital-embodied technological change and
sector-specific disembodied technological change in a multi-sector,
multi-factor general equilibrium model of investment-specific technological
change. The effect of capital-embodied technological change on relative
wages, which is represented by the first term in equation \eqref{eq: D.Wf/Wg-Qf},
is related to the factor complementarity effect in equation \eqref{eq: D.Wf/Wg-Lf}.
The effect of capital-embodied technological change on labor shares,
which is represented by the first term in equation \eqref{eq: D.WfLf/PY-Qf},
is related to the relative price effect and the factor complementarity
effect in equation \eqref{eq: D.WfLf/PY-Wf}. The first term in equation
\eqref{eq: D.WfLf/PY-Qf} can be divided into two parts, one in the
first line and the other in the second line. The two parts capture
the effects through changes in wages and rental prices, respectively.
Most elements of the matrix $\Upsilon^{q}$ consist of the aggregate
elasticities of substitution $\epsilon_{\ell_{f}\ell_{g}}^{\mathcal{C}}$,
and the others consist of the factor shares of income $\Lambda_{\ell_{f}}$.
Equations \eqref{eq: D.Wf/Wg-Qf} and \eqref{eq: D.WfLf/PY-Qf} show
that the aggregate elasticities of substitution are sufficient statistics
for quantifying the effects of technological change on the relative
wages and income shares of different types of labor if the income
shares of each factor are known.

We now consider two examples to demonstrate how the direction and
magnitude of the effects of technological change depend on the aggregate
elasticities of substitution among different types of capital and
labor.
\begin{cor}
\label{cor: Karabarbounis-Neiman}Suppose that production requires
only one type of capital, $k$, and one type of labor, $\ell$. Equation
\eqref{eq: D.WfLf/PY-Qf} becomes the \citet{Karabarbounis_Neiman_QJE14}
model.
\[
d\ln\Lambda_{\ell}=\left(1-\epsilon\right)\left(\frac{1-\Lambda_{\ell}}{\Lambda_{\ell}}\right)d\ln q+\sum_{n=1}^{N}\zeta_{n}\left[\left(1-\epsilon\right)\left(\frac{1-\Lambda_{\ell}}{\Lambda_{\ell}}\right)+\left(1-\frac{1}{1-\eta}\right)\left(\frac{\Lambda_{\ell}-\lambda_{\ell n}}{\Lambda_{\ell}}\right)\right]d\ln A_{n}.
\]
\end{cor}
This equation implies that the direction of the effect of capital-embodied
technological progress depends entirely on whether the aggregate elasticity
of substitution between capital and labor exceeds one. The labor share
can decline with capital-embodied technological progress if the aggregate
elasticity exceeds one, and the rate of decline becomes greater as
the aggregate elasticity increases. The magnitude of the effect also
becomes greater as the labor share decreases. The effect of sector-specific
disembodied technological change can be divided into two parts. The
first part in the square bracket takes the same form as the first
term and captures the effect through changes in wages. The second
part arises from the direct effect denoted by $\psi^{w}$ in equation
\eqref{eq: D.WfLf/PY-Wf} and is unlikely to be large unless factor
shares differ widely across sectors. Thus, the relative magnitude
of the first and second terms depends on that of the two types of
technological change.
\begin{cor}
\label{cor: D.Wh/Wu}Suppose that production requires capital equipment,
$k_{e}$, capital structures, $k_{s}$, skilled labor, $\ell_{h}$,
and unskilled labor, $\ell_{u}$, the aggregate elasticities of substitution
satisfy $\epsilon_{\ell_{h}k_{s}}^{\mathcal{C}}=\epsilon_{\ell_{u}k_{s}}^{\mathcal{C}}=1$
and $\epsilon_{\ell_{u}\ell_{h}}^{\mathcal{C}}-\epsilon_{\ell_{h}\ell_{u}}^{\mathcal{C}}=\epsilon_{\ell_{h}k_{e}}^{\mathcal{C}}-\epsilon_{\ell_{u}k_{e}}^{\mathcal{C}}$,
and there is no technological change embodied in capital structure,
as in \citet{Krusell_Ohanian_RiosRull_Violante_EM00}. Changes in
the skill premium due to technological change are given by
\begin{multline*}
d\ln\left(\frac{w_{h}}{w_{u}}\right)=\left(\frac{\epsilon_{\ell_{u}k_{e}}^{\mathcal{C}}-\epsilon_{\ell_{h}k_{e}}^{\mathcal{C}}}{\overline{\epsilon}_{\ell_{h}\ell_{u}}^{\mathcal{C}}}\right)\left(\frac{\Lambda_{k_{e}}+\Lambda_{\ell_{h}}+\Lambda_{\ell_{u}}}{\Lambda_{\ell_{h}}+\Lambda_{\ell_{u}}}\right)d\ln q_{e}\\
+\sum_{n=1}^{N}\left[-\left(\frac{\epsilon_{\ell_{u}k_{e}}^{\mathcal{C}}-\epsilon_{\ell_{h}k_{e}}^{\mathcal{C}}}{\overline{\epsilon}_{\ell_{h}\ell_{u}}^{\mathcal{C}}}\right)\left(\frac{\zeta_{n}}{\Lambda_{\ell_{h}}+\Lambda_{\ell_{u}}}\right)+\frac{1}{\overline{\epsilon}_{\ell_{h}\ell_{u}}^{\mathcal{C}}}\left(1-\frac{1}{1-\eta}\right)\left(\frac{\zeta_{n}\lambda_{\ell_{u}n}}{\Lambda_{\ell_{u}}}-\frac{\zeta_{n}\lambda_{\ell_{h}n}}{\Lambda_{\ell_{h}}}\right)\right]d\ln A_{n},
\end{multline*}
where $\overline{\epsilon}_{\ell_{h}\ell_{u}}^{\mathcal{C}}=[\Lambda_{\ell_{h}}/(\Lambda_{\ell_{h}}+\Lambda_{\ell_{u}})]\epsilon_{\ell_{h}\ell_{u}}^{\mathcal{C}}+[\Lambda_{\ell_{u}}/(\Lambda_{\ell_{h}}+\Lambda_{\ell_{u}})]\epsilon_{\ell_{u}\ell_{h}}^{\mathcal{C}}$,
which is known as the \citet*{McFadden_RES63} elasticity of substitution.
\end{cor}
This equation is a multi-sector, multi-factor general equilibrium
version of the \citet*{Krusell_Ohanian_RiosRull_Violante_EM00} model.
It implies that the direction and magnitude of the effect of capital-embodied
technological progress on the skill wage gap depend on the presence
and degree of capital\textendash skill complementarity. The skill
wage gap can increase with capital-embodied technological progress
if capital equipment is more complementary to skilled labor than to
unskilled labor (i.e., $\epsilon_{\ell_{u}k_{e}}^{\mathcal{F}}>\epsilon_{\ell_{h}k_{e}}^{\mathcal{F}}$).
The rate of the increase becomes greater as the difference between
the two elasticities or the income share of capital equipment increases.
The effect of sector-specific disembodied technological change on
the skill wage gap can be divided into two parts. The first part in
the square bracket takes a similar form to the effect of capital-embodied
technological change and captures the effect through a change in capital
equipment. However, this has the opposite sign to that of the first
term because sector-specific disembodied technological progress reduces
the need for the use of capital equipment. The second part is related
to the direct effect denoted by $\psi^{\ell}$ in equation \eqref{eq: D.Wf/Wg-Lf}.
The sign of the first part is negative in the presence of capital\textendash skill
complementarity, whereas that of the second part is likely to be positive.
In this setting, sector-specific disembodied technological progress
is unlikely to account for the widening of the skill wage gap.

\section{Data\label{sec: data}}

We have shown in Propositions \ref{prop: D.WfLf/PY-Lf} to \ref{prop: D.Wf/Wg-Qf}
that the aggregate elasticities of substitution and the factor shares
of income determine the direction and magnitude of changes in relative
wages and income shares of different types of labor attributable to
(i) changes in specific inputs, under which other inputs are held
constant, and (ii) changes in technologies, under which all inputs
are allowed to vary. This section begins by describing the data sources,
sample, and variables used to calculate the factor shares of income
and to estimate the aggregate elasticities of substitution. It then
documents trends in the relative prices, relative quantities, and
income shares of factors. The section ends by looking at the substitutability
between ICCT capital and the four types of labor.

\subsection{Data sources, sample, and variables\label{subsec: sample}}

The data used in the analysis are mainly from the EU KLEMS database,
which collects detailed and internationally comparable information
on the prices and quantities of capital and labor in major OECD countries.
Our analysis is based on the March 2008 version because it contains
the longest time series covering the 1980s, during which there were
dramatic changes in technology and inequality. For the years 1980
to 2005, we include all countries and years in the sample for which
required data are available. Our sample comprises 11 OECD countries:
Australia, Austria, the Czech Republic, Denmark, Finland, Germany,
Italy, Japan, the Netherlands, the United Kingdom, and the United
States. The economy of each country is divided into five sectors:
primary, manufacturing, utilities, construction, and services. The
sample comprises four non-primary sectors and includes 258 country-year
observations for each sector.\footnote{On average in our sample, the male share of employment is higher in
the construction and utilities sectors than in any subsector of manufacturing
or services.}

Labor is divided into skilled and unskilled labor, each of which is
further divided into male and female labor. Skilled labor consists
of workers who have completed college, and unskilled labor consists
of workers who do not enter or complete college. We calculate wages
by gender and skill by dividing the total labor compensation by the
total hours worked. When we calculate wages and hours worked, we adjust
for changes in the age and education composition of the labor force
over time. Appendix \ref{subsec: wages=000026hours} provides the
details of the adjustment procedure.

Capital is divided into ICCT and non-ICCT capital. ICCT capital consists
of computing equipment, communications equipment, and software. Non-ICCT
capital consists of transport equipment, other machinery and equipment,
non-residential structures and infrastructures, residential structures,
and other assets. We calculate the rental price of capital \citep{Jorgenson_AERPP63},
based on either the internal or external rate of return. The former
approach has the advantage of maintaining consistency between national
income and production accounts, whereas the latter approach has the
advantage of not requiring the assumption of competitive markets.
We employ the latter approach to examine the robustness of our results.
Appendix \ref{subsec: rental_price} provides the details of the calculation
procedure.

All variables measured in monetary values are converted into U.S.
dollars by the purchasing power parity index and deflated by the gross
value-added deflator with 1995 as the base year. The purchasing power
parity index is obtained from the 1997 Productivity Level Database.

In addition, the \citet{Barro_Lee_JDE13} database is used to calculate
the population size by gender and skill. The collective bargaining
coverage, the strictness of employment protection legislation, and
the level of minimum wages are obtained from OECD.Stat.\footnote{The collective bargaining coverage is measured by the percentage of
employees with the right to bargain. The strictness of employment
protection legislation is measured in terms of the regulation of individual
and collective dismissals of workers on regular contracts and the
regulation for hiring workers on temporary contracts.}

\subsection{Factor prices, quantities, and shares\label{subsec: trend}}

\begin{figure}[H]
\caption{Gender and skill premia\label{fig: premia_oecd}}

\begin{centering}
\subfloat[Male vs. female\label{fig: premia_oecd_gender}]{
\centering{}\includegraphics[scale=0.6]{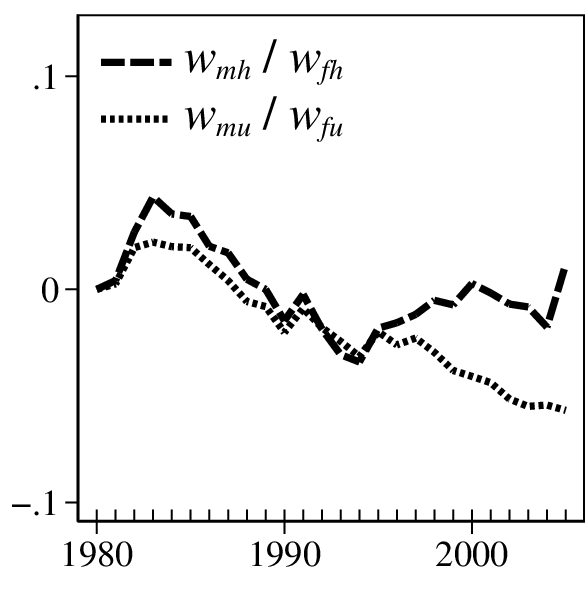}}\qquad{}\subfloat[Skilled vs. unskilled\label{fig: premia_oecd_skill}]{
\centering{}\includegraphics[scale=0.6]{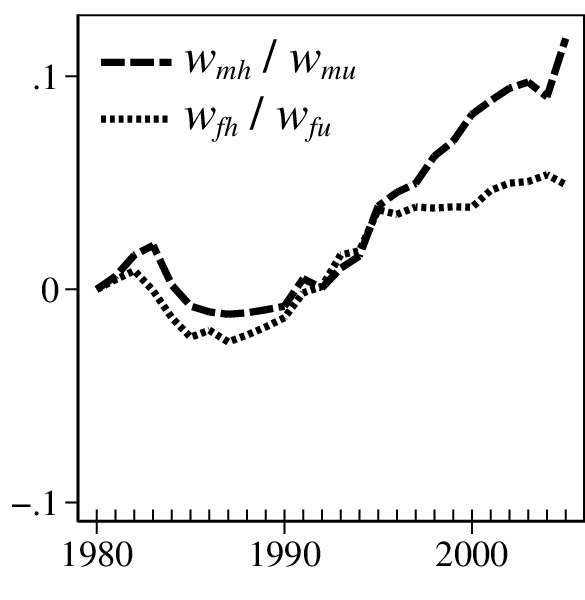}}
\par\end{centering}
\begin{singlespace}
\raggedright{}{\footnotesize\textit{Notes}}{\footnotesize : The wages
of male skilled, female skilled, male unskilled, and female unskilled
labor are denoted by $w_{mh}$, $w_{fh}$, $w_{mu}$, and $w_{fu}$,
respectively. All the series are logarithmically transformed and normalized
to zero in the year 1980. The 1980 values of the skilled gender wage
gap, $w_{mh}/w_{fh}$, unskilled gender wage gap, $w_{mu}/w_{fu}$,
male skill wage gap, $w_{mh}/w_{mu}$, and female skill wage gap,
$w_{fh}/w_{fu}$, are 1.35, 1.43, 1.59, and 1.67, respectively.}{\footnotesize\par}
\end{singlespace}
\end{figure}

\begin{figure}[H]
\caption{Gender and skill premia in the United States\label{fig: premia_us}}

\begin{centering}
\subfloat[Male vs. female\label{fig: premia_us_gender}]{
\centering{}\includegraphics[scale=0.6]{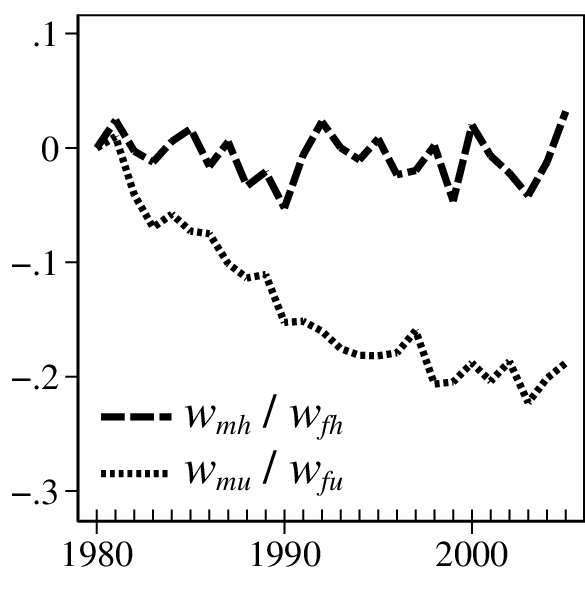}}\qquad{}\subfloat[Skilled vs. unskilled\label{fig: premia_us_skill}]{
\centering{}\includegraphics[scale=0.6]{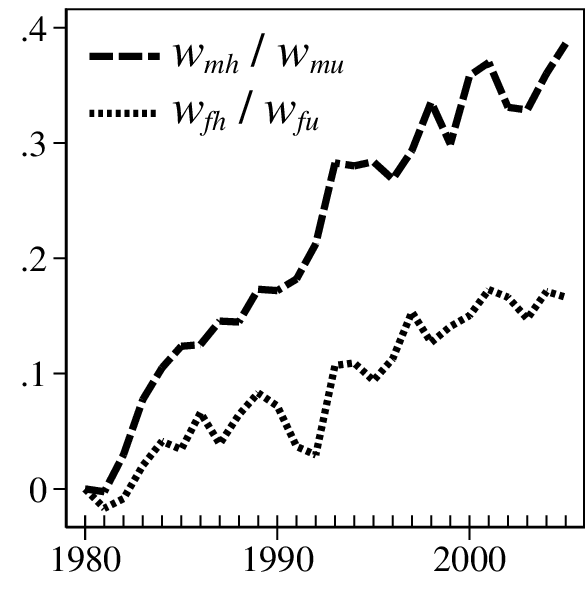}}
\par\end{centering}
\begin{singlespace}
\raggedright{}{\footnotesize\textit{Notes}}{\footnotesize : The wages
of male skilled, female skilled, male unskilled, and female unskilled
labor are denoted by $w_{mh}$, $w_{fh}$, $w_{mu}$, and $w_{fu}$,
respectively. All the series are logarithmically transformed and normalized
to zero in the year 1980. The 1980 values of the skilled gender wage
gap, $w_{mh}/w_{fh}$, unskilled gender wage gap, $w_{mu}/w_{fu}$,
male skill wage gap, $w_{mh}/w_{mu}$, and female skill wage gap,
$w_{fh}/w_{fu}$, are 1.49, 1.69, 1.39, and 1.57, respectively.}{\footnotesize\par}
\end{singlespace}
\end{figure}

During the period between the years 1980 and 2005, there was a difference
in the trends of the gender wage gap between skilled and unskilled
workers in OECD countries (Figure \ref{fig: premia_oecd_gender}).
The male\textendash female wage gap declined among unskilled workers,
but not among skilled workers. At the same time, there was a difference
in the trends of the skill wage gap between male and female workers
in OECD countries (Figure \ref{fig: premia_oecd_skill}). The skilled\textendash unskilled
wage gap increased among male workers after a slight drop in the early
1980s, while it did not increase among female workers except for the
period between the years 1987 and 1995. Taken together, the rate of
decrease in the male\textendash female wage gap was greater among
unskilled workers than among skilled workers, while the rate of increase
in the skilled\textendash unskilled wage gap was greater among male
workers than among female workers. This finding is perhaps surprising
but not inconsistent with the fact that the rate of decline in the
gender wage gap was greater at the middle or bottom than at the top
of the wage distribution in the United States from the 1980s to the
2000s \citep{Blau_Kahn_JEL17}. The patterns of changes in gender
and skill premia are more visible in the United States (Figures \ref{fig: premia_us_gender}
and \ref{fig: premia_us_skill}).\footnote{Since the mid-2000s, the widening of the skill wage gap has largely
stabilized, while the narrowing of the gender wage gap has decelerated
\citep*{Heathcote_Perri_Violante_Zhang_RED23}.} Across most countries in our sample, the rate of decline in the gender
wage gap tends to be greater among unskilled workers than among skilled
workers, while the rate of increase in the skill wage gap tends to
be greater among male workers than among female workers.\footnote{This pattern is observed in Australia, the Czech Republic, Denmark,
Germany, Japan, and the United States. For the rest of the countries,
there is no notable difference in the magnitude of changes in the
gender wage gap between skilled and unskilled workers and in the skill
wage gap between male and female workers.}

\begin{figure}[H]
\caption{Relative labor quantities\label{fig: labor_oecd}}

\begin{centering}
\subfloat[Male vs. female\label{fig: labor_oecd_gender}]{
\centering{}\includegraphics[scale=0.6]{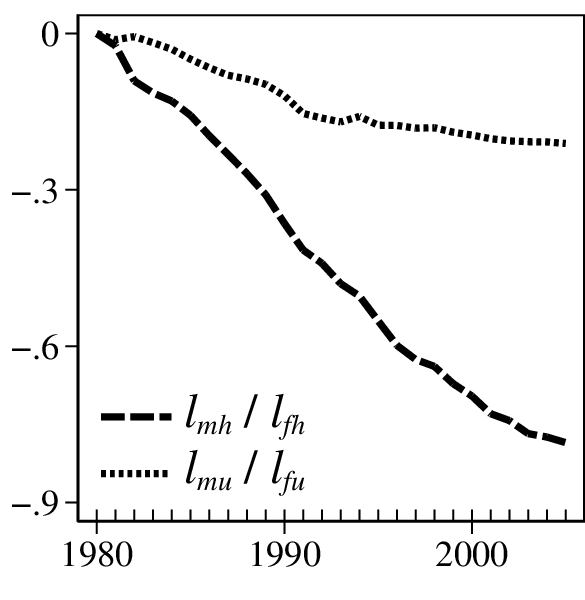}}\qquad{}\subfloat[Skilled vs. unskilled\label{fig: labor_oecd_skill}]{
\centering{}\includegraphics[scale=0.6]{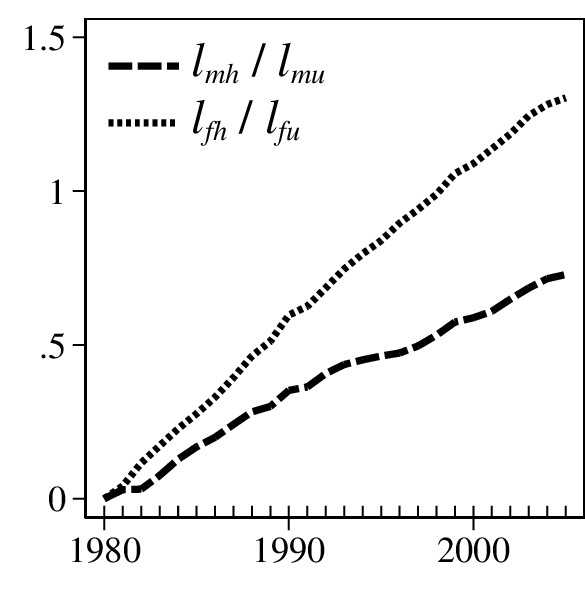}}
\par\end{centering}
\begin{singlespace}
\raggedright{}{\footnotesize\textit{Notes}}{\footnotesize : The quantities
of male skilled, female skilled, male unskilled, and female unskilled
labor are denoted by $\ell_{mh}$, $\ell_{fh}$, $\ell_{mu}$, and
$\ell_{fu}$, respectively. All the series are logarithmically transformed
and normalized to zero in the year 1980. The 1980 values of the skilled
male\textendash female ratio, $\ell_{mh}/\ell_{fh}$, unskilled male\textendash female
ratio, $\ell_{mu}/\ell_{fu}$, male skilled\textendash unskilled ratio,
$\ell_{mh}/\ell_{mu}$, and female skilled\textendash unskilled ratio,
$\ell_{fh}/\ell_{fu}$, are 4.66, 1.77, 0.15, and 0.09, respectively.}{\footnotesize\par}
\end{singlespace}
\end{figure}

Turning to the relative quantities of labor, there was a decline in
the quantity of male labor relative to female labor, and an increase
in the quantity of skilled labor relative to unskilled labor. The
rate of decline in the relative quantity of male labor was much greater
for skilled labor than for unskilled labor (Figure \ref{fig: labor_oecd_gender}).
At the same time, the rate of increase in the relative quantity of
skilled labor was much greater for female labor than for male labor
(Figure \ref{fig: labor_oecd_skill}). These observations suggest
that the relative supply of labor alone could not explain the differences
in the trends of the gender wage gap across skill groups and of the
skill wage gap across gender groups.

\begin{figure}[H]
\caption{ICCT and non-ICCT capital}

\begin{centering}
\subfloat[Investment prices\label{fig: rental_price}]{
\centering{}\includegraphics[scale=0.6]{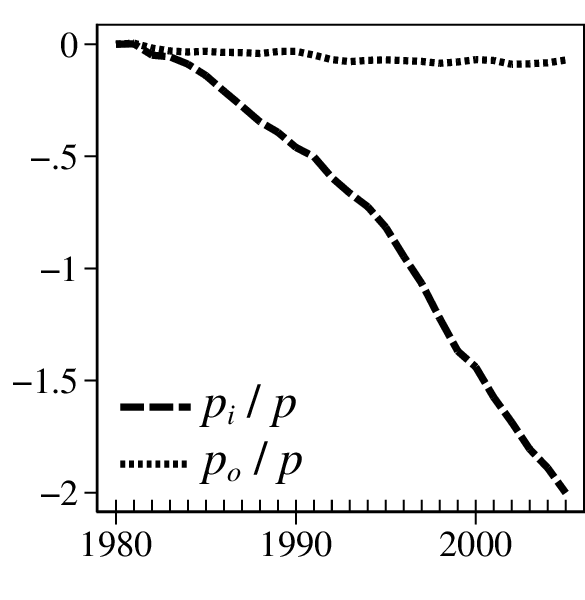}}\qquad{}\subfloat[Quantities\label{fig: capital}]{
\centering{}\includegraphics[scale=0.6]{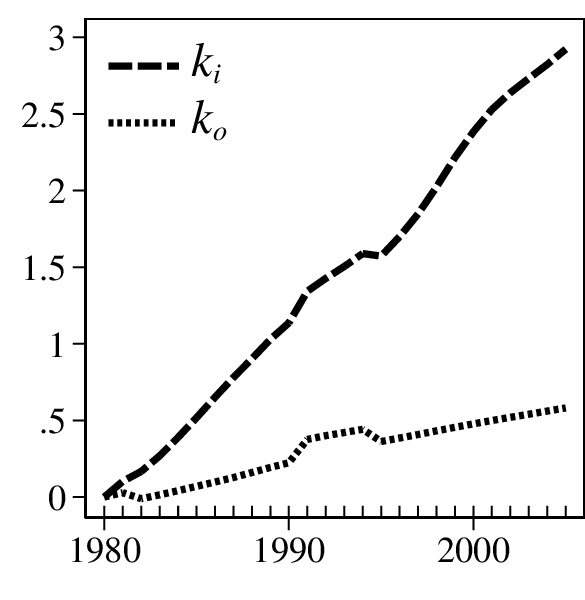}}
\par\end{centering}
\begin{singlespace}
\raggedright{}{\footnotesize\textit{Notes}}{\footnotesize : The investment
prices of ICCT and non-ICCT capital are denoted by $p_{i}$ and $p_{o}$,
respectively, and the output price is denoted by $p$. The quantities
of ICCT and non-ICCT capital are denoted by $k_{i}$ and $k_{o}$,
respectively. All the series are logarithmically transformed and normalized
to zero in the year 1980.}{\footnotesize\par}
\end{singlespace}
\end{figure}

The trends in the rental prices of capital differ significantly between
ICCT and non-ICCT capital (Figure \ref{fig: rental_price}). The rental
price of ICCT capital fell dramatically, but that of non-ICCT capital
remained almost unchanged. Meanwhile, the quantities of both ICCT
and non-ICCT capital increased. However, the rate of increase in ICCT
capital was far greater than that in non-ICCT capital (Figure \ref{fig: capital}).
These observations indicate substantial progress in ICCT during the
period.\footnote{Since the mid-2000s, the deepening of ICT capital has slowed markedly,
while the decline in ICT investment prices has flattened considerably
\citep*{Byrne_Corrado_IPM17,Sichel_ARE19,Goldin_Koutroumpis_Lafond_Winkler_JEL24}.} Appendix \ref{subsec: nonICT} shows that even if non-ICCT equipment
is distinguished from non-ICCT structures, there is no significant
difference in the trends of the prices and quantities between them.

\begin{figure}[h]
\caption{Income shares of the four types of labor\label{fig: labor_shares}}

\begin{centering}
\includegraphics[scale=0.6]{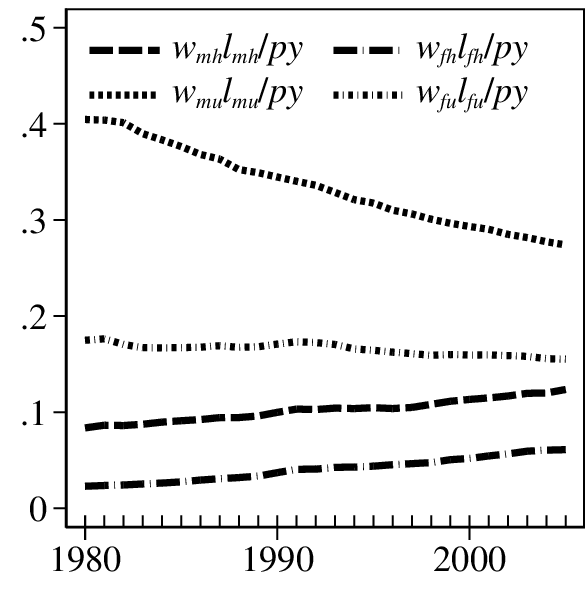}
\par\end{centering}
\begin{singlespace}
\raggedright{}{\footnotesize\textit{Notes}}{\footnotesize : The income
shares of male skilled labor, female skilled labor, male unskilled
labor, and female unskilled labor are denoted by $w_{mh}\ell_{mh}/py$,
$w_{fh}\ell_{fh}/py$, $w_{mu}\ell_{mu}/py$, and $w_{fu}\ell_{fu}/py$,
respectively.}{\footnotesize\par}
\end{singlespace}
\end{figure}

The direction and magnitude of changes in labor shares vary by gender
and skill (Figure \ref{fig: labor_shares}). The income shares of
male and female unskilled labor decreased, while the income shares
of male and female skilled labor increased. The magnitude of the decline
in the income share of male unskilled labor is substantially greater
than that of the changes in the income shares of the other three types
of labor. Taken together, the decline in the labor share is attributable
to the fall in the income share of male unskilled labor. This finding
is consistent with evidence of a decline in the income share of routine
labor \citep*{Eden_Gaggl_RED18}.

\subsection{Factor substitution\label{subsec: preliminary}}

Before we specify the sectoral production function, we conduct a preliminary
analysis to acquire a general idea of substitutability between ICCT
capital and the four types of labor. For labor of type $f$ in sector
$n$, we estimate the following equation:
\begin{equation}
\ln\left(\frac{k_{i,njt}}{\ell_{f,njt}}\right)=\varepsilon_{f,n}\ln\left(\frac{w_{f,njt}}{r_{i,njt}}\right)+\alpha_{nj}+\nu_{njt},\label{eq: Ki/Lf}
\end{equation}
where $\alpha_{nj}$ is a country-specific unobserved effect, $\nu_{njt}$
is the error term, and $j$ and $t$ are indices for countries and
years. The parameter of interest is $\varepsilon_{f,n}$, which captures
the percentage change in the capital\textendash labor ratio associated
with a one percent change in the relative price of labor to capital.
The greater $\varepsilon_{f,n}$ is, the more substitutable labor
of type $f$ is with ICCT capital in sector $n$.

We address two econometric issues: (i) unobserved heterogeneity and
(ii) simultaneity. First, we control for country-specific unobserved
effects by first differencing. In doing so, we take 5-year differences
to deal with the measurement error problem. Then, we make use of two
supply shifters as instrumental variables to identify equation \eqref{eq: Ki/Lf}
regarding the relative demand for factors. The first instrument is
the lagged population size of each demographic group. The second instrument
is the lagged price of ICCT investment relative to consumption, which
measures investment-specific technology accelerating capital accumulation.

\begin{sidewaystable}
\caption{Substitutability between ICCT capital and the four types of labor\label{tab: preliminary}}

\begin{centering}
\resizebox{1\textheight}{!}{%
\begin{tabular}{lcccccccccccccc}
\hline 
 & \multicolumn{4}{c}{OLS} &  & \multicolumn{4}{c}{First-difference GMM1} &  & \multicolumn{4}{c}{First-difference GMM2}\tabularnewline
 & $\varepsilon_{fh}$ & $\varepsilon_{mh}$ & $\varepsilon_{fu}$ & $\varepsilon_{mu}$ &  & $\varepsilon_{fh}$ & $\varepsilon_{mh}$ & $\varepsilon_{fu}$ & $\varepsilon_{mu}$ &  & $\varepsilon_{fh}$ & $\varepsilon_{mh}$ & $\varepsilon_{fu}$ & $\varsigma_{mu}$\tabularnewline
\cline{1-5}\cline{7-10}\cline{12-15}
 & \multicolumn{4}{c}{1980\textendash 2005} &  & \multicolumn{4}{c}{1980\textendash 2005} &  & \multicolumn{4}{c}{1980\textendash 2005}\tabularnewline
Manufacturing & 0.737 & 1.092 & 1.135 & 1.225 &  & 0.592 & 1.182 & 1.537 & 1.613 &  & 0.562 & 0.986 & 1.305 & 1.339\tabularnewline
 & (0.136) & (0.144) & (0.179) & (0.154) &  & \{0.378, 0.829\} & \{1.041, 1.364\} & \{1.298, 1.823\} & \{1.392, 1.893\} &  & \{0.373, 0.751\} & \{0.832, 1.160\} & \{1.050, 1.607\} & \{1.113, 1.604\}\tabularnewline
Utilities & 0.369 & 0.869 & 0.941 & 1.207 &  & 0.472 & 1.088 & 1.283 & 1.513 &  & 0.482 & 0.975 & 1.156 & 1.319\tabularnewline
 & (0.234) & (0.159) & (0.058) & (0.143) &  & \{0.321, 0.689\} & \{0.841, 1.513\} & \{1.044, 1.683\} & \{1.235, 2.009\} &  & \{0.347, 0.647\} & \{0.696, 1.339\} & \{0.886, 1.504\} & \{1.029, 1.692\}\tabularnewline
Construction & 0.728 & 1.016 & 1.015 & 1.004 &  & 1.107 & 1.828 & 1.819 & 2.027 &  & 0.923 & 1.373 & 1.456 & 1.583\tabularnewline
 & (0.338) & (0.445) & (0.201) & (0.254) &  & \{0.413, 2.047\} & \{1.109, 2.865\} & \{1.160, 2.753\} & \{1.368, 3.024\} &  & \{0.454, 1.558\} & \{0.874, 2.053\} & \{0.970, 2.123\} & \{1.092, 2.277\}\tabularnewline
Services & 0.652 & 0.929 & 0.993 & 1.137 &  & 0.713 & 1.109 & 1.350 & 1.554 &  & 0.687 & 0.963 & 1.189 & 1.306\tabularnewline
 & (0.148) & (0.106) & (0.079) & (0.077) &  & \{0.592, 0.858\} & \{0.978, 1.302\} & \{1.193, 1.573\} & \{1.344, 1.886\} &  & \{0.577, 0.798\} & \{0.843, 1.100\} & \{1.052, 1.345\} & \{1.136, 1.508\}\tabularnewline
\cline{2-5}\cline{7-10}\cline{12-15}
 & \multicolumn{4}{c}{1980\textendash 1995} &  & \multicolumn{4}{c}{1980\textendash 1995} &  & \multicolumn{4}{c}{1980\textendash 1995}\tabularnewline
Manufacturing & 0.969 & 1.092 & 0.962 & 1.062 &  & 0.599 & 1.564 & 1.967 & 2.232 &  & 0.592 & 1.398 & 1.836 & 2.039\tabularnewline
 & (0.349) & (0.370) & (0.264) & (0.294) &  & \{0.199, 1.009\} & \{1.200, 2.109\} & \{1.531, 2.545\} & \{1.650, 3.247\} &  & \{0.161, 0.996\} & \{1.083, 1.853\} & \{1.447, 2.366\} & \{1.542, 2.903\}\tabularnewline
Utilities & 0.277 & 0.581 & 0.751 & 0.987 &  & 0.490 & 1.397 & 1.586 & 2.047 &  & 0.545 & 1.540 & 1.743 & 2.222\tabularnewline
 & (0.376) & (0.302) & (0.126) & (0.217) &  & \{0.090, 1.000\} & \{0.899, 3.007\} & \{1.110, 2.709\} & \{1.401, 4.071\} &  & \{0.180, 0.929\} & \{1.152, 2.289\} & \{1.329, 2.498\} & \{1.643, 3.481\}\tabularnewline
Construction & 0.283 & 0.409 & 0.465 & 0.305 &  & 1.137 & 2.712 & 2.474 & 3.046 &  & 1.041 & 2.340 & 2.156 & 2.672\tabularnewline
 & (0.640) & (0.716) & (0.304) & (0.319) &  & \{0.023, 2.622\} & \{1.292, 5.372\} & \{1.251, 4.577\} & \{1.737, 5.588\} &  & \{\textendash 0.063, 2.410\} & \{0.985, 4.474\} & \{1.089, 3.820\} & \{1.571, 4.577\}\tabularnewline
Services & 0.492 & 0.588 & 0.784 & 0.783 &  & 0.662 & 1.450 & 1.668 & 2.170 &  & 0.655 & 1.441 & 1.668 & 2.152\tabularnewline
 & (0.348) & (0.253) & (0.114) & (0.146) &  & \{0.339, 0.950\} & \{1.172, 2.079\} & \{1.388, 2.254\} & \{1.695, 3.397\} &  & \{0.349, 0.928\} & \{1.176, 2.010\} & \{1.403, 2.158\} & \{1.716, 3.061\}\tabularnewline
\cline{2-5}\cline{7-10}\cline{12-15}
 & \multicolumn{4}{c}{1990\textendash 2005} &  & \multicolumn{4}{c}{1990\textendash 2005} &  & \multicolumn{4}{c}{1990\textendash 2005}\tabularnewline
Manufacturing & 0.765 & 1.048 & 0.880 & 0.965 &  & 0.576 & 0.986 & 1.298 & 1.296 &  & 0.548 & 0.891 & 1.180 & 1.183\tabularnewline
 & (0.152) & (0.164) & (0.185) & (0.160) &  & \{0.415, 0.763\} & \{0.872, 1.144\} & \{1.104, 1.579\} & \{1.134, 1.531\} &  & \{0.401, 0.700\} & \{0.753, 1.050\} & \{0.955, 1.461\} & \{0.990, 1.415\}\tabularnewline
Utilities & 0.281 & 0.735 & 0.656 & 0.968 &  & 0.440 & 0.950 & 1.118 & 1.260 &  & 0.457 & 0.877 & 1.037 & 1.157\tabularnewline
 & (0.278) & (0.113) & (0.160) & (0.161) &  & \{0.293, 0.630\} & \{0.700, 1.287\} & \{0.895, 1.435\} & \{1.034, 1.587\} &  & \{0.319, 0.629\} & \{0.603, 1.242\} & \{0.775, 1.386\} & \{0.891, 1.503\}\tabularnewline
Construction & 0.637 & 0.832 & 0.775 & 0.663 &  & 1.066 & 1.414 & 1.458 & 1.514 &  & 0.888 & 1.176 & 1.289 & 1.342\tabularnewline
 & (0.328) & (0.447) & (0.143) & (0.169) &  & \{0.440, 2.093\} & \{0.798, 2.335\} & \{0.885, 2.325\} & \{0.973, 2.345\} &  & \{0.488, 1.508\} & \{0.749, 1.789\} & \{0.848, 1.932\} & \{0.920, 1.971\}\tabularnewline
Services & 0.691 & 0.905 & 0.788 & 0.920 &  & 0.727 & 0.977 & 1.208 & 1.298 &  & 0.689 & 0.890 & 1.104 & 1.167\tabularnewline
 & (0.189) & (0.100) & (0.094) & (0.079) &  & \{0.580, 0.885\} & \{0.849, 1.153\} & \{1.057, 1.418\} & \{1.137, 1.543\} &  & \{0.564, 0.814\} & \{0.772, 1.019\} & \{0.978, 1.249\} & \{1.035, 1.322\}\tabularnewline
\hline 
\end{tabular}}
\par\end{centering}
\begin{singlespace}
\raggedright{}{\footnotesize\textit{Notes}}{\footnotesize : Estimates
of $\epsilon_{f,n}$ in equation \eqref{eq: Ki/Lf} for female skilled
labor ($f=fh$), male skilled labor ($f=mh$), female unskilled labor
($f=fu$), and male unskilled labor ($f=mu$) are reported. Standard
errors are in parentheses of the first four columns. The 95\% Anderson\textendash Rubin
confidence sets are in curly brackets of the last eight columns. Both
of them are robust to clustering at the country level. Lagged population
size and lagged ICCT investment price are used as instrumental variables
in the fifth to eighth columns and the ninth to the last columns,
respectively.}{\footnotesize\par}
\end{singlespace}
\end{sidewaystable}

Table \ref{tab: preliminary} presents the ordinary least squares
(OLS) and first-difference generalized method of moments (GMM) estimates
of $\varepsilon_{f,n}$ for female skilled labor ($f=fh$), male skilled
labor ($f=mh$), female unskilled labor ($f=fu$), and male unskilled
labor ($f=mu$) in manufacturing, utilities, construction, and services
sectors. Both estimates indicate that the degrees of substitution
between ICCT capital and the four types of labor are smaller in the
order of female skilled, male skilled, female unskilled, and male
unskilled labor for all sectors (i.e., $\varepsilon_{fh,n}<\varepsilon_{mh,n}<\varepsilon_{fu,n}<\varepsilon_{mu,n}$
for all $n$). The first-difference GMM estimates differ by the type
of labor more than the OLS estimates. In both cases, the differences
in the estimates by the type of labor are more significant for manufacturing,
utilities, and services sectors than for the construction sector.
Similar results are obtained for the 1980\textendash 1995 and 1990\textendash 2005
periods. Although the degrees of substitution between ICCT capital
and unskilled labor become smaller from the former to the latter period,
ICCT capital is more substitutable with male and female unskilled
labor than with male and female skilled labor. The results described
thus far are reinforced by the \citet*{Anderson_Rubin_AMS49} confidence
sets robust to weak instruments. The first-difference GMM estimates
remain largely unchanged regardless of whether the lagged population
size or the lagged investment price is used as an instrument. Overall,
ICCT capital is more complementary not only to skilled labor than
unskilled labor but also to female labor than male labor.

\section{Estimation\label{sec: estimation}}

This section describes the specification, identification, estimation,
and aggregation of the sectoral production function. We consider the
sectoral production function in which all four types of labor are
imperfect substitutes for one another and substitutable to varying
degrees with ICCT capital.

\subsection{Specification\label{subsec: specification}}

The economy consists of multiple sectors. In each sector, production
requires two types of capital (ICCT capital, $k_{i}$, and non-ICCT
capital, $k_{o}$) and four types of labor (male skilled labor, $\ell_{mh}$,
female skilled labor, $\ell_{fh}$, male unskilled labor, $\ell_{mu}$,
and female unskilled labor, $\ell_{fu}$). Building upon the analysis
by \citet{Fallon_Layard_JPE75} and \citet{Krusell_Ohanian_RiosRull_Violante_EM00}
and in Sections \ref{subsec: trend} and \ref{subsec: preliminary},
the sectoral production function is specified as
\begin{equation}
y_{n}=A_{n}k_{o,n}^{\theta_{o,n}}\left[\left(1-\theta_{mu,n}\right)B_{n}^{\sigma_{mu,n}}+\theta_{mu,n}\ell_{mu,n}^{\sigma_{mu,n}}\right]^{\frac{1-\theta_{o,n}}{\sigma_{mu,n}}},\label{eq: ces4}
\end{equation}
where
\begin{align}
B_{n}= & \left[\left(1-\theta_{fu,n}\right)C_{n}^{\sigma_{fu,n}}+\theta_{fu,n}\ell_{fu,n}^{\sigma_{fu,n}}\right]^{\frac{1}{\sigma_{fu,n}}},\label{eq: nestB}\\
C_{n}= & \left[\left(1-\theta_{mh,n}\right)D_{n}^{\sigma_{mh,n}}+\theta_{mh,n}\ell_{mh,n}^{\sigma_{mh,n}}\right]^{\frac{1}{\sigma_{mh,n}}},\label{eq: nestC}\\
D_{n}= & \left[\left(1-\theta_{fh,n}\right)k_{i,n}^{\sigma_{fh,n}}+\theta_{fh,n}\ell_{fh,n}^{\sigma_{fh,n}}\right]^{\frac{1}{\sigma_{fh,n}}}.\label{eq: nestD}
\end{align}
This production function involves four substitution parameters ($\sigma_{mh}$,
$\sigma_{fh}$, $\sigma_{mu}$, and $\sigma_{fu}$) that are less
than one and five share parameters ($\theta_{mh}$, $\theta_{fh}$,
$\theta_{mu}$, $\theta_{fu}$, and $\theta_{o}$) that lie between
zero and one for each sector. The production function exhibits capital\textendash skill
complementarity for male labor if $\sigma_{mu}>\sigma_{mh}$ and for
female labor if $\sigma_{fu}>\sigma_{fh}$, and capital\textendash gender
complementarity for skilled labor if $\sigma_{mh}>\sigma_{fh}$ and
for unskilled labor if $\sigma_{mu}>\sigma_{fu}$. We allow for different
degrees of capital\textendash skill complementarity for male and female
labor and different degrees of capital\textendash gender complementarity
for skilled and unskilled labor. In addition, we allow for different
degrees of capital\textendash skill\textendash gender complementarity
across sectors.

The multi-level nested CES production function \eqref{eq: ces4} is
not only analytically tractable enough to derive the general equilibrium
effects of technological change on the relative wages and income shares
of different types of labor but also flexible enough to allow for
different degrees of capital\textendash skill and capital\textendash gender
complementarities. There are three features in the specification of
the production function. First, ICCT capital is located inside the
CES function to allow for the possibility that technological change
embodied in ICCT capital can influence the gender wage gap and the
skill wage gap. There were dramatic changes in the price and quantity
of ICCT equipment (see Figure \ref{fig: capital} in the last section),
whereas there were no such changes in the prices and quantities of
non-ICCT equipment or structures (see Figure \ref{fig: capital'}
in Appendix \ref{subsec: nonICT}). Second, ICCT capital is placed
in the lowest nest in the four-level CES function to allow for the
possibility that the four types of labor may be substitutable to varying
degrees with ICCT capital. Otherwise, the rate of change in the gender
(skill) wage gap due to progress in ICCT would be the same for skilled
and unskilled (male and female) workers, which is not consistent with
the data. Finally, the production function is required to be concave
(i.e., the substitution parameters must be less than one) to satisfy
the economic principle that the relative price of skills should decrease
with a rise in the relative supply of skills. For this reason, skilled
labor is placed in a nest lower than that of unskilled labor, as in
\citet{Fallon_Layard_JPE75} and \citet{Krusell_Ohanian_RiosRull_Violante_EM00},
and female labor is placed in a nest lower than that of male labor
for each skill type. Our specification is consistent with key features
of the data described in Sections \ref{subsec: trend} and \ref{subsec: preliminary}.
Appendices \ref{subsec: specification1} and \ref{subsec: specification2}
provide further details on the assumptions, implications, and relevance
of alternative specifications.

\subsection{Identification\label{subsec: identification}}

Cost minimization implies that the marginal rate of technical substitution
equals the ratio of input prices irrespective of the degree of competition
in goods markets. For the purpose of our analysis, we make use of
five conditions, concerning (i) the relative price of female skilled
labor to ICCT capital, (ii) the skilled gender wage gap, (iii) the
female skill wage gap, (iv) the unskilled gender wage gap, and (v)
the male skill wage gap, to estimate the production function parameters.
The conditions lead to the following first-difference equations:
\begin{align}
\Delta\ln\left(\frac{w_{fh,n}}{r_{i,n}}\right)= & -\left(1-\sigma_{fh,n}\right)\Delta\ln\left(\frac{\ell_{fh,n}}{k_{i,n}}\right)+\Delta v_{fh,n},\label{eq: D.Wfh/Ri}\\
\Delta\ln\left(\frac{w_{mh,n}}{w_{fh,,n}}\right)= & -\left(1-\sigma_{mh,n}\right)\Delta\ln\ell_{mh,n}+\left(1-\sigma_{fh,n}\right)\Delta\ln\ell_{fh,n}\nonumber \\
 & -\left(\sigma_{mh,n}-\sigma_{fh,n}\right)\Delta\ln\widetilde{D}_{n}+\Delta v_{mh,n},\label{eq: D.Wmh/Wfh}\\
\Delta\ln\left(\frac{w_{fh,n}}{w_{fu,n}}\right)= & -\left(1-\sigma_{fh,n}\right)\Delta\ln\ell_{fh,n}+\left(1-\sigma_{fu,n}\right)\Delta\ln\ell_{fu,n}\nonumber \\
 & +\left(\sigma_{mh,n}-\sigma_{fh,n}\right)\Delta\ln\widetilde{D}_{n}+\left(\sigma_{fu,n}-\sigma_{mh,n}\right)\Delta\ln\widetilde{C}_{n}+\Delta v_{fu,n},\label{eq: D.Wfh/Wfu}\\
\Delta\ln\left(\frac{w_{mu,n}}{w_{fu,n}}\right)= & -\left(1-\sigma_{mu,n}\right)\Delta\ln\ell_{mu,n}+\left(1-\sigma_{fu,n}\right)\Delta\ln\ell_{fu,n}\nonumber \\
 & -\left(\sigma_{mu,n}-\sigma_{fu,n}\right)\Delta\ln\widetilde{B}_{n}+\Delta v_{mu1,n},\label{eq: D.Wmu/Wfu}\\
\Delta\ln\left(\frac{w_{mh,n}}{w_{mu,n}}\right)= & \left(1-\sigma_{mu,n}\right)\Delta\ln\ell_{mu,n}-\left(1-\sigma_{mh,n}\right)\Delta\ln\ell_{mh,n}\nonumber \\
 & +\left(\sigma_{fu,n}-\sigma_{mh,n}\right)\Delta\ln\widetilde{C}_{n}+\left(\sigma_{mu,n}-\sigma_{fu,n}\right)\Delta\ln\widetilde{B}_{n}+\Delta v_{mu2,n}.\label{eq: D.Wmh/Wmu}
\end{align}
where the error terms, $\Delta v$, are added to account for measurement
errors and shocks to non-neutral disembodied technologies, as detailed
in Appendix \ref{subsec: error_terms}, and the log changes of the
CES aggregates are approximated by the weighted average of the log
changes of factor inputs. It can be shown by Taylor expansion that
\begin{align*}
\Delta\ln D_{n}\simeq & \left(1-\varphi_{fh,n}\right)\Delta\ln k_{i,n}+\varphi_{fh,n}\Delta\ln\ell_{fh,n}\equiv\Delta\ln\widetilde{D}_{n},\\
\Delta\ln C_{n}\simeq & \left(1-\varphi_{mh,n}\right)\Delta\ln\widetilde{D}_{n}+\varphi_{mh,n}\Delta\ln\ell_{mh,n}\equiv\Delta\ln\widetilde{C}_{n},\\
\Delta\ln B_{n}\simeq & \left(1-\varphi_{fu,n}\right)\Delta\ln\widetilde{C}_{n}+\varphi_{fu,n}\Delta\ln\ell_{fu,n}\equiv\Delta\ln\widetilde{B}_{n},
\end{align*}
where the weights can be calculated directly from the data as $\varphi_{fh,n}=w_{fh,n}\ell_{fh,n}/(r_{i,n}k_{i,n}+w_{fh,n}\ell_{fh,n})$,
$\varphi_{mh,n}=w_{mh,n}\ell_{mh,n}/(r_{i,n}k_{i,n}+w_{fh,n}\ell_{fh,n}+w_{mh,n}\ell_{mh,n})$,
and $\varphi_{fu,n}=w_{fu,n}\ell_{fu,n}/(r_{i,n}k_{i,n}+w_{fh,n}\ell_{fh,n}+w_{mh,n}\ell_{mh,n}+w_{fu,n}\ell_{fu,n})$.\footnote{As a result of these approximations, the share parameters are eliminated
from the system of equations \eqref{eq: D.Wfh/Ri}\textendash \eqref{eq: D.Wmh/Wmu}.
The substitution parameters can be estimated solely using equations
\eqref{eq: D.Wfh/Ri}\textendash \eqref{eq: D.Wmh/Wmu}. However,
it is unnecessary to estimate the share parameters in our analysis
because they are not required to compute the aggregate elasticities
of substitution. Alternatively, the share and substitution parameters
can be estimated sequentially using the levels and log differences
of the marginal-rate-of-technical substitution equations. We confirm
that the approximations do not affect the estimates of the substitution
parameters by comparing the results obtained from both approaches.} Because all time-invariant factors are eliminated by first differencing,
the estimates of the system of equations \eqref{eq: D.Wfh/Ri}\textendash \eqref{eq: D.Wmh/Wmu}
are immune to the omitted variable bias due to unobserved country-specific
characteristics. Differences are taken over five years to avoid the
attenuation bias due to measurement error.

Because prices and quantities are jointly determined in markets, identifying
equations \eqref{eq: D.Wfh/Ri}\textendash \eqref{eq: D.Wmh/Wmu}
for factor demand requires using supply shifters as instrumental variables.
The instruments used are the lagged population size of each demographic
group and the lagged price of ICCT investment relative to consumption.
This means that we exploit the variation in past birth rates and university
enrollment capacity or the variation in investment-specific technology
across countries over time by taking advantage of cross-country panel
data. With either of the instruments, it is straightforward to see
that the substitution parameters $\sigma_{fh}$, $\sigma_{mh}$, $\sigma_{fu}$,
and $\sigma_{mu}$ can be identified sequentially from equations \eqref{eq: D.Wfh/Ri}\textendash \eqref{eq: D.Wmh/Wmu}
in the order of those in equations \eqref{eq: nestD}, \eqref{eq: nestC},
\eqref{eq: nestB}, and \eqref{eq: ces4} from the lowest to the highest
nest. The error terms are presumably correlated across equations \eqref{eq: D.Wfh/Ri}\textendash \eqref{eq: D.Wmh/Wmu}.
The substitution parameters in the system of equations \eqref{eq: D.Wfh/Ri}\textendash \eqref{eq: D.Wmh/Wmu}
are estimated jointly by GMM to improve efficiency. Standard errors
are clustered at the country level to account for heteroscedasticity
and serial correlation. In addition, we control for other factors
such as labor market institutions, disembodied factor-biased technological
change, and markups.

\subsection{Aggregation\label{subsec: aggregation}}

When the sectoral production function is of the nested CES form \eqref{eq: ces4},
the aggregate elasticity of substitution \eqref{eq: morishima_c}
can be expressed as the weighted average of the substitution parameters
in production and consumption:
\begin{align}
\epsilon_{\ell_{f}\ell_{g}}^{\mathcal{C}}= & \sum_{n=1}^{N}\pi_{\ell_{f}\ell_{g},n}^{fh}\left(\frac{1}{1-\sigma_{fh,n}}\right)+\sum_{n=1}^{N}\pi_{\ell_{f}\ell_{g},n}^{mh}\left(\frac{1}{1-\sigma_{mh,n}}\right)+\sum_{n=1}^{N}\pi_{\ell_{f}\ell_{g},n}^{fu}\left(\frac{1}{1-\sigma_{fu,n}}\right)+\sum_{n=1}^{N}\pi_{\ell_{f}\ell_{g},n}^{mu}\left(\frac{1}{1-\sigma_{mu,n}}\right)\nonumber \\
 & +\sum_{n=1}^{N}\pi_{\ell_{f}\ell_{g},n}^{o}+\sum_{n=1}^{N}\pi_{\ell_{f}\ell_{g},n}^{c}\left(\frac{1}{1-\eta}\right),\label{eq: morishima_c''}
\end{align}
where the sum of the weights on the substitution parameters adds up
to one. The weights on the substitution parameters depend on the income
shares of each factor, as detailed in Appendix \ref{subsec: weights}.
The elasticity of substitution in consumption is set at $1/(1-\eta)=0.9$,
as in \citet*{Baqaee_Farhi_JEEA19}.

After computing the aggregate elasticities of substitution using the
estimated substitution parameters, we can evaluate the effects of
technological change on, as well as the contributions of factor inputs
to, the relative wages and income shares of the four types of labor.
The direction and magnitude of the effects of technological change
embodied in ICCT capital depend on the presence and degree of capital\textendash skill
or capital\textendash gender complementarity (i.e., the sign and magnitude
of differences between the aggregate elasticities, $\epsilon_{\ell_{f}k_{i}}^{\mathcal{C}}-\epsilon_{\ell_{g}k_{i}}^{\mathcal{C}}$)
and the rate of progress in ICCT. The degrees of capital\textendash skill
and capital\textendash gender complementarities are approximately
proportional to the differences in the substitution parameters and
the income share of ICCT capital relative to the income shares of
each type of labor. Appendix \ref{subsec: GE_effect_diff} provides
the exact expressions for the effects of technological change on the
relative wages and income shares of the four types of labor.

\section{Results\label{sec: results}}

This section begins by presenting the estimates of sectoral production
function parameters and assessing the robustness of the estimates.
It then reports the estimates of the aggregate elasticities of substitution
and the quantitative contributions of factor inputs to the relative
wages and income shares of the four types of labor. The section ends
with findings on the effects of technological change.

\subsection{Sectoral results}

\subsubsection{Production function estimates}

Table \ref{tab: baseline} presents the GMM estimates of the substitution
parameters in the production functions for manufacturing, utilities,
construction and services sectors. For comparison it also presents
corresponding estimates under the assumption that the aggregate production
function takes the form of equation \eqref{eq: ces4}. Significant
differences are observed in the estimates of the four substitution
parameters for three out of four sectors. For manufacturing, utilities,
and services sectors, the null hypotheses of $\sigma_{fh}=\sigma_{mh}$,
$\sigma_{mh}=\sigma_{fu}$, and $\sigma_{fu}=\sigma_{mu}$ are rejected
at the 2 percent significance level (see Panel D of Table \ref{tab: specification}
in Appendix \ref{subsec: specification1}). The estimates suggest
that ICCT capital is more complementary not only to skilled labor
than unskilled labor but also to female labor than male labor (i.e.,
$\sigma_{fh}<\sigma_{mh}<\sigma_{fu}<\sigma_{mu}$). Moreover, the
null hypothesis that the substitution parameters are equal across
the three sectors is rejected at the 2 percent significance level.\footnote{The null hypotheses that the substitution parameters are equal between
the manufacturing and utilities sectors, between the utilities and
services sectors, and between the manufacturing and services sectors
are rejected at the 10 percent, 3 percent, and 13 percent significance
levels, respectively.} For the construction sector, the substitution parameters are estimated
under the null hypothesis of equal parameters (i.e., $\sigma_{fh}=\sigma_{mh}=\sigma_{fu}=\sigma_{mu}$)
because the null hypothesis cannot be rejected at the 39 percent significance
level.\footnote{On average in our sample, the shares of college graduates and women
each account for less than 10 percent in the construction sector.
This pattern is consistent with the absence of capital\textendash skill
and capital\textendash gender complementarities.} These results are not an artifact of an arbitrary nesting structure
of the CES production function. The relative magnitude of the estimated
degrees of factor substitution is consistent with that reported in
Table \ref{tab: preliminary}. The estimates remain similar regardless
of whether the lagged population size or the lagged investment price
is used as an instrument. The estimates obtained with the lagged population
size are used in the subsequent analyses because they are more precise.
When the substitution parameters are estimated under the assumption
that the aggregate production function for the non-primary sector
exists, the parameter estimates differ from those obtained for the
manufacturing, utilities, and construction sectors.

\begin{table}[h]
\caption{Production function estimates for each sector\label{tab: baseline}}

\begin{centering}
\begin{tabular}{lr@{\extracolsep{0pt}.}lr@{\extracolsep{0pt}.}lr@{\extracolsep{0pt}.}lr@{\extracolsep{0pt}.}lr@{\extracolsep{0pt}.}lr@{\extracolsep{0pt}.}lr@{\extracolsep{0pt}.}lr@{\extracolsep{0pt}.}lr@{\extracolsep{0pt}.}l}
\hline 
 & \multicolumn{8}{c}{GMM1} & \multicolumn{2}{c}{} & \multicolumn{8}{c}{GMM2}\tabularnewline
 & \multicolumn{2}{c}{$\sigma_{fh}$} & \multicolumn{2}{c}{$\sigma_{mh}$} & \multicolumn{2}{c}{$\sigma_{fu}$} & \multicolumn{2}{c}{$\sigma_{mu}$} & \multicolumn{2}{c}{} & \multicolumn{2}{c}{$\sigma_{fh}$} & \multicolumn{2}{c}{$\sigma_{mh}$} & \multicolumn{2}{c}{$\sigma_{fu}$} & \multicolumn{2}{c}{$\sigma_{mu}$}\tabularnewline
\cline{2-9}\cline{12-19}
Manufacturing & \textendash 0&689 & 0&288 & 0&600 & 0&776 & \multicolumn{2}{c}{} & \textendash 0&780 & 0&136 & 0&540 & 0&693\tabularnewline
 & (0&319) & (0&046) & (0&043) & (0&054) & \multicolumn{2}{c}{} & (0&303) & (0&074) & (0&069) & (0&052)\tabularnewline
Utilities & \textendash 0&962 & 0&184 & 0&404 & 0&631 & \multicolumn{2}{c}{} & \textendash 1&076 & 0&064 & 0&329 & 0&551\tabularnewline
 & (0&362) & (0&104) & (0&063) & (0&058) & \multicolumn{2}{c}{} & (0&324) & (0&158) & (0&084) & (0&051)\tabularnewline
Construction & 0&628 & 0&628 & 0&628 & 0&628 & \multicolumn{2}{c}{} & 0&506 & 0&506 & 0&506 & 0&506\tabularnewline
 & (0&077) & (0&077) & (0&077) & (0&077) & \multicolumn{2}{c}{} & (0&092) & (0&092) & (0&092) & (0&092)\tabularnewline
Services & \textendash 0&413 & 0&340 & 0&598 & 0&861 & \multicolumn{2}{c}{} & \textendash 0&456 & 0&203 & 0&561 & 0&789\tabularnewline
 & (0&134) & (0&067) & (0&034) & (0&048) & \multicolumn{2}{c}{} & (0&119) & (0&075) & (0&036) & (0&058)\tabularnewline
\cline{2-9}\cline{12-19}
Non-primary & \textendash 0&477 & 0&333 & 0&604 & 0&847 & \multicolumn{2}{c}{} & \textendash 0&517 & 0&193 & 0&560 & 0&784\tabularnewline
 & (0&171) & (0&061) & (0&036) & (0&047) & \multicolumn{2}{c}{} & (0&145) & (0&071) & (0&041) & (0&044)\tabularnewline
\hline 
\end{tabular}
\par\end{centering}
\raggedright{}{\footnotesize\textit{Notes}}{\footnotesize : Standard
errors in parentheses are clustered at the country level. Lagged population
size and lagged ICCT investment price are used as instrumental variables
in the first to fourth columns and the fifth to the last columns,
respectively. The production function parameters for the construction
sector are estimated under the restriction of $\sigma_{fh}=\sigma_{mh}=\sigma_{fu}=\sigma_{mu}$
because the restriction cannot be rejected.}{\footnotesize\par}
\end{table}

\subsubsection{Robustness checks}

Table \ref{tab: robustness} presents the robustness of the estimates
to controlling for the influence of labor market institutions, goods
market imperfections, and disembodied factor-biased technological
change. The first three rows report the estimates of the substitution
parameters after controlling for the influence of labor market institutions.
We allow for the possibility that the actual wage may deviate from
the competitive wage depending on the collective bargaining coverage,
the strictness of employment protection legislation, and the presence
and level of minimum wages. Specifically, we add the collective bargaining
coverage and the employment protection legislation in log-difference
form and the presence of minimum wages and its interaction with their
level in first-difference form to equations \eqref{eq: D.Wfh/Ri}\textendash \eqref{eq: D.Wmh/Wmu}.
The relative magnitude of the estimated substitution parameters remains
unchanged with the additional controls. Although the first three rows
of Table \ref{tab: robustness} report the estimates when the three
sets of variables are added individually to avoid the multicollinearity
problem, the relative magnitude of the estimated substitution parameters
remains unchanged even if they are added jointly.

The fourth row of Table \ref{tab: robustness} reports the estimates
of the substitution parameters after taking into account imperfect
competition in goods markets. We can relax the assumption of competitive
markets by recalculating the rental price of capital based on the
external rate of return because all estimating equations hold irrespective
of the degree of markup. The estimates of the substitution parameters
remain essentially unchanged.

The last row of Table \ref{tab: robustness} reports the estimates
of the substitution parameters after controlling for disembodied factor-biased
technological change. We take this into account by adding trend polynomials
with country-specific coefficients in equations \eqref{eq: D.Wfh/Ri}\textendash \eqref{eq: D.Wmh/Wmu}.
Trend polynomials capture changes in the direction and magnitude of
disembodied technological change, while country-specific coefficients
capture cross-country differences in the speed and timing of disembodied
technological change. We choose the order of the polynomials to fit
the trends in relative wages for each sector and country. Basically,
we include trend polynomials when they are statistically significant
at the 10 percent level. The estimates of the substitution parameters
remain essentially unchanged. The results are robust to country-specific
nonlinear time trends regardless of their causes even though the trends
could be attributable in part to other unobserved factors such as
discrimination and social norms.

\begin{table}[ph]
\caption{Production function estimates with additional controls\label{tab: robustness}}

\begin{centering}
\resizebox{1\textwidth}{!}{%
\begin{tabular}{lr@{\extracolsep{0pt}.}lr@{\extracolsep{0pt}.}lr@{\extracolsep{0pt}.}lr@{\extracolsep{0pt}.}lcr@{\extracolsep{0pt}.}lr@{\extracolsep{0pt}.}lr@{\extracolsep{0pt}.}lr@{\extracolsep{0pt}.}l}
\hline 
\multirow{1}{*}{} & \multicolumn{8}{c}{Manufacturing sector} &  & \multicolumn{8}{c}{Construction sector}\tabularnewline
 & \multicolumn{2}{c}{$\sigma_{fh}$} & \multicolumn{2}{c}{$\sigma_{mh}$} & \multicolumn{2}{c}{$\sigma_{fu}$} & \multicolumn{2}{c}{$\sigma_{mu}$} &  & \multicolumn{2}{c}{$\sigma_{fh}$} & \multicolumn{2}{c}{$\sigma_{mh}$} & \multicolumn{2}{c}{$\sigma_{fu}$} & \multicolumn{2}{c}{$\sigma_{mu}$}\tabularnewline
\cline{2-9}\cline{11-18}
Collective bargaining & \textendash 0&601 & 0&307 & 0&626 & 0&720 &  & 0&676 & 0&676 & 0&676 & 0&676\tabularnewline
\quad{}coverage & (0&408) & (0&074) & (0&047) & (0&065) &  & (0&074) & (0&074) & (0&074) & (0&074)\tabularnewline
Employment protection & \textendash 0&855 & 0&166 & 0&553 & 0&707 &  & 0&496 & 0&496 & 0&496 & 0&496\tabularnewline
\quad{}legislation & (0&328) & (0&056) & (0&055) & (0&065) &  & (0&082) & (0&082) & (0&082) & (0&082)\tabularnewline
Minimum wages & \textendash 0&668 & 0&387 & 0&635 & 0&784 &  & 0&592 & 0&592 & 0&592 & 0&592\tabularnewline
 & (0&439) & (0&057) & (0&053) & (0&094) &  & (0&071) & (0&071) & (0&071) & (0&071)\tabularnewline
Goods market & \textendash 0&928 & 0&191 & 0&559 & 0&733 &  & 0&518 & 0&518 & 0&518 & 0&518\tabularnewline
\quad{}imperfections & (0&368) & (0&061) & (0&049) & (0&058) &  & (0&087) & (0&087) & (0&087) & (0&087)\tabularnewline
Disembodied factor-biased & \textendash 0&722 & 0&336 & 0&542 & 0&854 &  & 0&674 & 0&674 & 0&674 & 0&674\tabularnewline
\quad{}technological change & (0&141) & (0&047) & (0&028) & (0&030) &  & (0&060) & (0&060) & (0&060) & (0&060)\tabularnewline
\cline{2-9}\cline{11-18}
 & \multicolumn{8}{c}{Utilities sector} &  & \multicolumn{8}{c}{Services sector}\tabularnewline
 & \multicolumn{2}{c}{$\sigma_{fh}$} & \multicolumn{2}{c}{$\sigma_{mh}$} & \multicolumn{2}{c}{$\sigma_{fu}$} & \multicolumn{2}{c}{$\sigma_{mu}$} &  & \multicolumn{2}{c}{$\sigma_{fh}$} & \multicolumn{2}{c}{$\sigma_{mh}$} & \multicolumn{2}{c}{$\sigma_{fu}$} & \multicolumn{2}{c}{$\sigma_{mu}$}\tabularnewline
\cline{2-9}\cline{11-18}
Collective bargaining & \textendash 0&891 & 0&204 & 0&375 & 0&566 &  & \textendash 0&458 & 0&282 & 0&584 & 0&809\tabularnewline
\quad{}coverage & (0&421) & (0&139) & (0&083) & (0&054) &  & (0&186) & (0&074) & (0&040) & (0&064)\tabularnewline
Employment protection & \textendash 1&344 & 0&152 & 0&348 & 0&594 &  & \textendash 0&563 & 0&246 & 0&567 & 0&828\tabularnewline
\quad{}legislation & (0&400) & (0&113) & (0&054) & (0&042) &  & (0&159) & (0&065) & (0&041) & (0&060)\tabularnewline
Minimum wages & \textendash 1&117 & 0&211 & 0&388 & 0&636 &  & \textendash 0&418 & 0&361 & 0&604 & 0&872\tabularnewline
 & (0&365) & (0&109) & (0&066) & (0&063) &  & (0&139) & (0&081) & (0&040) & (0&059)\tabularnewline
Goods market & \textendash 1&200 & 0&076 & 0&330 & 0&565 &  & \textendash 0&626 & 0&212 & 0&518 & 0&791\tabularnewline
\quad{}imperfections & (0&363) & (0&088) & (0&064) & (0&054) &  & (0&169) & (0&073) & (0&034) & (0&047)\tabularnewline
Disembodied factor-biased & \textendash 0&661 & 0&439 & 0&705 & 0&908 &  & \textendash 0&371 & 0&473 & 0&627 & 0&899\tabularnewline
\quad{}technological change & (0&246) & (0&132) & (0&081) & (0&050) &  & (0&095) & (0&086) & (0&049) & (0&043)\tabularnewline
\hline 
\end{tabular}}
\par\end{centering}
\begin{singlespace}
\raggedright{}{\footnotesize\textit{Notes}}{\footnotesize : Standard
errors in parentheses are clustered at the country level. Lagged population
size is used as an instrumental variable. The production function
parameters for the construction sector are estimated under the restriction
of $\sigma_{fh}=\sigma_{mh}=\sigma_{fu}=\sigma_{mu}$ because it cannot
be rejected.}{\footnotesize\par}
\end{singlespace}
\end{table}

\subsection{Aggregate results}

\subsubsection{Elasticities of substitution}

Table \ref{tab: elasticity} presents the estimates of the elasticities
of substitution in the aggregate production and cost functions. The
estimated elasticities of substitution between ICCT capital and the
four types of labor are greater in the order of male unskilled, female
unskilled, male skilled, and female skilled labor in both cases. The
elasticity estimates imply that the expansion of ICCT equipment would
narrow the gender wage gap and widen the skill wage gap. In this sense,
the estimates are consistent with the literature that points to technological
advances as causes of changes in the structure of wages and employment
\citep*{Autor_Levy_Murnane_QJE03,Black_SpitzOener_RESTAT10}. The
magnitude of the factor complementarity effects associated with a
rise in ICCT capital is directly proportional to the differences in
the elasticities of substitution between ICCT capital and the four
types of labor as well as the rate of increase in ICCT capital. The
elasticity estimates imply that the effects of ICCT capital on the
gender (skill) wage gap would be greater for skilled (female) labor
than for unskilled (male) labor. All else held constant, a one percent
increase in ICCT capital would reduce the skilled (unskilled) gender
wage gap by 0.30 (0.02) percent and raise the male (female) skill
wage gap by 0.08 (0.35) percent. At the same time, the magnitude of
the relative quantity effects is inversely proportional to the elasticities
of labor\textendash labor substitution and directly proportional to
the rates of increases in the quantities of labor. The elasticity
estimates imply that the effects of labor quantities would be greater
in the order of female skilled, male skilled, female unskilled, and
male unskilled labor.

\begin{table}[h]
\caption{Aggregate elasticities of substitution\label{tab: elasticity}}

\begin{centering}
\resizebox{1\textwidth}{!}{%
\begin{tabular}{r@{\extracolsep{0pt}.}lr@{\extracolsep{0pt}.}lr@{\extracolsep{0pt}.}lr@{\extracolsep{0pt}.}lr@{\extracolsep{0pt}.}lr@{\extracolsep{0pt}.}lr@{\extracolsep{0pt}.}lr@{\extracolsep{0pt}.}lr@{\extracolsep{0pt}.}lr@{\extracolsep{0pt}.}lr@{\extracolsep{0pt}.}lr@{\extracolsep{0pt}.}lr@{\extracolsep{0pt}.}lr@{\extracolsep{0pt}.}lr@{\extracolsep{0pt}.}l}
\hline 
\multicolumn{14}{c}{$\epsilon_{\ell_{f}\ell_{g}}^{\mathcal{F}}$} & \multicolumn{2}{c}{} & \multicolumn{14}{c}{$\epsilon_{\ell_{f}\ell_{g}}^{\mathcal{C}}$}\tabularnewline
\cline{1-14}\cline{17-30}
\multicolumn{2}{c}{} & \multicolumn{2}{c}{$k_{i}$} & \multicolumn{2}{c}{$\ell_{fh}$} & \multicolumn{2}{c}{$\ell_{mh}$} & \multicolumn{2}{c}{$\ell_{fu}$} & \multicolumn{2}{c}{$\ell_{mu}$} & \multicolumn{2}{c}{$k_{o}$} & \multicolumn{2}{c}{} & \multicolumn{2}{c}{} & \multicolumn{2}{c}{$k_{i}$} & \multicolumn{2}{c}{$\ell_{fh}$} & \multicolumn{2}{c}{$\ell_{mh}$} & \multicolumn{2}{c}{$\ell_{fu}$} & \multicolumn{2}{c}{$\ell_{mu}$} & \multicolumn{2}{c}{$k_{o}$}\tabularnewline
\multicolumn{2}{c}{$k_{i}$} & \multicolumn{2}{c}{} & 0&78 & 1&52 & 2&49 & 5&66 & 1&00 & \multicolumn{2}{c}{} & \multicolumn{2}{c}{$k_{i}$} & \multicolumn{2}{c}{} & 0&79 & 1&53 & 2&50 & 5&86 & 1&00\tabularnewline
\multicolumn{2}{c}{} & \multicolumn{2}{c}{} & (0&04) & (0&08) & (0&14) & (0&82) & (0&00) & \multicolumn{2}{c}{} & \multicolumn{2}{c}{} & \multicolumn{2}{c}{} & (0&04) & (0&09) & (0&14) & (0&99) & (0&00)\tabularnewline
\multicolumn{2}{c}{$\ell_{fh}$} & 0&76 & \multicolumn{2}{c}{} & 1&54 & 2&48 & 6&22 & 1&00 & \multicolumn{2}{c}{} & \multicolumn{2}{c}{$\ell_{fh}$} & 0&75 & \multicolumn{2}{c}{} & 1&52 & 2&47 & 5&93 & 1&00\tabularnewline
\multicolumn{2}{c}{} & (0&04) & \multicolumn{2}{c}{} & (0&09) & (0&14) & (1&44) & (0&00) & \multicolumn{2}{c}{} & \multicolumn{2}{c}{} & (0&04) & \multicolumn{2}{c}{} & (0&08) & (0&13) & (1&04) & (0&00)\tabularnewline
\multicolumn{2}{c}{$\ell_{mh}$} & 0&99 & 1&06 & \multicolumn{2}{c}{} & 2&50 & 5&80 & 1&00 & \multicolumn{2}{c}{} & \multicolumn{2}{c}{$\ell_{mh}$} & 1&12 & 1&20 & \multicolumn{2}{c}{} & 2&51 & 5&86 & 1&00\tabularnewline
\multicolumn{2}{c}{} & (0&04) & (0&05) & \multicolumn{2}{c}{} & (0&14) & (0&93) & (0&00) & \multicolumn{2}{c}{} & \multicolumn{2}{c}{} & (0&05) & (0&07) & \multicolumn{2}{c}{} & (0&14) & (0&99) & (0&00)\tabularnewline
\multicolumn{2}{c}{$\ell_{fu}$} & 1&05 & 1&12 & 1&96 & \multicolumn{2}{c}{} & 5&90 & 1&00 & \multicolumn{2}{c}{} & \multicolumn{2}{c}{$\ell_{fu}$} & 1&31 & 1&41 & 2&07 & \multicolumn{2}{c}{} & 5&89 & 1&00\tabularnewline
\multicolumn{2}{c}{} & (0&05) & (0&05) & (0&10) & \multicolumn{2}{c}{} & (1&01) & (0&00) & \multicolumn{2}{c}{} & \multicolumn{2}{c}{} & (0&06) & (0&07) & (0&10) & \multicolumn{2}{c}{} & (1&01) & (0&00)\tabularnewline
\multicolumn{2}{c}{$\ell_{mu}$} & 1&07 & 1&16 & 2&26 & 3&44 & \multicolumn{2}{c}{} & 1&00 & \multicolumn{2}{c}{} & \multicolumn{2}{c}{$\ell_{mu}$} & 1&67 & 1&83 & 3&07 & 4&11 & \multicolumn{2}{c}{} & 1&00\tabularnewline
\multicolumn{2}{c}{} & (0&05) & (0&06) & (0&13) & (0&27) & \multicolumn{2}{c}{} & (0&00) & \multicolumn{2}{c}{} & \multicolumn{2}{c}{} & (0&14) & (0&18) & (0&34) & (0&53) & \multicolumn{2}{c}{} & (0&00)\tabularnewline
\multicolumn{2}{c}{$k_{o}$} & 1&03 & 1&10 & 1&76 & 2&03 & 1&73 & \multicolumn{2}{c}{} & \multicolumn{2}{c}{} & \multicolumn{2}{c}{$k_{o}$} & 1&40 & 1&53 & 2&33 & 2&93 & 3&49 & \multicolumn{2}{c}{}\tabularnewline
\multicolumn{2}{c}{} & (0&04) & (0&05) & (0&07) & (0&07) & (0&04) & \multicolumn{2}{c}{} & \multicolumn{2}{c}{} & \multicolumn{2}{c}{} & (0&09) & (0&11) & (0&19) & (0&29) & (0&51) & \multicolumn{2}{c}{}\tabularnewline
\hline 
\end{tabular}}
\par\end{centering}
\begin{singlespace}
\raggedright{}{\footnotesize\textit{Notes}}{\footnotesize : The elasticities
of substitution are computed using equations \eqref{eq: morishima_y}
and \eqref{eq: morishima_c} of Lemma \ref{prop: epsilon_f =000026 epsilon_c}
and evaluated at the sample means across all countries and years.
Standard errors in parentheses are computed using bootstrap with 500
replications.}{\footnotesize\par}
\end{singlespace}
\end{table}

The estimated elasticities of labor\textendash labor substitution
in the aggregate production (cost) function range from 1.1 to 6.2
(1.2 to 5.9). The elasticity estimates are within the range of standard
estimates in the literature, although they are not strictly comparable
to those in other studies, such as \citet*{Acemoglu_Autor_Lyle_JPE04}
and \citet*{Johnson_Keane_JoLE13}, because of differences in the
degree of classification of labor. The estimated aggregate elasticities
of substitution between female skilled labor and the other three types
of labor are greater in the order of male unskilled, female unskilled,
and male skilled labor. Both female skilled and unskilled labor are
more complementary to male skilled labor than to male unskilled labor.

The \citet{Morishima_KH67} elasticities of substitution are asymmetric
by definition. The differences in the estimated elasticities of relative
demand for capital with respect to wages are greater than those in
the estimated elasticities of relative demand for labor with respect
to the rental prices of capital. However, in both cases, ICCT capital
is more complementary not only to skilled labor than unskilled labor
but also to female labor than male labor.

The estimated elasticity of substitution between ICCT capital and
male unskilled labor in the aggregate cost (production) function exceeds
one at $\epsilon_{\ell_{mu}k_{i}}^{\mathcal{C}}=1.67$ ($\epsilon_{\ell_{mu}k_{i}}^{\mathcal{F}}=1.07$).
ICCT capital is consistently more substitutable with male unskilled
labor than with the other three types of labor. The estimates suggest
that the income share of male unskilled labor is likely to decline
with a fall in the price or a rise in the quantity of ICCT capital.

\subsubsection{Contributions of factor inputs}

\paragraph{Gender and skill premia}

Table \ref{tab: relative_wages} presents the quantitative contributions
of specific factor inputs to changes in the gender premium by skill
and the skill premium by gender. The first column reports the observed
changes in gender and skill premia, while the second column reports
the sum of the changes attributable to factor inputs. These results
are obtained without adjusting any parameters to match relative wages
in the aggregate economy. The rate of decline in the gender wage gap
is greater among unskilled workers than among skilled workers, while
the rate of increase in the skill wage gap is greater among male workers
than among female workers. From this point of view, the observed pattern
of changes in gender and skill premia is consistent with the total
contribution of factor inputs.

\begin{table}[h]
\caption{Contributions of factor inputs to gender and skill premia, 1980\textendash 2005\label{tab: relative_wages}}

\begin{centering}
\resizebox{1\textwidth}{!}{%
\begin{tabular}{lr@{\extracolsep{0pt}.}lr@{\extracolsep{0pt}.}lr@{\extracolsep{0pt}.}lr@{\extracolsep{0pt}.}lr@{\extracolsep{0pt}.}lr@{\extracolsep{0pt}.}lr@{\extracolsep{0pt}.}lr@{\extracolsep{0pt}.}lr@{\extracolsep{0pt}.}lr@{\extracolsep{0pt}.}l}
\hline 
 & \multicolumn{2}{c}{Data} & \multicolumn{2}{c}{Total} & \multicolumn{2}{c}{FCE} & \multicolumn{2}{c}{RQE} & \multicolumn{2}{c}{$\Delta\ln k_{i}$} & \multicolumn{2}{c}{$\Delta\ln\ell_{fh}$} & \multicolumn{2}{c}{$\Delta\ln\ell_{mh}$} & \multicolumn{2}{c}{$\Delta\ln\ell_{fu}$} & \multicolumn{2}{c}{$\Delta\ln\ell_{mu}$} & \multicolumn{2}{c}{$\Delta\ln k_{o}$}\tabularnewline
\cline{2-21}
\multirow{2}{*}{$\Delta\ln\left(\frac{w_{mh}}{w_{fh}}\right)$} & 0&012 & 0&070 & \textendash 0&885 & 0&954 & \textendash 0&884 & 1&423 & \textendash 0&469 & \textendash 0&001 & 0&000 & 0&000\tabularnewline
 & \multicolumn{2}{c}{} & (0&050) & (0&070) & (0&047) & (0&070) & (0&067) & (0&027) & (0&001) & (0&000) & (0&000)\tabularnewline
\multirow{2}{*}{$\Delta\ln\left(\frac{w_{mu}}{w_{fu}}\right)$} & \textendash 0&057 & \textendash 0&103 & \textendash 0&164 & 0&061 & \textendash 0&070 & \textendash 0&045 & \textendash 0&049 & 0&059 & 0&001 & 0&000\tabularnewline
 & \multicolumn{2}{c}{} & (0&015) & (0&013) & (0&005) & (0&005) & (0&004) & (0&004) & (0&004) & (0&000) & (0&000)\tabularnewline
\multirow{2}{*}{$\Delta\ln\left(\frac{w_{mh}}{w_{mu}}\right)$} & 0&118 & 0&052 & 0&373 & \textendash 0&321 & 0&222 & 0&129 & \textendash 0&320 & 0&022 & \textendash 0&001 & 0&000\tabularnewline
 & \multicolumn{2}{c}{} & (0&019) & (0&023) & (0&019) & (0&014) & (0&010) & (0&019) & (0&002) & (0&000) & (0&000)\tabularnewline
\multirow{2}{*}{$\Delta\ln\left(\frac{w_{fh}}{w_{fu}}\right)$} & 0&049 & \textendash 0&121 & 1&136 & \textendash 1&257 & 1&036 & \textendash 1&340 & 0&100 & 0&082 & 0&000 & 0&000\tabularnewline
 & \multicolumn{2}{c}{} & (0&046) & (0&071) & (0&059) & (0&069) & (0&062) & (0&013) & (0&005) & (0&000) & (0&000)\tabularnewline
\cline{2-21}
\multirow{2}{*}{$\Delta\ln\left(\frac{\overline{w}_{M}}{\overline{w}_{F}}\right)$} & \textendash 0&083 & \textendash 0&111 & \textendash 0&307 & 0&196 & \textendash 0&228 & 0&147 & \textendash 0&094 & 0&063 & 0&001 & 0&000\tabularnewline
 & \multicolumn{2}{c}{} & (0&020) & (0&022) & (0&012) & (0&017) & (0&014) & (0&006) & (0&004) & (0&000) & (0&000)\tabularnewline
\multirow{2}{*}{$\Delta\ln\left(\frac{\overline{w}_{H}}{\overline{w}_{U}}\right)$} & 0&081 & \textendash 0&017 & 0&574 & \textendash 0&590 & 0&454 & \textendash 0&356 & \textendash 0&172 & 0&058 & 0&000 & 0&000\tabularnewline
 & \multicolumn{2}{c}{} & (0&022) & (0&030) & (0&028) & (0&024) & (0&016) & (0&011) & (0&002) & (0&000) & (0&000)\tabularnewline
\hline 
\end{tabular}}
\par\end{centering}
\raggedright{}{\footnotesize\textit{Notes}}{\footnotesize : The first
column reports the observed changes in the logarithms of relative
wages. The second column reports the sum of the factor complementarity
effect (FCE) in the third column and the relative quantity effect
(RQE) in the fourth column, or equivalently, the sum of the individual
contributions of factor inputs in the fifth to last columns. The figures
in the third to last columns are computed using equation \eqref{eq: D.Wf/Wg-Lf}
of Proposition \ref{prop: D.Wf/Wg-Lf}. Standard errors in parentheses
are computed using bootstrap with 500 replications.}{\footnotesize\par}
\end{table}

The third and fourth columns report the changes attributable to the
factor complementarity effect and the relative quantity effect. The
fifth to last columns report the changes attributable to each factor
input. This period shows a substantial increase in ICCT capital and
female skilled labor, a moderate increase in non-ICCT capital and
male skilled labor, a small increase in female unskilled labor, and
a small decrease in male unskilled labor. The increase in ICCT capital
contributes to narrowing the gender wage gap and widening the skill
wage gap. Meanwhile, the increase in male and female skilled labor
contributes to widening the gender wage gap and narrowing the skill
wage gap. Consequently, whether the gender (skill) wage gap narrows
(widens) depends on the outcome of the race between progress in ICCT
and advances in female employment (educational attainment). These
results hold not only for the four wage gaps (skilled gender wage
gap, unskilled gender wage gap, male skill wage gap, and female skill
wage gap) reported in the first four rows, but also for the two aggregate
wage gaps (aggregate gender wage gap and aggregate skill wage gap)
reported in the last two rows.

The factor complementarity effect associated with a rise in ICCT capital
is greater for the skilled gender wage gap and the female skill wage
gap, reflecting the result that ICCT capital is significantly more
complementary to female skilled labor than to the other three types
of labor. If all else were held constant, the expansion of ICCT equipment
during the period would have reduced the skilled gender wage gap by
88 percent and raised the female skill wage gap by 104 percent. At
the same time, the relative quantity effect is greater for the skilled
gender wage gap and the female skill wage gap, reflecting the fact
that the rate of increase in female skilled labor is significantly
greater than that in the other three types of labor. If all else were
held constant, a fall in the relative quantity of male skilled labor
to female skilled labor during the period would have raised the skilled
gender wage gap by 95 ($=(1.423-0.469)\times100$) percent, and a
rise in the relative quantity of female skilled labor to female unskilled
labor would have reduced the female skill wage gap by 126 ($=(\text{\textendash}1.340+0.082)\times100$)
percent. Consequently, the skilled gender wage gap did not narrow
as the two large effects canceled out. Similarly, the female skill
wage gap did not widen as the two large effects canceled out. Meanwhile,
the unskilled gender wage gap narrowed as the factor complementarity
effect exceeded the relative quantity effect. Similarly, the male
skill wage gap widened as the factor complementarity effect exceeded
the relative quantity effect. A rise in female skilled labor also
increased the male skill wage gap as female skilled labor was more
complementary to male skilled labor than to male unskilled labor.
Given the results that the factor complementarity effects on the skilled
gender wage gap and the female skill wage gap are so large, if there
were no change in the relative supply of female skilled labor, the
skilled gender wage gap would have narrowed more than the unskilled
gender wage gap, and the female skill wage gap would have widened
more than the male skill wage gap. Taken together, a smaller rate
of decline in the gender wage gap for skilled workers than for unskilled
workers and a smaller rate of increase in the skill wage gap for female
workers than for male workers are attributable not to a lack of demand
for female skilled labor but to an abundance in the supply of female
skilled labor.

\paragraph{Labor shares}

Changes in the labor share can be decomposed into those in the income
shares of the four types of labor $\ell_{f}$ for $f\in L=\left\{ mh,fh,mu,fu\right\} $,
as shown in equation \eqref{eq: D.WfLf/PY}.
\begin{equation}
\Delta\ln\left(\frac{\sum_{f\in L}w_{f}\ell_{f}}{py}\right)=\sum_{f\in L}\left(\frac{w_{f}\ell_{f}}{\sum_{f\in L}w_{f}\ell_{f}}\right)\Delta\ln\left(\frac{w_{f}\ell_{f}}{py}\right),\label{eq: D.WL/Y}
\end{equation}
where $py=\sum_{f\in L}w_{f}\ell_{f}+r_{i}k_{i}+r_{o}k_{o}$. We evaluate
the weights on the rates of changes in the income shares of each type
of labor at the sample means across all countries and years. Changes
in the labor share between the years 1980 and 2005 can be decomposed
as
\begin{eqnarray*}
\Delta\ln\left(\frac{\sum_{f\in L}w_{f}\ell_{f}}{py}\right) & = & \underset{\textrm{male skilled}}{\underbrace{0.165\times0.457}}+\underset{\textrm{female skilled}}{\underbrace{0.066\times1.230}}+\underset{\textrm{male unskilled}}{\underbrace{0.513\times\left(-0.389\right)}}+\underset{\textrm{female unskilled}}{\underbrace{0.256\times\left(-0.122\right)}}\\
 & = & 0.075+0.081-0.200-0.031.
\end{eqnarray*}
The labor share did not decline uniformly across gender and skill
groups. The income shares of male and female skilled labor increased,
while the income shares of male and female unskilled labor declined.
The reason for the decline in the labor share is that the majority
of labor was unskilled to whom the share of income decreased. The
decline in the labor share is attributable almost entirely to a fall
in the income share of male unskilled labor.

\begin{table}[h]
\caption{Contributions of factor inputs to labor shares, 1980\textendash 2005\label{tab: labor_shares}}

\begin{centering}
\resizebox{1\textwidth}{!}{%
\begin{tabular}{lr@{\extracolsep{0pt}.}lr@{\extracolsep{0pt}.}lr@{\extracolsep{0pt}.}lr@{\extracolsep{0pt}.}lr@{\extracolsep{0pt}.}lr@{\extracolsep{0pt}.}lr@{\extracolsep{0pt}.}lr@{\extracolsep{0pt}.}lr@{\extracolsep{0pt}.}lr@{\extracolsep{0pt}.}l}
\hline 
 & \multicolumn{2}{c}{Data} & \multicolumn{2}{c}{Total} & \multicolumn{2}{c}{FCE} & \multicolumn{2}{c}{RQE} & \multicolumn{2}{c}{$\Delta\ln k_{i}$} & \multicolumn{2}{c}{$\Delta\ln\ell_{fh}$} & \multicolumn{2}{c}{$\Delta\ln\ell_{mh}$} & \multicolumn{2}{c}{$\Delta\ln\ell_{fu}$} & \multicolumn{2}{c}{$\Delta\ln\ell_{mu}$} & \multicolumn{2}{c}{$\Delta\ln k_{o}$}\tabularnewline
\cline{2-21}
\multirow{2}{*}{$\Delta\ln\left(\frac{w_{mh}\ell_{mh}}{py}\right)$} & 0&457 & 0&438 & 0&147 & 0&291 & 0&092 & 0&050 & 0&312 & \textendash 0&019 & 0&003 & 0&000\tabularnewline
 & \multicolumn{2}{c}{} & (0&015) & (0&020) & (0&011) & (0&015) & (0&010) & (0&016) & (0&001) & (0&000) & (0&000)\tabularnewline
\multirow{2}{*}{$\Delta\ln\left(\frac{w_{fh}\ell_{fh}}{py}\right)$} & 1&230 & 1&153 & 1&030 & 0&123 & 0&976 & 0&134 & 0&059 & \textendash 0&018 & 0&003 & 0&000\tabularnewline
 & \multicolumn{2}{c}{} & (0&047) & (0&067) & (0&051) & (0&071) & (0&060) & (0&012) & (0&002) & (0&000) & (0&000)\tabularnewline
\multirow{2}{*}{$\Delta\ln\left(\frac{w_{mu}\ell_{mu}}{py}\right)$} & \textendash 0&389 & \textendash 0&343 & \textendash 0&259 & \textendash 0&084 & \textendash 0&130 & \textendash 0&079 & \textendash 0&091 & \textendash 0&041 & \textendash 0&003 & 0&000\tabularnewline
 & \multicolumn{2}{c}{} & (0&011) & (0&008) & (0&009) & (0&004) & (0&003) & (0&003) & (0&001) & (0&000) & (0&000)\tabularnewline
\multirow{2}{*}{$\Delta\ln\left(\frac{w_{fu}\ell_{fu}}{py}\right)$} & \textendash 0&122 & \textendash 0&029 & \textendash 0&086 & 0&058 & \textendash 0&060 & \textendash 0&034 & \textendash 0&042 & 0&104 & 0&003 & 0&000\tabularnewline
 & \multicolumn{2}{c}{} & (0&009) & (0&009) & (0&007) & (0&005) & (0&003) & (0&003) & (0&003) & (0&000) & (0&000)\tabularnewline
\cline{2-21}
\multirow{2}{*}{$\Delta\ln\left(\frac{\sum_{f}w_{f}\ell_{f}}{py}\right)$} & \textendash 0&075 & \textendash 0&035 & \textendash 0&063 & 0&028 & \textendash 0&003 & \textendash 0&032 & \textendash 0&002 & 0&001 & 0&000 & 0&000\tabularnewline
 & \multicolumn{2}{c}{} & (0&005) & (0&003) & (0&002) & (0&007) & (0&002) & (0&001) & (0&000) & (0&000) & (0&000)\tabularnewline
\hline 
\end{tabular}}
\par\end{centering}
\begin{singlespace}
\raggedright{}{\footnotesize\textit{Notes}}{\footnotesize : The first
column reports the observed changes in the logarithms of labor shares.
The second column reports the sum of the factor complementarity effect
(FCE) in the third column and the relative quantity effect (RQE) in
the fourth column, or equivalently, the sum of the individual contributions
of factor inputs in the fifth to last columns. The figures in the
third to last columns are computed using equation \eqref{eq: D.WfLf/PY-Lf}
of Proposition \ref{prop: D.WfLf/PY-Lf}. Standard errors in parentheses
are computed using bootstrap with 500 replications.}{\footnotesize\par}
\end{singlespace}
\end{table}

Table \ref{tab: labor_shares} presents the quantitative contributions
of specific factor inputs to changes in the income shares of the four
types of labor. The first column reports the observed changes in the
income shares of the four types of labor, while the second column
reports the sum of the changes attributable to factor inputs. These
results are obtained without adjusting any parameters to match labor
shares in the aggregate economy. The observed changes in the income
shares of the four types of labor align with the total contribution
of factor inputs. The total contribution to changes in the income
shares of male and female skilled labor are positive and statistically
significant, while that in the income shares of male and female unskilled
labor are negative and statistically significant.

The third and fourth columns report the changes attributable to the
factor complementarity effect and the relative quantity effect. The
two effects work in the same direction for the income shares of male
and female skilled labor and male unskilled labor. The former is much
greater than the latter for the income shares of female skilled labor
and male unskilled labor, although the converse is true for the income
share of male skilled labor. The two effects work in opposite directions
for the income share of female unskilled labor. However, the direction
of the factor complementarity effect aligns with that of the actual
change for the income shares of all types of labor. Overall, the factor
complementarity effect is essential for understanding changes in the
income shares of different types of labor. Both the factor complementarity
effect and the relative quantity effect are obscured when aggregated
across different types of labor, because the two effects tend to be
positive for skilled labor but negative for unskilled labor.

The fifth to last columns report the changes attributable to each
factor input. The expansion of ICCT equipment contributes significantly
to the increase in the income shares of male and female skilled labor
and the decrease in the income shares of male and female unskilled
labor. The magnitude of the effect depends on the extent to which
the rate of increase in the wages of male or female skilled (unskilled)
labor is greater (smaller) than that in the nominal output. A rise
in male and female skilled labor also contributes modestly to changes
in labor shares.

\subsubsection{Effects of technological change}

The first three columns of Table \ref{tab: GE_effect} present the
effects of three types of technological change on the relative wages
and income shares of the four types of labor. The effects of technological
change in Table \ref{tab: GE_effect} are measured in levels to facilitate
interpretation of the magnitude of the effects on each relative wage
and income share, whereas the contributions of factor inputs in Tables
\ref{tab: relative_wages} and \ref{tab: labor_shares} are measured
in logarithms to facilitate comparison of the contributions to the
four wage gaps or four income shares. The first to third columns report
the changes attributable to technological change embodied in ICCT
capital, technological change embodied in non-ICCT capital, and sector-specific
disembodied technological change, respectively. The effects of technological
change embodied in ICCT capital are significant, whereas those of
technological change embodied in non-ICCT capital and of sector-specific
disembodied technological change are marginal. These results hold
not only for the four wage gaps and the four labor shares but also
for the aggregate gender wage gap and the aggregate skill wage gap.
The direction of the effects of technological change embodied in ICCT
capital aligns with that of the effects of the expansion of ICCT equipment
in Tables \ref{tab: relative_wages} and \ref{tab: labor_shares},
reflecting the presence of capital\textendash skill\textendash gender
complementarity. The results indicate that a fall in the relative
price of ICCT investment owing to technological progress contributes
to a narrowing of the gender wage gap, a widening of the skill wage
gap, an increase in the income share of skilled labor, and a decline
in the income share of unskilled labor. The magnitude of the effect
on relative wages is largest for female skilled labor in percentage-point
terms, while that of the effect on labor shares is largest for male
unskilled labor. The small effects of technological change embodied
in non-ICCT capital reflect the lack of significant decrease in the
relative price of non-ICCT investment and of complementarity among
non-ICCT capital, skills, and gender. The small effects of sector-specific
disembodied technological change reflect the lack of significant change
in sector-specific disembodied technology.

\begin{table}[h]
\caption{Effects of technological change, 1980\textendash 2005\label{tab: GE_effect}}

\begin{centering}
\begin{tabular}{lr@{\extracolsep{0pt}.}lr@{\extracolsep{0pt}.}lr@{\extracolsep{0pt}.}lr@{\extracolsep{0pt}.}lr@{\extracolsep{0pt}.}lr@{\extracolsep{0pt}.}lr@{\extracolsep{0pt}.}l}
\hline 
 & \multicolumn{2}{c}{$\Delta\ln q_{i}$} & \multicolumn{2}{c}{$\Delta\ln q_{o}$} & \multicolumn{2}{c}{$\Delta\ln A_{n}$} & \multicolumn{2}{c}{} & \multicolumn{6}{c}{$\Delta\ln q_{i}$}\tabularnewline
\qquad{}$1/\gamma$ & \multicolumn{2}{c}{0} & \multicolumn{2}{c}{0} & \multicolumn{2}{c}{0} & \multicolumn{2}{c}{} & \multicolumn{2}{c}{1/2} & \multicolumn{2}{c}{1} & \multicolumn{2}{c}{2}\tabularnewline
\cline{1-7}\cline{10-15}
 & \multicolumn{6}{c}{Relative wages} & \multicolumn{2}{c}{} & \multicolumn{6}{c}{Relative wages}\tabularnewline
\multirow{2}{*}{$\Delta\left(\frac{w_{mh}}{w_{fh}}\right)$} & \textendash 0&886 & \textendash 0&015 & 0&048 & \multicolumn{2}{c}{} & \textendash 0&628 & \textendash 0&486 & \textendash 0&333\tabularnewline
 & (0&052) & (0&001) & (0&003) & \multicolumn{2}{c}{} & (0&042) & (0&036) & (0&027)\tabularnewline
\multirow{2}{*}{$\Delta\left(\frac{w_{mu}}{w_{fu}}\right)$} & \textendash 0&074 & \textendash 0&001 & 0&003 & \multicolumn{2}{c}{} & \textendash 0&090 & \textendash 0&096 & \textendash 0&098\tabularnewline
 & (0&006) & (0&000) & (0&001) & \multicolumn{2}{c}{} & (0&008) & (0&010) & (0&012)\tabularnewline
\multirow{2}{*}{$\Delta\left(\frac{w_{mh}}{w_{mu}}\right)$} & 0&262 & 0&004 & \textendash 0&015 & \multicolumn{2}{c}{} & 0&272 & 0&264 & 0&239\tabularnewline
 & (0&014) & (0&000) & (0&001) & \multicolumn{2}{c}{} & (0&013) & (0&013) & (0&015)\tabularnewline
\multirow{2}{*}{$\Delta\left(\frac{w_{fh}}{w_{fu}}\right)$} & 1&283 & 0&021 & \textendash 0&070 & \multicolumn{2}{c}{} & 0&958 & 0&767 & 0&550\tabularnewline
 & (0&053) & (0&001) & (0&003) & \multicolumn{2}{c}{} & (0&044) & (0&039) & (0&031)\tabularnewline
\cline{2-7}\cline{10-15}
\multirow{2}{*}{$\Delta\left(\frac{\overline{w}_{M}}{\overline{w}_{F}}\right)$} & \textendash 0&249 & \textendash 0&004 & 0&013 & \multicolumn{2}{c}{} & \textendash 0&226 & \textendash 0&210 & \textendash 0&186\tabularnewline
 & (0&014) & (0&000) & (0&001) & \multicolumn{2}{c}{} & (0&016) & (0&016) & (0&017)\tabularnewline
\multirow{2}{*}{$\Delta\left(\frac{\overline{w}_{H}}{\overline{w}_{U}}\right)$} & 0&564 & 0&009 & \textendash 0&031 & \multicolumn{2}{c}{} & 0&465 & 0&401 & 0&319\tabularnewline
 & (0&013) & (0&000) & (0&001) & \multicolumn{2}{c}{} & (0&015) & (0&016) & (0&017)\tabularnewline
\cline{2-7}\cline{10-15}
 & \multicolumn{6}{c}{Labor shares} & \multicolumn{2}{c}{} & \multicolumn{6}{c}{Labor shares}\tabularnewline
\multirow{2}{*}{$\Delta\left(\frac{w_{mh}\ell_{mh}}{py}\right)$} & 0&006 & 0&000 & 0&000 & \multicolumn{2}{c}{} & 0&010 & 0&014 & 0&019\tabularnewline
 & (0&001) & (0&000) & (0&000) & \multicolumn{2}{c}{} & (0&001) & (0&001) & (0&001)\tabularnewline
\multirow{2}{*}{$\Delta\left(\frac{w_{fh}\ell_{fh}}{py}\right)$} & 0&017 & 0&000 & \textendash 0&001 & \multicolumn{2}{c}{} & 0&019 & 0&020 & 0&022\tabularnewline
 & (0&001) & (0&000) & (0&000) & \multicolumn{2}{c}{} & (0&001) & (0&001) & (0&001)\tabularnewline
\multirow{2}{*}{$\Delta\left(\frac{w_{mu}\ell_{mu}}{py}\right)$} & \textendash 0&039 & \textendash 0&001 & 0&002 & \multicolumn{2}{c}{} & \textendash 0&056 & \textendash 0&070 & \textendash 0&091\tabularnewline
 & (0&002) & (0&000) & (0&000) & \multicolumn{2}{c}{} & (0&003) & (0&005) & (0&007)\tabularnewline
\multirow{2}{*}{$\Delta\left(\frac{w_{fu}\ell_{fu}}{py}\right)$} & \textendash 0&008 & 0&000 & 0&001 & \multicolumn{2}{c}{} & \textendash 0&008 & \textendash 0&007 & \textendash 0&004\tabularnewline
 & (0&001) & (0&000) & (0&000) & \multicolumn{2}{c}{} & (0&001) & (0&001) & (0&002)\tabularnewline
\cline{2-7}\cline{10-15}
\multirow{2}{*}{$\Delta\left(\frac{\sum_{f}w_{f}\ell_{f}}{py}\right)$} & \textendash 0&001 & 0&000 & 0&000 & \multicolumn{2}{c}{} & \textendash 0&006 & \textendash 0&009 & \textendash 0&013\tabularnewline
 & (0&004) & (0&000) & (0&000) & \multicolumn{2}{c}{} & (0&004) & (0&004) & (0&004)\tabularnewline
\hline 
\end{tabular}
\par\end{centering}
\begin{singlespace}
\raggedright{}{\footnotesize\textit{Notes}}{\footnotesize : The changes
in the levels of relative wages and labor shares attributable to technological
change in the first to third columns are computed using equations
\eqref{eq: D.Wf/Wg-Qf} and \eqref{eq: D.WfLf/PY-Qf} of Proposition
\ref{prop: D.Wf/Wg-Qf}, those in the fourth to last columns are computed
using equations \eqref{eq: D.Wf/Wg-Qf'} and \eqref{eq: D.WfLf/PY-Qf'}
of Proposition \ref{prop: D.Wf/Wg-Qf'} for each value of the labor
supply elasticity $1/\gamma$. Standard errors in parentheses are
computed using bootstrap with 500 replications.}{\footnotesize\par}
\end{singlespace}
\end{table}

The last three columns of Table \ref{tab: GE_effect} presents the
effects of technological change embodied in ICCT capital on the relative
wages and income shares of the four types of labor for each value
of the labor supply elasticity. For ease of reference, the first column
reproduces the first column of Table \ref{tab: GE_effect}, which
corresponds to the case in which the labor supply elasticity is zero.
Given the difficulty of selecting a single value of the aggregate
hours elasticity \citep*{Chetty_Guren_Manoli_Weber_AERPP11,Keane_Rogerson_JEL12},
the second to fourth columns show how the effects vary according to
the labor supply elasticity in the range of 0.5 to 2. The effects
of technological change embodied in ICCT capital on the skilled gender
wage gap and the female skill wage gap are greatest in absolute value
when the supply of labor is exogenous and become smaller in absolute
value as the supply of labor is more elastic. Progress in ICCT raises
the value of the marginal product of female skilled labor relative
to other types of labor, and thus, the relative wages of female skilled
labor. If the supply of labor is endogenous, the quantities of male
(female) skilled labor relative to female skilled (unskilled) labor
can decrease (increase) in response to progress in ICCT. The decrease
(increase) in the supply of male (female) skilled labor relative to
female skilled (unskilled) labor moderates a fall (rise) in the skilled
gender (female skill) wage gap. Such feedback effects become stronger
as the supply of labor is more elastic. However, even if the labor
supply elasticity is high at two, the effects of technological change
embodied in ICCT capital on the skilled gender wage gap and the female
skill wage gap remain greater than their observed changes. Moreover,
the effects of technological change embodied in ICCT capital on the
unskilled gender wage gap and the male skill wage gap do not vary
substantially depending on the value of the labor supply elasticity.
Even if the labor supply elasticity is high at two, the effects of
technological change embodied in ICCT capital on the unskilled gender
wage gap and the male skill wage gap are greater than their observed
changes. All else held constant, when the supply of labor is unit
elastic, the fall in the relative price of ICCT capital would have
reduced the skilled (unskilled) gender wage gap by 49 (10) percentage
points and raised the male (female) skill wage gap by 26 (77) percentage
points.

The effects of technological change embodied in ICCT capital on the
income shares of the four types of labor tend to be smallest in absolute
value when the supply of labor is exogenous and become greater in
absolute value as the supply of labor is more elastic. The rate of
change in the income share of labor of type $f$ can be decomposed
as $\Delta\ln\Lambda_{f}=\Delta\ln(w_{f}/p)+\Delta\ln\ell_{f}-\Delta\ln y$.
As the labor supply elasticity increases, the rate of increase in
equilibrium wages becomes smaller, whereas the rates of increases
in equilibrium quantities of labor and output become greater. The
effects of progress in ICCT on labor shares become greater (smaller)
as the labor supply elasticity increases if the rate of increase in
the equilibrium quantities of labor is large (small) relative to the
rate of decrease in the equilibrium wages plus the rate of increase
in the equilibrium quantities of output. The effects on the income
shares of male and female skilled labor increase due to a greater
increase in the equilibrium quantities of labor, whereas that on the
income shares of male unskilled labor decrease due to a smaller increase
in the equilibrium quantities of labor. The effect on the income share
of female unskilled labor is insensitive to the value of the labor
supply elasticity. All else held constant, the fall in the relative
price of ICCT capital would have raised the income share of male (female)
skilled labor by 1.4 (2.0) percentage points and reduced the income
share of male (female) unskilled labor by 7.0 (0.7) percentage points
in the case in which the supply of labor is unit elastic. In this
case, the fall in the relative price of ICCT capital accounts for
34 (53) percent of the increase in the income share of male (female)
skilled labor and 53 (34) percent of the decline in the income share
of male (female) unskilled labor. The effect of technological change
embodied in ICCT capital also becomes more visible as the labor supply
elasticity increases, although it is obscured when aggregated across
different types of labor.

Appendix \ref{sec: additional} presents three sets of additional
results on the effects of technological change. First, the magnitude
of the effects of technological change embodied in ICCT capital remains
largely unchanged regardless of whether labor is perfectly mobile
across sectors (Table \ref{tab: GE_effect_immobile}). Second, for
all countries, technological change embodied in ICCT capital narrows
the gender wage gap and widens the skill wage gap, while it increases
the income share of skilled labor and decreases the income share of
unskilled labor (Table \ref{tab: GE_effect_by_country}). Finally,
the magnitude of the changes in relative wages and income shares attributable
to ICCT capital may differ depending on whether they are obtained
by estimating and aggregating sectoral production function parameters
or by assuming and estimating an aggregate production function (Table
\ref{tab: GE_effect_aggregate}).

\section{Conclusion\label{sec: conclusion}}

We developed a unified framework to evaluate the effects of capital-embodied
technological change on, as well as the contributions of factor inputs
to, the relative wages and income shares of different types of labor.
We applied this framework to advanced economies consisting of multiple
sectors, where two types of capital (ICCT and non-ICCT capital) and
four types of labor (male skilled, female skilled, male unskilled,
and female unskilled labor) are used to produce goods and services.

Our results show that the expansion of ICCT equipment results in an
increase in the skill wage gap and a decline in the gender wage gap,
while the increased supply of male and female skilled labor prevents
an increase in the skill wage gap and a decline in the gender wage
gap. The observed patterns of changes in the four wage gaps (skilled
gender wage gap, unskilled gender wage gap, male skill wage gap, and
female skill wage gap) can be explained in terms of the race between
progress in ICCT and advances in educational attainment and female
employment. The unskilled gender wage gap is more likely to decline
than the skilled gender wage gap, while the male skill wage gap is
more likely to increase than the female skill wage gap. The reason
for this is not because the demand for female skilled labor is insufficient
but because the supply of female skilled labor becomes abundant. Our
results further show that the effects of technological change embodied
in ICCT equipment are sufficiently large to account for the narrowing
of the unskilled gender wage gap, the widening of the male skill wage
gap, and about half of the decline in the income share of male unskilled
labor that accounts for most of the decline in the labor share.

Our findings have implications for policies aimed at promoting female
employment, technological progress, and educational attainment. The
role of these policies is similar in that they could help raise national
income but different in that they could widen or narrow the gender
wage gap and the skill wage gap. The male\textendash female wage gap
would decrease as a result of policies that raise the relative demand
for female labor (e.g., tax credits for and grant funding to research
and development), whereas it would increase as a result of policies
that raise the relative supply of female labor (e.g., maternity leave,
childcare facilities, and the equal employment opportunity law). At
the same time, the skilled\textendash unskilled wage gap would decrease
as a result of policies that raise the relative supply of skilled
labor (e.g., tuition exemption and subsidies), whereas it would increase
as a result of policies that raise the relative demand for skilled
labor (e.g., tax credits for and grant funding to research and development).
Our findings suggest not only that policies promoting educational
attainment would be effective in reducing the skill wage gap but also
that policies promoting technological progress would be effective
in reducing the gender wage gap.

\clearpage{}

\bibliographystyle{econ}
\bibliography{../../references}

\clearpage{}

\appendix
\begin{center}
{\Large\textbf{Appendix}}{\Large\par}
\par\end{center}

\setcounter{equation}{0} \renewcommand{\theequation}{A\arabic{equation}} \renewcommand{\theHequation}{A\arabic{equation}}

\setcounter{table}{0} \renewcommand{\thetable}{A\arabic{table}}

\setcounter{figure}{0} \renewcommand{\thefigure}{A\arabic{figure}}

\section{Extension\label{sec: extension}}

\subsection{Endogenous labor supply}

We have shown how the equilibrium relative wages and income shares
of different types of labor can change in response to technological
change under the assumption that the supply of labor is exogenous.
In such a case, the equilibrium relative quantities of labor are invariant
to technological change. However, if the supply of labor is endogenous,
the equilibrium relative quantities of labor can change in response
to a change in the equilibrium relative wages due to technological
change. In such a general case, the equilibrium relative wages and
income shares can undergo further changes as a consequence of changes
in the relative quantities of labor. Here, we extend the framework
to allow for endogenous labor supply.

We consider a representative household that has preferences over sequences
of consumption $\{c_{fnt}\}_{t=0}^{\infty}$ and hours worked $\{l_{ft}\}_{t=0}^{\infty}$
characterized by
\[
\sum_{t=0}^{\infty}\beta^{t}\sum_{f=1}^{F_{\ell}}\mu_{f}\left[\mathcal{U}\left(c_{ft}\right)-\chi_{f}\frac{l_{ft}^{1+\gamma}}{1+\gamma}\right],
\]
where $\chi_{f}$ denotes the weight on the disutility of hours worked
by an individual of type $f$.
\begin{prop}
\label{prop: D.Wf/Wg-Qf'}Suppose that labor is supplied endogenously.
Changes in relative wages due to technological change are given by
\begin{equation}
d\ln\left(\frac{w_{f}}{w_{g}}\right)=\sum_{h=1}^{F_{k}}\left(\Xi_{\left(f,h\right)}^{q}-\Xi_{\left(g,h\right)}^{q}\right)d\ln q_{h}+\sum_{n=1}^{N}\left(\Xi_{\left(f,n\right)}^{A}-\Xi_{\left(g,n\right)}^{A}\right)d\ln A_{n}\quad\text{for }f,g=1,\ldots,F_{\ell},\label{eq: D.Wf/Wg-Qf'}
\end{equation}
where $\Xi_{\left(f,g\right)}^{q}$ is the $fg$-th element of the
$F_{\ell}\times F_{k}$ matrix $\Xi^{q}$, and $\Xi_{\left(f,n\right)}^{A}$
is the $fn$-th element of the $F_{\ell}\times N$ matrix $\Xi^{A}$.
The matrices $\Xi^{q}$ and $\Xi^{A}$ are described in Appendix \ref{sec: proofs}.
Changes in labor shares due to technological change are given by
\begin{multline}
d\ln\Lambda_{\ell_{f}}=\sum_{h=1}^{F_{k}}\left\{ \sum_{g\neq f}^{F}\left(1-\epsilon_{\ell_{g}\ell_{f}}^{\mathcal{C}}\right)\Lambda_{\ell_{g}}\Xi_{\left(f,h\right)}^{q}+\sum_{g\neq f}^{F_{\ell}}\left[-\left(1-\epsilon_{\ell_{f}\ell_{g}}^{\mathcal{C}}\right)\Lambda_{\ell_{g}}+\sum_{s\neq f,g}^{F}\left(\epsilon_{\ell_{f}\ell_{g}}^{\mathcal{C}}-\epsilon_{\ell_{s}\ell_{g}}^{\mathcal{C}}\right)\Lambda_{\ell_{s}}\right]\Xi_{\left(g,h\right)}^{q}\right.\\
\left.-\left[-\left(1-\epsilon_{\ell_{f}k_{h}}^{\mathcal{C}}\right)\Lambda_{k_{h}}+\sum_{s\neq f,h}^{F}\left(\epsilon_{\ell_{f}k_{h}}^{\mathcal{C}}-\epsilon_{\ell_{s}k_{h}}^{\mathcal{C}}\right)\Lambda_{\ell_{s}}\right]\right\} d\ln q_{h}+\sum_{n=1}^{N}\left\{ \Phi_{\left(f,n\right)}^{w}+\sum_{g\neq f}^{F}\left(1-\epsilon_{\ell_{g}\ell_{f}}^{\mathcal{C}}\right)\Lambda_{\ell_{g}}\Xi_{\left(f,n\right)}^{A}\right.\\
\left.+\sum_{g\neq f}^{F_{\ell}}\left[-\left(1-\epsilon_{\ell_{f}\ell_{g}}^{\mathcal{C}}\right)\Lambda_{\ell_{g}}+\sum_{s\neq f,g}^{F}\left(\epsilon_{\ell_{f}\ell_{g}}^{\mathcal{C}}-\epsilon_{\ell_{s}\ell_{g}}^{\mathcal{C}}\right)\Lambda_{\ell_{s}}\right]\Xi_{\left(g,n\right)}^{A}\right\} d\ln A_{n}\quad\text{for }f=1,\ldots,F_{\ell}.\label{eq: D.WfLf/PY-Qf'}
\end{multline}
\end{prop}
Equations \eqref{eq: D.Wf/Wg-Qf'} and \eqref{eq: D.WfLf/PY-Qf'}
take similar forms to equations \eqref{eq: D.Wf/Wg-Qf} and \eqref{eq: D.WfLf/PY-Qf},
respectively, and nest them as special cases with zero elasticity
of labor supply. Most elements of the matrix $\Xi^{q}$ consist of
the aggregate elasticities of substitution $\epsilon_{\ell_{f}\ell_{g}}^{\mathcal{C}}$
only or both the aggregate elasticities of substitution and the aggregate
elasticity of labor supply $1/\gamma$, and the others consist of
the factor shares of income $\Lambda_{\ell_{f}}$. Equations \eqref{eq: D.Wf/Wg-Qf'}
and \eqref{eq: D.WfLf/PY-Qf'} show that the aggregate elasticities
of substitution and the aggregate elasticity of labor supply are sufficient
statistics for quantifying the effects of technological change on
the relative wages and income shares of different types of labor if
the income shares of each factor are known.

Again, we consider two examples to demonstrate how the direction and
magnitude of the effects of technological change depend on the aggregate
elasticities of substitution among different types of capital and
labor and the aggregate elasticity of labor supply. While Corollary
\ref{cor: Karabarbounis-Neiman} holds true regardless of whether
the supply of labor is exogenous or endogenous, Corollary \ref{cor: D.Wh/Wu}
can be generalized as follows.
\begin{cor}
\label{cor:  D.Wh/Wu'}Suppose that production requires capital equipment,
$k_{e}$, capital structures, $k_{s}$, skilled labor, $\ell_{h}$,
and unskilled labor, $\ell_{u}$, the aggregate elasticities of substitution
satisfy $\epsilon_{\ell_{h}k_{s}}^{\mathcal{C}}=\epsilon_{\ell_{u}k_{s}}^{\mathcal{C}}=1$
and $\epsilon_{\ell_{u}\ell_{h}}^{\mathcal{C}}-\epsilon_{\ell_{h}\ell_{u}}^{\mathcal{C}}=\epsilon_{\ell_{h}k_{e}}^{\mathcal{C}}-\epsilon_{\ell_{u}k_{e}}^{\mathcal{C}}$,
and there is no technological change embodied in capital structure,
as in \citet{Krusell_Ohanian_RiosRull_Violante_EM00}. Changes in
the skill premium due to technological change are given by
\begin{multline*}
d\ln\left(\frac{w_{h}}{w_{u}}\right)=\left(\frac{\epsilon_{\ell_{u}k_{e}}^{\mathcal{C}}-\epsilon_{\ell_{h}k_{e}}^{\mathcal{C}}}{\overline{\epsilon}_{\ell_{h}\ell_{u}}^{\mathcal{C}}+1\left/\gamma\right.}\right)\left(\frac{\Lambda_{k_{e}}+\Lambda_{\ell_{h}}+\Lambda_{\ell_{u}}}{\Lambda_{\ell_{h}}+\Lambda_{\ell_{u}}}\right)d\ln q_{e}\\
+\sum_{n=1}^{N}\left[\left(\frac{1}{\overline{\epsilon}_{\ell_{h}\ell_{u}}^{\mathcal{C}}+1\left/\gamma\right.}\right)\left(1-\frac{1}{1-\eta}\right)\left(\frac{\zeta_{n}\lambda_{\ell_{u}n}}{\Lambda_{\ell_{u}}}-\frac{\zeta_{n}\lambda_{\ell_{h}n}}{\Lambda_{\ell_{h}}}\right)-\left(\frac{\epsilon_{\ell_{u}k_{e}}^{\mathcal{C}}-\epsilon_{\ell_{h}k_{e}}^{\mathcal{C}}}{\overline{\epsilon}_{\ell_{h}\ell_{u}}^{\mathcal{C}}+1\left/\gamma\right.}\right)\left(\frac{\zeta_{n}}{\Lambda_{\ell_{h}}+\Lambda_{\ell_{u}}}\right)\right]d\ln A_{n}.
\end{multline*}
\end{cor}
The direction of the effect of technological progress embodied in
capital equipment on the skill wage gap remains the same regardless
of whether the supply of labor is exogenous or endogenous. However,
the magnitude of the effect becomes smaller as the labor supply elasticity
increases.

\subsection{Imperfect labor mobility}

We have shown how the equilibrium relative wages and income shares
of different types of labor can change in response to technological
change under the assumption that labor is perfectly mobile across
sectors. This assumption may be reasonable in the long run and useful
for expressing the aggregate output as a function of the aggregate
factor inputs. However, it may be costly for workers to switch jobs
across sectors \citep{Lee_Wolpin_EM06}, and sectoral wages are not
fully equalized in the data. Here, we relax the assumption of perfect
labor mobility across sectors and consider the equilibrium in which
the demand for factor $f$ equals the supply of factor $f$ for each
sector: $\ell_{fn}=\mu_{f}l_{fn}$. In this case, the equilibrium
wages vary across sectors. The gender (skill) wage gap is measured
by the ratio of the average wages of male (skilled) labor to those
of female (unskilled) labor. The average wages of labor $f$ is defined
as $\overline{w}_{f}=\sum_{n=1}^{N}w_{fn}\ell_{fn}/\sum_{n=1}^{N}\ell_{fn}$,
where $w_{fn}$ is the wages of labor $f$ in sector $n$. The income
share of factor $f$ in the aggregate economy can be written as $\Lambda_{\ell_{f}}=\sum_{n=1}^{N}\Lambda_{\ell_{f}n}$,
where the income share of sector $n$'s factor $f$ in the aggregate
economy is defined as $\Lambda_{\ell_{f}n}=w_{fn}\ell_{fn}/py$.
\begin{prop}
\label{prop: D.Wf/Wg-Qf''}Suppose that labor is not perfectly mobile
across sectors. Changes in the relative wages due to technological
change are given by
\begin{multline}
d\ln\left(\frac{\overline{w}_{f}}{\overline{w}_{g}}\right)=\sum_{h=1}^{F_{k}}\sum_{n=1}^{N}\left(\frac{\Lambda_{\ell_{f}n}}{\Lambda_{\ell_{f}}}\Omega_{f\left(n,h\right)}^{q}-\frac{\Lambda_{\ell_{g}n}}{\Lambda_{\ell_{g}}}\Omega_{g\left(n,h\right)}^{q}\right)d\ln q_{h}\\
+\sum_{n=1}^{N}\sum_{m=1}^{N}\left(\frac{\Lambda_{\ell_{f}n}}{\Lambda_{\ell_{f}}}\Omega_{f\left(n,m\right)}^{A}-\frac{\Lambda_{\ell_{g}n}}{\Lambda_{\ell_{g}}}\Omega_{g\left(n,m\right)}^{A}\right)d\ln A_{m}\quad\text{for }f,g=1,\ldots,F_{\ell},\label{eq: D.Wf/Wg-Qf''}
\end{multline}
where $\Omega_{f(n,g)}^{q}$ is the $ng$-th element of the $N\times F_{k}$
matrix $\Omega_{f}^{q}$, and $\Omega_{f(n,m)}^{A}$ is the $nm$-th
element of the $N\times N$ matrix $\Omega_{f}^{A}$. The matrices
$\Omega_{f}^{q}$ and $\Omega_{f}^{A}$ are described in the proof
below. Changes in labor shares due to technological change are given
by
\begin{multline}
d\ln\Lambda_{\ell_{f}}=\sum_{h=1}^{F_{k}}\sum_{n=1}^{N}\sum_{m=1}^{N}\frac{\Lambda_{\ell_{f}n}}{\Lambda_{\ell_{f}}}\left(\sum_{g=1}^{F_{\ell}}\Pi_{UL,fg\left(n,m\right)}^{w}\Omega_{g\left(m,h\right)}^{q}-\Pi_{UR,fh\left(n,m\right)}^{w}\right)d\ln q_{h}\\
+\sum_{n=1}^{N}\sum_{m=1}^{N}\frac{\Lambda_{\ell_{f}n}}{\Lambda_{\ell_{f}}}\left(\sum_{g=1}^{F_{\ell}}\sum_{l=1}^{N}\Pi_{UL,fg\left(n,l\right)}^{w}\Omega_{g\left(l,m\right)}^{A}+\Theta_{f\left(n,m\right)}^{w}\right)d\ln A_{m}\quad\text{for }f=1,\ldots,F_{\ell},\label{eq: D.WfLf/PY-Qf''}
\end{multline}
where $\Pi_{UR,fg(n,m)}^{w}$ is the $nm$-th element of the $N\times N$
matrix $\Pi_{UR,fg}^{w}$, $\Pi_{UL,fg(n,m)}^{w}$ is the $nm$-th
element of the $N\times N$ matrix $\Pi_{UL,fg}^{w}$, and $\Theta_{f}^{w}$
is the $N\times N$ matrix whose $nm$-th element is given by $\Theta_{f(n,m)}^{w}=-[1-1/(1-\eta)](I_{(n,m)}-\zeta_{m})$.
The matrices $\Pi_{UR,fg}^{w}$ and $\Pi_{UL,fg}^{w}$ are described
in in the proof below.
\end{prop}
Most elements of the matrix $\Omega_{f}^{q}$ consist of the sectoral
elasticities of substitution defined as $\epsilon_{\ell_{fn}\ell_{gm}}^{\mathcal{C}}=\partial\ln(\ell_{fn}/\ell_{gm})/\partial\ln(w_{gm}/w_{fn})$,
and the others are the factor shares of income $\Lambda_{\ell_{f}n}$.
All elements of the matrices $\Pi_{UR,fg}^{w}$ and $\Pi_{UL,fg}^{w}$
consist of the sectoral elasticities of substitution $\epsilon_{\ell_{fn}\ell_{gm}}^{\mathcal{C}}$
and the factor shares of income $\Lambda_{\ell_{f}n}$. Equations
\eqref{eq: D.Wf/Wg-Qf''} and \eqref{eq: D.WfLf/PY-Qf''} show that
the sectoral elasticities of substitution are sufficient statistics
for quantifying the effects of capital-embodied technological change
on the relative wages and income shares of different types of labor
if the income shares of each factor are known.

\section{Proofs\label{sec: proofs}}

\paragraph{Proof of Lemma \ref{prop: epsilon_f =000026 epsilon_c}}

Using the facts that $\partial\ln\mathcal{F}(\ell_{1},...,\ell_{F})/\partial\ln\ell_{f}=\Lambda_{\ell_{f}}$
and $\partial\ln\mathcal{C}(w_{1},...,w_{F},y)$ $/\partial\ln w_{f}=\Lambda_{\ell_{f}}$,
the aggregate elasticities of substitution can be rewritten as
\begin{align}
\frac{1}{\epsilon_{\ell_{f}\ell_{g}}^{\mathcal{F}}}= & \frac{\partial\ln\Lambda_{\ell_{f}}}{\partial\ln\ell_{g}}-\frac{\partial\ln\Lambda_{\ell_{g}}}{\partial\ln\ell_{g}}+1,\label{eq: morishima_y'}\\
\epsilon_{\ell_{f}\ell_{g}}^{\mathcal{C}}= & \frac{\partial\ln\Lambda_{\ell_{f}}}{\partial\ln w_{g}}-\frac{\partial\ln\Lambda_{\ell_{g}}}{\partial\ln w_{g}}+1.\label{eq: morishima_c'''}
\end{align}

The first-order Taylor expansion of the factor share, $\Lambda_{\ell_{f}}=\sum_{n=1}^{N}\zeta_{n}\lambda_{\ell_{f}n}$,
yields
\begin{equation}
\Delta\ln\Lambda_{\ell_{f}}=\sum_{n=1}^{N}\frac{\zeta_{n}\lambda_{\ell_{f}n}}{\Lambda_{\ell_{f}}}\Delta\ln\zeta_{n}+\sum_{n=1}^{N}\frac{\zeta_{n}\lambda_{\ell_{f}n}}{\Lambda_{\ell_{f}}}\Delta\ln\lambda_{\ell_{f}n}.\label{eq: D.WfLf/Y}
\end{equation}
The first-order Taylor expansion of the factor share in sector $n$,
$\lambda_{\ell_{f}n}=w_{f}\ell_{fn}/p_{n}y_{n}$, yields
\begin{equation}
\Delta\ln\lambda_{\ell_{f}n}=\sum_{g=1}^{F}\frac{\partial\ln\lambda_{\ell_{f}n}}{\partial\ln w_{g}}\Delta\ln w_{g}.\label{eq: D.WfLfn/PnYn-Wf}
\end{equation}
The first-order Taylor expansion of the expenditure share \eqref{eq: D.PnYn/PY}
yields
\begin{equation}
\Delta\ln\zeta_{n}=-\frac{\eta}{1-\eta}\Delta\ln p_{n}+\frac{\eta}{1-\eta}\sum_{m=1}^{N}\zeta_{m}\Delta\ln p_{m}.\label{eq: D.PnYn/PY-Pn}
\end{equation}
The first-order Taylor expansion of the profit maximizing condition,
$p_{n}=\mathcal{\mathcal{\widetilde{C}}}_{n}(w_{1},\ldots,w_{F})/A_{n}$,
yields
\begin{equation}
\Delta\ln p_{n}=\sum_{f=1}^{F}\lambda_{\ell_{f}n}\Delta\ln w_{f}-\Delta\ln A_{n}.\label{eq: D.Pn}
\end{equation}
Substituting this into equation \eqref{eq: D.PnYn/PY-Pn} yields
\begin{equation}
\Delta\ln\zeta_{n}=\left(1-\frac{1}{1-\eta}\right)\left[\sum_{f=1}^{F}\left(\lambda_{\ell_{f}n}-\Lambda_{\ell_{f}}\right)\Delta\ln w_{f}-\Delta\ln A_{n}+\sum_{m=1}^{N}\zeta_{m}\Delta\ln A_{m}\right].\label{eq: D.PnYn/PY-Wf}
\end{equation}
Substituting equations \eqref{eq: D.WfLfn/PnYn-Wf} and \eqref{eq: D.PnYn/PY-Wf}
into equation \eqref{eq: D.WfLf/Y} yields
\begin{equation}
\Delta\ln\Lambda_{\ell_{f}}=\sum_{g=1}^{F}\Psi_{\left(f,g\right)}^{w}\Delta\ln w_{g}+\psi_{\left(f\right)}^{w}=\sum_{g=1}^{F}\Psi_{\left(f,g\right)}^{w}\Delta\ln\left(\frac{w_{g}}{p}\right)+\psi_{\left(f\right)}^{w}.\label{eq: D.WfLf/PY-Wf'}
\end{equation}
The second equality comes from the fact that equation \eqref{eq: D.WfLf/PY-Wf'}
is homogeneous of degree zero in factor prices because $\sum_{g=1}^{F}\Psi_{\left(f,g\right)}^{w}=0$.
By the definition of factor shares, $\Delta\ln w_{f}=\Delta\ln\Lambda_{\ell_{f}}-\Delta\ln\ell_{f}+\Delta\ln(py)$.
Substituting this into equation \eqref{eq: D.WfLf/PY-Wf'} yields
\[
\Delta\ln\Lambda_{\ell_{f}}=\sum_{g=1}^{F}\Psi_{\left(f,g\right)}^{w}\Delta\ln\Lambda_{\ell_{g}}-\sum_{g=1}^{F}\Psi_{\left(f,g\right)}^{w}\Delta\ln\ell_{g}+\psi_{\left(f\right)}^{w}.
\]
Solving the system of equations for $\Delta\ln\Lambda_{\ell_{f}}$
yields
\begin{equation}
\Delta\ln\Lambda_{\ell_{f}}=\sum_{g=1}^{F}\Psi_{\left(f,g\right)}^{\ell}\Delta\ln\ell_{g}+\psi_{\left(f\right)}^{\ell}.\label{eq: D.WfLf/PY-Lf'}
\end{equation}

Equation \eqref{eq: D.WfLf/PY-Lf'} implies $\partial\ln\Lambda_{\ell_{f}}/\partial\ln\ell_{g}=\Psi_{\left(f,g\right)}^{\ell}$,
while equation \eqref{eq: D.WfLf/PY-Wf'} implies $\partial\ln\Lambda_{\ell_{f}}/\partial\ln w_{g}=\Psi_{\left(f,g\right)}^{w}$.
Substituting these into equations \eqref{eq: morishima_y'} and \eqref{eq: morishima_c'''}
yields equations \eqref{eq: morishima_y} and \eqref{eq: morishima_c}.

\paragraph{Proof of Proposition \ref{prop: D.WfLf/PY-Lf}}

Using the fact that $\sum_{f=1}^{F}\Lambda_{\ell_{f}}\Psi_{\left(f,g\right)}^{\ell}=0$,
equation \eqref{eq: D.WfLf/PY-Lf'} can be rearranged as
\begin{multline*}
\Delta\ln\Lambda_{\ell_{f}}=\sum_{g\neq f}^{F}\left[1-\left(\Psi_{\left(g,f\right)}^{\ell}-\Psi_{\left(f,f\right)}^{\ell}+1\right)\right]\Lambda_{\ell_{g}}\Delta\ln\ell_{f}+\sum_{g\neq f}^{F}\left[\left(\Psi_{\left(f,g\right)}^{\ell}-\Psi_{\left(g,g\right)}^{\ell}+1\right)-1\right]\Lambda_{\ell_{g}}\Delta\ln\ell_{g}\\
+\sum_{g\neq f}^{F}\sum_{h\neq f,g}^{F}\left[\left(\Psi_{\left(f,g\right)}^{\ell}-\Psi_{\left(g,g\right)}^{\ell}+1\right)-\left(\Psi_{\left(h,g\right)}^{\ell}-\Psi_{\left(g,g\right)}^{\ell}+1\right)\right]\Lambda_{\ell_{h}}\Delta\ln\ell_{g}+\psi_{\left(f\right)}^{\ell}.
\end{multline*}
Substituting equation \eqref{eq: morishima_y} into this yields equation
\eqref{eq: D.WfLf/PY-Lf}.

Similarly, using the fact that $\sum_{f=1}^{F}\Lambda_{\ell_{f}}\Psi_{\left(f,g\right)}^{w}=0$,
equation \eqref{eq: D.WfLf/PY-Wf'} can be rearranged as
\begin{multline*}
\Delta\ln\Lambda_{\ell_{f}}=\sum_{g\neq f}^{F}\left[1-\left(\Psi_{\left(g,f\right)}^{w}-\Psi_{\left(f,f\right)}^{w}+1\right)\right]\Lambda_{\ell_{g}}\Delta\ln w_{f}+\sum_{g\neq f}^{F}\left[\left(\Psi_{\left(f,g\right)}^{w}-\Psi_{\left(g,g\right)}^{w}+1\right)-1\right]\Lambda_{\ell_{g}}\Delta\ln w_{g}\\
+\sum_{g\neq f}^{F}\sum_{h\neq f,g}^{F}\left[\left(\Psi_{\left(f,g\right)}^{w}-\Psi_{\left(g,g\right)}^{w}+1\right)-\left(\Psi_{\left(h,g\right)}^{w}-\Psi_{\left(g,g\right)}^{w}+1\right)\right]\Lambda_{\ell_{h}}\Delta\ln w_{g}+\psi_{\left(f\right)}^{w}.
\end{multline*}
Substituting equation \eqref{eq: morishima_c} into this yields equation
\eqref{eq: D.WfLf/PY-Wf}.

\paragraph{Proof of Proposition \ref{prop: D.Wf/Wg-Lf}}

By definition, the log change of relative factor shares can be written
as
\begin{equation}
\Delta\ln\left(\frac{\Lambda_{\ell_{f}}}{\Lambda_{\ell_{g}}}\right)=\Delta\ln\left(\frac{w_{f}}{w_{g}}\right)+\Delta\ln\left(\frac{\ell_{f}}{\ell_{g}}\right).\label{eq: D.WfLf/WgLg}
\end{equation}
Substituting equation \eqref{eq: D.WfLf/PY-Lf'} into equation \eqref{eq: D.WfLf/WgLg}
yields
\[
\Delta\ln\left(\frac{w_{f}}{w_{g}}\right)=\sum_{h=1}^{F}\left[\Psi_{\left(f,h\right)}^{\ell}-\Psi_{\left(h,h\right)}^{\ell}+1-\left(\Psi_{\left(g,h\right)}^{\ell}-\Psi_{\left(h,h\right)}^{\ell}+1\right)\right]\Delta\ln\ell_{h}-\Delta\ln\ell_{f}+\Delta\ln\ell_{g}+\psi_{\left(f\right)}^{\ell}-\psi_{\left(g\right)}^{\ell}.
\]
Using equation \eqref{eq: morishima_y}, this equation can be rewritten
as equation \eqref{eq: D.Wf/Wg-Lf}.

Substituting \eqref{eq: D.WfLf/PY-Wf'} into equation \eqref{eq: D.WfLf/WgLg}
yields
\[
\Delta\ln\left(\frac{\ell_{f}}{\ell_{g}}\right)=\sum_{h=1}^{F}\left[\Psi_{\left(f,h\right)}^{w}-\Psi_{\left(h,h\right)}^{w}+1-\left(\Psi_{\left(g,h\right)}^{w}-\Psi_{\left(h,h\right)}^{w}+1\right)\right]\Delta\ln w_{h}-\Delta\ln w_{f}+\Delta\ln w_{g}+\psi_{\left(f\right)}^{w}-\psi_{\left(g\right)}^{w}.
\]
Using equation \eqref{eq: morishima_c}, this equation can be rewritten
as equation \eqref{eq: D.Lf/Lg-Wf}.

\paragraph{Proof of Proposition \ref{prop: D.Wf/Wg-Qf}}

The Euler equation \eqref{eq: Euler} can be derived from the household's
problem when the interest rate is defined as $\iota_{t}=(1/\beta)[(\partial\mathcal{U}(c_{t-1})/\partial c_{t-1})/(\partial\mathcal{U}(c_{t})/\partial c_{t})]-1$.
It follows that the log change in the rental price of capital between
steady states are given by
\begin{equation}
d\ln\left(\frac{r_{f}}{p}\right)=-d\ln q_{f}.\label{eq: D.Rf/P}
\end{equation}
Subtracting $d\ln p$ from both sides of equation \eqref{eq: D.Pn}
and substituting equation \eqref{eq: D.Rf/P} into this yields
\begin{equation}
d\ln\left(\frac{p_{n}}{p}\right)=\sum_{f=1}^{F_{\ell}}\lambda_{\ell_{f}n}d\ln\left(\frac{w_{f}}{p}\right)-\sum_{f=1}^{F_{k}}\lambda_{k_{f}n}d\ln q_{f}-d\ln A_{n}.\label{eq: D.Pn/P-Wf=000026Qf}
\end{equation}
The first-order Taylor expansion of the aggregate price, $p=(\sum_{n=1}^{N}\theta_{cn}^{1/(1-\eta)}p_{n}^{-\eta/(1-\eta)})^{-(1-\eta)/\eta}$,
yields $\sum_{n=1}^{N}\zeta_{n}d\ln(p_{n}/p)=0$. Substituting equation
\eqref{eq: D.Pn/P-Wf=000026Qf} into this yields
\begin{equation}
\sum_{f=1}^{F_{\ell}}\Lambda_{\ell_{f}}d\ln\left(\frac{w_{f}}{p}\right)=\sum_{f=1}^{F_{k}}\Lambda_{k_{f}}d\ln q_{f}+\sum_{n=1}^{N}\zeta_{n}d\ln A_{n}.\label{eq: D.Wf/P-Qf}
\end{equation}

Equation \eqref{eq: D.WfLf/PY-Wf'} can be rewritten as
\[
d\ln\Lambda_{\ell_{f}}=\sum_{g=1}^{F}\Psi_{\left(f,g\right)}^{w}d\ln\left(\frac{w_{g}}{p}\right)+\sum_{n=1}^{N}\Phi_{\left(f,n\right)}^{w}d\ln A_{n},
\]
where $\Phi^{w}$ is the $F\times N$ matrix whose $fn$-th element
is given by $\Phi_{\left(f,n\right)}^{w}=-[1-1/(1-\eta)](\zeta_{n}\lambda_{\ell_{f}n}/\Lambda_{\ell_{f}}-\zeta_{n})$.
Let the matrix $\Psi^{w}$ be partitioned as
\[
\Psi^{w}=\left[\begin{array}{cc}
\Psi_{UL}^{w} & \Psi_{UR}^{w}\\
\Psi_{LL}^{w} & \Psi_{LR}^{w}
\end{array}\right],
\]
where $\Psi_{UL}^{w}$, $\Psi_{UR}^{w}$, $\Psi_{LL}^{w}$, and $\Psi_{LR}^{w}$
are $F_{\ell}\times F_{\ell}$, $F_{\ell}\times F_{k}$, $F_{k}\times F_{\ell}$,
and $F_{k}\times F_{k}$ matrices, respectively. The equation can
be further rewritten as
\[
d\ln\Lambda_{\ell_{f}}=\sum_{g=1}^{F_{\ell}}\Psi_{UL\left(f,g\right)}^{w}d\ln\left(\frac{w_{g}}{p}\right)+\sum_{g=1}^{F_{k}}\Psi_{UR\left(f,g\right)}^{w}d\ln\left(\frac{r_{g}}{p}\right)+\sum_{n=1}^{N}\Phi_{\left(f,n\right)}^{w}d\ln A_{n}\quad\text{for }f=1,\ldots,F_{\ell}.
\]
 Substituting equation \eqref{eq: D.Rf/P} into this yields
\begin{equation}
d\ln\Lambda_{\ell_{f}}=\sum_{g=1}^{F_{\ell}}\Psi_{UL\left(f,g\right)}^{w}d\ln\left(\frac{w_{g}}{p}\right)-\sum_{g=1}^{F_{k}}\Psi_{UR\left(f,g\right)}^{w}d\ln q_{g}+\sum_{n=1}^{N}\Phi_{\left(f,n\right)}^{w}d\ln A_{n}\quad\text{for }f=1,\ldots,F_{\ell}.\label{eq: D.WfLf/Y-Wf=000026Qf}
\end{equation}
Substituting $\Lambda_{\ell_{f}}=w_{f}\ell_{f}/py$ into this yields
\begin{multline}
\sum_{g=1}^{F_{\ell}}\left(\Psi_{UL\left(f,g\right)}^{w}-I_{\left(f,g\right)}\right)d\ln\left(\frac{w_{g}}{p}\right)=\sum_{g=1}^{F_{\ell}}I_{\left(f,g\right)}d\ln\ell_{g}\\
+\sum_{g=1}^{F_{k}}\Psi_{UR\left(f,g\right)}^{w}d\ln q_{g}-\sum_{n=1}^{N}\Phi_{\left(f,n\right)}^{w}d\ln A_{n}-d\ln y\quad\text{for }f=1,\ldots,F_{\ell}.\label{eq: D.Wf/P-Qf=000026Lf=000026Y}
\end{multline}
Subtracting the equation for $f=F_{\ell}$ from that for any $f\in\{1,\ldots,F_{\ell}-1$\}
yields
\begin{multline}
\sum_{g=1}^{F_{\ell}}\left(\Psi_{UL\left(f,g\right)}^{w}-\Psi_{UL\left(F_{\ell},g\right)}^{w}-I_{\left(f,g\right)}+I_{\left(F_{\ell},g\right)}\right)d\ln\left(\frac{w_{g}}{p}\right)=\sum_{g=1}^{F_{\ell}}\left(I_{\left(f,g\right)}-I_{\left(F_{\ell},g\right)}\right)d\ln\ell_{g}\\
+\sum_{g=1}^{F_{k}}\left(\Psi_{UR\left(f,g\right)}^{w}-\Psi_{UR\left(F_{\ell},g\right)}^{w}\right)d\ln q_{g}-\sum_{n=1}^{N}\left(\Phi_{\left(f,n\right)}^{w}-\Phi_{\left(F_{\ell},n\right)}^{w}\right)d\ln A_{n}.\label{eq: D.Wf/P-Qf=000026Lf}
\end{multline}

Solving the system of equations \eqref{eq: D.Wf/P-Qf} and \eqref{eq: D.Wf/P-Qf=000026Lf}
for $d\ln(w_{f}/p)$ and setting $d\ln\ell_{f}=0$ yields
\begin{equation}
d\ln\left(\frac{w_{f}}{p}\right)=\sum_{g=1}^{F_{k}}\Upsilon_{\left(f,g\right)}^{q}d\ln q_{g}+\sum_{n=1}^{N}\Upsilon_{\left(f,n\right)}^{A}d\ln A_{n}\quad\text{for }f=1,\ldots,F_{\ell},\label{eq: D.Wf/P-Qf'}
\end{equation}
where $\Upsilon_{(f,g)}^{q}$ is the $fg$-th element of the $F_{\ell}\times F_{k}$
matrix{\footnotesize
\begin{align*}
\Upsilon^{q}= & \left[\begin{array}{ccccc}
-\epsilon_{\ell_{F_{\ell}}\ell_{1}}^{\mathcal{C}} & \epsilon_{\ell_{1}\ell_{2}}^{\mathcal{C}}-\epsilon_{\ell_{F_{\ell}}\ell_{2}}^{\mathcal{C}} & \cdots & \epsilon_{\ell_{1}\ell_{F_{\ell}-1}}^{\mathcal{C}}-\epsilon_{\ell_{F_{\ell}}\ell_{F_{\ell}-1}}^{\mathcal{C}} & \epsilon_{\ell_{1}\ell_{F_{\ell}}}^{\mathcal{C}}\\
\epsilon_{\ell_{2}\ell_{1}}^{\mathcal{C}}-\epsilon_{\ell_{F_{\ell}}\ell_{1}}^{\mathcal{C}} & -\epsilon_{\ell_{F_{\ell}}\ell_{2}}^{\mathcal{C}} & \cdots & \epsilon_{\ell_{2}\ell_{F_{\ell}-1}}^{\mathcal{C}}-\epsilon_{\ell_{F_{\ell}}\ell_{F_{\ell}-1}}^{\mathcal{C}} & \epsilon_{\ell_{2}\ell_{F_{\ell}}}^{\mathcal{C}}\\
\vdots & \vdots & \ddots & \vdots & \vdots\\
\epsilon_{\ell_{F_{\ell}-1}\ell_{1}}^{\mathcal{C}}-\epsilon_{\ell_{F_{\ell}}\ell_{1}}^{\mathcal{C}} & \epsilon_{\ell_{F_{\ell}-1}\ell_{2}}^{\mathcal{C}}-\epsilon_{\ell_{F_{\ell}}\ell_{2}}^{\mathcal{C}} & \cdots & -\epsilon_{\ell_{F_{\ell}}\ell_{F_{\ell}-1}}^{\mathcal{C}} & \epsilon_{\ell_{F_{\ell}-1}\ell_{F_{\ell}}}^{\mathcal{C}}\\
\Lambda_{\ell_{1}} & \Lambda_{\ell_{2}} & \cdots & \Lambda_{\ell_{F_{\ell}-1}} & \Lambda_{\ell_{F_{\ell}}}
\end{array}\right]^{-1}\\
 & \times\left[\begin{array}{ccc}
\epsilon_{\ell_{1}k_{1}}^{\mathcal{C}}-\epsilon_{\ell_{F_{\ell}}k_{1}}^{\mathcal{C}} & \cdots & \epsilon_{\ell_{1}k_{F_{k}}}^{\mathcal{C}}-\epsilon_{\ell_{F_{\ell}}k_{F_{k}}}^{\mathcal{C}}\\
\vdots & \vdots & \vdots\\
\epsilon_{\ell_{F_{\ell}-1}k_{1}}^{\mathcal{C}}-\epsilon_{\ell_{F_{\ell}}k_{1}}^{\mathcal{C}} & \cdots & \epsilon_{\ell_{F_{\ell}-1}k_{F_{k}}}^{\mathcal{C}}-\epsilon_{\ell_{F_{\ell}}k_{F_{k}}}^{\mathcal{C}}\\
\Lambda_{k_{1}} & \cdots & \Lambda_{k_{F_{k}}}
\end{array}\right],
\end{align*}
}and $\Upsilon_{(f,n)}^{A}$ is the $fn$-th element of the $F_{\ell}\times N$
matrix{\footnotesize
\begin{align*}
\Upsilon^{A}= & -\left[\begin{array}{ccccc}
-\epsilon_{\ell_{F_{\ell}}\ell_{1}}^{\mathcal{C}} & \epsilon_{\ell_{1}\ell_{2}}^{\mathcal{C}}-\epsilon_{\ell_{F_{\ell}}\ell_{2}}^{\mathcal{C}} & \cdots & \epsilon_{\ell_{1}\ell_{F_{\ell}-1}}^{\mathcal{C}}-\epsilon_{\ell_{F_{\ell}}\ell_{F_{\ell}-1}}^{\mathcal{C}} & \epsilon_{\ell_{1}\ell_{F_{\ell}}}^{\mathcal{C}}\\
\epsilon_{\ell_{2}\ell_{1}}^{\mathcal{C}}-\epsilon_{\ell_{F_{\ell}}\ell_{1}}^{\mathcal{C}} & -\epsilon_{\ell_{F_{\ell}}\ell_{2}}^{\mathcal{C}} & \cdots & \epsilon_{\ell_{2}\ell_{F_{\ell}-1}}^{\mathcal{C}}-\epsilon_{\ell_{F_{\ell}}\ell_{F_{\ell}-1}}^{\mathcal{C}} & \epsilon_{\ell_{2}\ell_{F_{\ell}}}^{\mathcal{C}}\\
\vdots & \vdots & \ddots & \vdots & \vdots\\
\epsilon_{\ell_{F_{\ell}-1}\ell_{1}}^{\mathcal{C}}-\epsilon_{\ell_{F_{\ell}}\ell_{1}}^{\mathcal{C}} & \epsilon_{\ell_{F_{\ell}-1}\ell_{2}}^{\mathcal{C}}-\epsilon_{\ell_{F_{\ell}}\ell_{2}}^{\mathcal{C}} & \cdots & -\epsilon_{\ell_{F_{\ell}}\ell_{F_{\ell}-1}}^{\mathcal{C}} & \epsilon_{\ell_{F_{\ell}-1}\ell_{F_{\ell}}}^{\mathcal{C}}\\
\Lambda_{\ell_{1}} & \Lambda_{\ell_{2}} & \cdots & \Lambda_{\ell_{F_{\ell}-1}} & \Lambda_{\ell_{F_{\ell}}}
\end{array}\right]^{-1}\\
 & \times\left[\begin{array}{ccc}
\Phi_{\left(1,1\right)}^{w}-\Phi_{\left(F_{\ell},1\right)}^{w} & \cdots & \Phi_{\left(1,N\right)}^{w}-\Phi_{\left(F_{\ell},N\right)}^{w}\\
\vdots & \vdots & \vdots\\
\Phi_{\left(F_{\ell}-1,1\right)}^{w}-\Phi_{\left(F_{\ell},1\right)}^{w} & \cdots & \Phi_{\left(F_{\ell}-1,N\right)}^{w}-\Phi_{\left(F_{\ell},N\right)}^{w}\\
\zeta_{1} & \cdots & \zeta_{N}
\end{array}\right].
\end{align*}
}Equation \eqref{eq: D.Wf/P-Qf'} immediately implies equation \eqref{eq: D.Wf/Wg-Qf}.
Substituting equation \eqref{eq: D.Wf/P-Qf'} into equation \eqref{eq: D.WfLf/Y-Wf=000026Qf}
yields equation \eqref{eq: D.WfLf/PY-Qf}.

\paragraph{Proof of Proposition \ref{prop: D.Wf/Wg-Qf'}}

The labor supply function can be derived from the household's problem
as $l_{f}=(\partial\mathcal{U}(c)/\partial c)^{1/\gamma}\chi_{f}^{-1/\gamma}(w_{f}/p)^{1/\gamma}$.
Substituting the labor market clearing condition, $\ell_{f}=\mu_{f}l_{f}$,
into this yields $\ell_{f}=\mu_{f}(\partial\mathcal{U}(c)/\partial c)^{1/\gamma}\chi_{f}^{-1/\gamma}(w_{f}/p)^{1/\gamma}$.
The log-linearization yields
\[
d\ln\ell_{f}=\frac{c}{\gamma}\frac{\partial^{2}\mathcal{U}\left(c\right)\left/\partial c^{2}\right.}{\partial\mathcal{U}\left(c\right)\left/\partial c\right.}d\ln c+\frac{1}{\gamma}\sum_{g=1}^{F_{\ell}}I_{\left(f,g\right)}d\ln\left(\frac{w_{g}}{p}\right).
\]
Substituting this into equation \eqref{eq: D.Wf/P-Qf=000026Lf=000026Y}
yields
\begin{multline*}
\sum_{g=1}^{F_{\ell}}\left[\Psi_{UL\left(f,g\right)}^{w}-\left(\frac{1+\gamma}{\gamma}\right)I_{\left(f,g\right)}\right]d\ln\left(\frac{w_{g}}{p}\right)=-d\ln y+\frac{c}{\gamma}\frac{\partial^{2}\mathcal{U}\left(c\right)\left/\partial c^{2}\right.}{\partial\mathcal{U}\left(c\right)\left/\partial c\right.}d\ln c\\
+\sum_{g=1}^{F_{k}}\Psi_{UR\left(f,g\right)}^{w}d\ln q_{g}-\sum_{n=1}^{N}\Phi_{\left(f,n\right)}^{w}d\ln A_{n}\quad\text{for }f=1,\ldots,F_{\ell}.
\end{multline*}
Subtracting the equation for $f=F_{\ell}$ from that for any $f\in\{1,\ldots,F_{\ell}-1$\}
yields
\begin{multline}
\sum_{g=1}^{F_{\ell}}\left[\left(\Psi_{UL\left(f,g\right)}^{w}-\Psi_{UL\left(F_{\ell},g\right)}^{w}\right)-\left(I_{\left(f,g\right)}-I_{\left(F_{\ell},g\right)}\right)\left(\frac{1+\gamma}{\gamma}\right)\right]d\ln\left(\frac{w_{g}}{p}\right)\\
=\sum_{g=1}^{F_{k}}\left(\Psi_{UR\left(f,g\right)}^{w}-\Psi_{UR\left(F_{\ell},g\right)}^{w}\right)d\ln q_{g}-\sum_{n=1}^{N}\left(\Phi_{\left(f,n\right)}^{w}-\Phi_{\left(F_{\ell},n\right)}^{w}\right)d\ln A_{n}.\label{eq: D.Wf/P-Qf=000026Lf'}
\end{multline}

Solving the system of equations \eqref{eq: D.Wf/P-Qf} and \eqref{eq: D.Wf/P-Qf=000026Lf'}
for $d\ln(w_{f}/p)$ yields
\begin{equation}
d\ln\left(\frac{w_{f}}{p}\right)=\sum_{g=1}^{F_{k}}\Xi_{\left(f,g\right)}^{q}d\ln q_{g}+\sum_{n=1}^{N}\Xi_{\left(f,n\right)}^{A}d\ln A_{n}\quad\text{for }f=1,\ldots,F_{\ell},\label{eq: D.Wf/P-Qf''}
\end{equation}
where $\Xi_{(f,g)}^{q}$ is the $fg$-th element of the $F_{\ell}\times F_{k}$
matrix{\footnotesize
\begin{align*}
\Xi^{q}= & \left[\begin{array}{ccccc}
-\epsilon_{\ell_{F_{\ell}}\ell_{1}}^{\mathcal{C}}-\frac{1}{\gamma} & \epsilon_{\ell_{1}\ell_{2}}^{\mathcal{C}}-\epsilon_{\ell_{F_{\ell}}\ell_{2}}^{\mathcal{C}} & \cdots & \epsilon_{\ell_{1}\ell_{F_{\ell}-1}}^{\mathcal{C}}-\epsilon_{\ell_{F_{\ell}}\ell_{F_{\ell}-1}}^{\mathcal{C}} & \epsilon_{\ell_{1}\ell_{F_{\ell}}}^{\mathcal{C}}+\frac{1}{\gamma}\\
\epsilon_{\ell_{2}\ell_{1}}^{\mathcal{C}}-\epsilon_{\ell_{F_{\ell}}\ell_{1}}^{\mathcal{C}} & -\epsilon_{\ell_{F_{\ell}}\ell_{2}}^{\mathcal{C}}-\frac{1}{\gamma} & \cdots & \epsilon_{\ell_{2}\ell_{F_{\ell}-1}}^{\mathcal{C}}-\epsilon_{\ell_{F_{\ell}}\ell_{F_{\ell}-1}}^{\mathcal{C}} & \epsilon_{\ell_{2}\ell_{F_{\ell}}}^{\mathcal{C}}+\frac{1}{\gamma}\\
\vdots & \vdots & \ddots & \vdots & \vdots\\
\epsilon_{\ell_{F_{\ell}-1}\ell_{1}}^{\mathcal{C}}-\epsilon_{\ell_{F_{\ell}}\ell_{1}}^{\mathcal{C}} & \epsilon_{\ell_{F_{\ell}-1}\ell_{2}}^{\mathcal{C}}-\epsilon_{\ell_{F_{\ell}}\ell_{2}}^{\mathcal{C}} & \cdots & -\epsilon_{\ell_{F_{\ell}}\ell_{F_{\ell}-1}}^{\mathcal{C}}-\frac{1}{\gamma} & \epsilon_{\ell_{F_{\ell}-1}\ell_{F_{\ell}}}^{\mathcal{C}}+\frac{1}{\gamma}\\
\Lambda_{\ell_{1}} & \Lambda_{\ell_{2}} & \cdots & \Lambda_{\ell_{F_{\ell}-1}} & \Lambda_{\ell_{F_{\ell}}}
\end{array}\right]^{-1}\\
 & \times\left[\begin{array}{ccc}
\epsilon_{\ell_{1}k_{1}}^{\mathcal{C}}-\epsilon_{\ell_{F_{\ell}}k_{1}}^{\mathcal{C}} & \cdots & \epsilon_{\ell_{1}k_{F_{k}}}^{\mathcal{C}}-\epsilon_{\ell_{F_{\ell}}k_{F_{k}}}^{\mathcal{C}}\\
\vdots & \vdots & \vdots\\
\epsilon_{\ell_{F_{\ell}-1}k_{1}}^{\mathcal{C}}-\epsilon_{\ell_{F_{\ell}}k_{1}}^{\mathcal{C}} & \cdots & \epsilon_{\ell_{F_{\ell}-1}k_{F_{k}}}^{\mathcal{C}}-\epsilon_{\ell_{F_{\ell}}k_{F_{k}}}^{\mathcal{C}}\\
\Lambda_{k_{1}} & \cdots & \Lambda_{k_{F_{k}}}
\end{array}\right],
\end{align*}
}and $\Xi_{(f,n)}^{A}$ is the $fn$-th element of the $F_{\ell}\times N$
matrix{\footnotesize
\begin{align*}
\Xi^{A}= & -\left[\begin{array}{ccccc}
-\epsilon_{\ell_{F_{\ell}}\ell_{1}}^{\mathcal{C}}-\frac{1}{\gamma} & \epsilon_{\ell_{1}\ell_{2}}^{\mathcal{C}}-\epsilon_{\ell_{F_{\ell}}\ell_{2}}^{\mathcal{C}} & \cdots & \epsilon_{\ell_{1}\ell_{F_{\ell}-1}}^{\mathcal{C}}-\epsilon_{\ell_{F_{\ell}}\ell_{F_{\ell}-1}}^{\mathcal{C}} & \epsilon_{\ell_{1}\ell_{F_{\ell}}}^{\mathcal{C}}+\frac{1}{\gamma}\\
\epsilon_{\ell_{2}\ell_{1}}^{\mathcal{C}}-\epsilon_{\ell_{F_{\ell}}\ell_{1}}^{\mathcal{C}} & -\epsilon_{\ell_{F_{\ell}}\ell_{2}}^{\mathcal{C}}-\frac{1}{\gamma} & \cdots & \epsilon_{\ell_{2}\ell_{F_{\ell}-1}}^{\mathcal{C}}-\epsilon_{\ell_{F_{\ell}}\ell_{F_{\ell}-1}}^{\mathcal{C}} & \epsilon_{\ell_{2}\ell_{F_{\ell}}}^{\mathcal{C}}+\frac{1}{\gamma}\\
\vdots & \vdots & \ddots & \vdots & \vdots\\
\epsilon_{\ell_{F_{\ell}-1}\ell_{1}}^{\mathcal{C}}-\epsilon_{\ell_{F_{\ell}}\ell_{1}}^{\mathcal{C}} & \epsilon_{\ell_{F_{\ell}-1}\ell_{2}}^{\mathcal{C}}-\epsilon_{\ell_{F_{\ell}}\ell_{2}}^{\mathcal{C}} & \cdots & -\epsilon_{\ell_{F_{\ell}}\ell_{F_{\ell}-1}}^{\mathcal{C}}-\frac{1}{\gamma} & \epsilon_{\ell_{F_{\ell}-1}\ell_{F_{\ell}}}^{\mathcal{C}}+\frac{1}{\gamma}\\
\Lambda_{\ell_{1}} & \Lambda_{\ell_{2}} & \cdots & \Lambda_{\ell_{F_{\ell}-1}} & \Lambda_{\ell_{F_{\ell}}}
\end{array}\right]^{-1}\\
 & \times\left[\begin{array}{ccc}
\Phi_{\left(1,1\right)}^{w}-\Phi_{\left(F_{\ell},1\right)}^{w} & \cdots & \Phi_{\left(1,N\right)}^{w}-\Phi_{\left(F_{\ell},N\right)}^{w}\\
\vdots & \vdots & \vdots\\
\Phi_{\left(F_{\ell}-1,1\right)}^{w}-\Phi_{\left(F_{\ell},1\right)}^{w} & \cdots & \Phi_{\left(F_{\ell}-1,N\right)}^{w}-\Phi_{\left(F_{\ell},N\right)}^{w}\\
\zeta_{1} & \cdots & \zeta_{N}
\end{array}\right].
\end{align*}
}Equation \eqref{eq: D.Wf/P-Qf''} immediately implies equation \eqref{eq: D.Wf/Wg-Qf'}.
Substituting equation \eqref{eq: D.Wf/P-Qf''} into equation \eqref{eq: D.WfLf/Y-Wf=000026Qf}
yields equation \eqref{eq: D.WfLf/PY-Qf'}.

\paragraph{Proof of Proposition \ref{prop: D.Wf/Wg-Qf''}}

The log-linearization of the profit maximizing condition, $p_{n}=\mathcal{\mathcal{\widetilde{C}}}_{n}(w_{1n},\ldots,w_{Fn})$
$/A_{n}$, yields
\begin{equation}
d\ln p_{n}=\sum_{f=1}^{F}\lambda_{\ell_{f}n}d\ln w_{fn}-d\ln A_{n}.\label{eq: D.Pn-Wfn}
\end{equation}
Subtracting $d\ln p$ from both sides of equation \eqref{eq: D.Pn-Wfn}
and substituting equation \eqref{eq: D.Rf/P} into this yields
\begin{equation}
d\ln\left(\frac{p_{n}}{p}\right)=\sum_{f=1}^{F_{\ell}}\lambda_{\ell_{f}n}d\ln\left(\frac{w_{fn}}{p}\right)-\sum_{f=1}^{F_{k}}\lambda_{k_{f}n}d\ln q_{f}-d\ln A_{n}.\label{eq: D.Pn/P-Wfn=000026Qf}
\end{equation}
The log-linearization of the aggregate price, $p=(\sum_{n=1}^{N}\theta_{cn}^{1/(1-\eta)}p_{n}^{-\eta/(1-\eta)})^{-(1-\eta)/\eta}$,
yields $\sum_{n=1}^{N}\zeta_{n}d\ln(p_{n}/p)$ $=0$. Substituting
equation \eqref{eq: D.Pn/P-Wfn=000026Qf} into this yields
\begin{equation}
\sum_{f=1}^{F_{\ell}}\sum_{n=1}^{N}\zeta_{n}\lambda_{\ell_{f}n}d\ln\left(\frac{w_{fn}}{p}\right)=\sum_{f=1}^{F_{k}}\Lambda_{k_{f}}d\ln q_{f}+\sum_{n=1}^{N}\zeta_{n}d\ln A_{n}.\label{eq: D.Wfn/P-Qf}
\end{equation}

The log-linearization of sector $n$\textquoteright s factor share,
$\Lambda_{\ell_{f}n}=\zeta_{n}\lambda_{\ell_{f}n}$, yields
\begin{equation}
d\ln\Lambda_{\ell_{f}n}=d\ln\zeta_{n}+d\ln\lambda_{\ell_{f}n}.\label{eq: D.WfnLfn/PY}
\end{equation}
The expenditure share takes the same form as equation \eqref{eq: D.PnYn/PY}.
The log-linearization of equation \eqref{eq: D.PnYn/PY} yields $d\ln\zeta_{n}=-[\eta/(1-\eta)]d\ln p_{n}+[\eta/(1-\eta)]\sum_{m=1}^{N}\zeta_{m}d\ln p_{m}$.
Substituting equation \eqref{eq: D.Pn-Wfn} into this yields
\begin{equation}
d\ln\zeta_{n}=\left(1-\frac{1}{1-\eta}\right)\left[\sum_{f=1}^{F}\left(\lambda_{\ell_{f}n}d\ln w_{fn}-\sum_{m=1}^{N}\zeta_{m}\lambda_{\ell_{f}m}d\ln w_{fm}\right)-\left(d\ln A_{n}-\sum_{m=1}^{N}\zeta_{m}d\ln A_{m}\right)\right].\label{eq: D.PnYn/PY-Wfn}
\end{equation}
The log-linearization of the factor share in sector $n$, $\lambda_{\ell_{f}n}=w_{fn}\ell_{fn}/p_{n}y_{n}$,
yields
\begin{equation}
d\ln\lambda_{\ell_{f}n}=\sum_{g=1}^{F}\frac{\partial\ln\lambda_{\ell_{f}n}}{\partial\ln w_{gn}}d\ln w_{gn}.\label{eq: D.WfnLfn/PnYn-Wfn}
\end{equation}
Substituting equations \eqref{eq: D.PnYn/PY-Wfn} and \eqref{eq: D.WfnLfn/PnYn-Wfn}
into equation \eqref{eq: D.WfnLfn/PY} yields
\begin{equation}
d\ln\Lambda_{\ell_{f}n}=\sum_{g=1}^{F}\sum_{m=1}^{N}\Pi_{fg\left(n,m\right)}^{w}d\ln\left(\frac{w_{gm}}{p}\right)+\sum_{m=1}^{N}\Theta_{f\left(n,m\right)}^{w}d\ln A_{m},\label{eq: D.WfnLfn/PY-Wfn}
\end{equation}
where $\Pi_{fg(n,m)}^{w}=I_{(n,m)}(\partial\ln\lambda_{\ell_{f}m}/\partial\ln w_{gm}+[1-1/(1-\eta)]\lambda_{\ell_{g}m})-[1-1/(1-\eta)]\zeta_{m}\lambda_{\ell_{g}m}$
and $\Theta_{f(n,m)}^{w}=-(1-1/(1-\eta))(I_{(n,m)}-\zeta_{m})$. Define
the $FN\times FN$ matrix{\footnotesize
\[
\Pi^{w}=\left[\begin{array}{cccccc}
\Pi_{11}^{w} & \cdots & \Pi_{1F_{\ell}}^{w} & \Pi_{1,F_{\ell}+1}^{w} & \cdots & \Pi_{1,F_{\ell}+F_{k}}^{w}\\
\vdots & \ddots & \vdots & \vdots & \ddots & \vdots\\
\Pi_{F_{\ell}1}^{w} & \cdots & \Pi_{F_{\ell}F_{\ell}}^{w} & \Pi_{F_{\ell},F_{\ell}+1}^{w} & \cdots & \Pi_{F_{\ell},F_{\ell}+F_{k}}^{w}\\
\Pi_{F_{\ell}+1,1}^{w} & \cdots & \Pi_{F_{\ell}+1,F_{\ell}}^{w} & \Pi_{F_{\ell}+1,F_{\ell}+1}^{w} & \cdots & \Pi_{F_{\ell}+1,F_{\ell}+F_{k}}^{w}\\
\vdots & \ddots & \vdots & \vdots & \ddots & \vdots\\
\Pi_{F_{\ell}+F_{k},1}^{w} & \cdots & \Pi_{F_{\ell}+F_{k},F_{\ell}}^{w} & \Pi_{F_{\ell}+F_{k},F_{\ell}+1}^{w} & \cdots & \Pi_{F_{\ell}+F_{k},F_{\ell}+F_{k}}^{w}
\end{array}\right],
\]
}whose $fg$-th block matrix is{\footnotesize
\[
\Pi_{fg}^{w}=\left[\begin{array}{ccc}
\Pi_{fg(1,1)}^{w} & \cdots & \Pi_{fg(1,N)}^{w}\\
\vdots & \ddots & \vdots\\
\Pi_{fg(N,1)}^{w} & \cdots & \Pi_{fg(N,N)}^{w}
\end{array}\right],
\]
}and partition it as{\footnotesize
\[
\Pi^{w}=\left[\begin{array}{cc}
\Pi_{UL}^{w} & \Pi_{UR}^{w}\\
\Pi_{LL}^{w} & \Pi_{LR}^{w}
\end{array}\right],
\]
}where $\Pi_{UL}^{w}$, $\Pi_{UR}^{w}$, $\Pi_{LL}^{w}$, and $\Pi_{LR}^{w}$
are $F_{\ell}N\times F_{\ell}N$, $F_{\ell}N\times F_{k}N$, $F_{k}N\times F_{\ell}N$,
and $F_{k}N\times F_{k}N$ matrices, respectively. The upper-left
and the upper-right submatrices are{\footnotesize
\begin{align*}
\Pi_{UL}^{w}= & \left[\begin{array}{ccc}
\Pi_{UL,11}^{w} & \cdots & \Pi_{UL,1F_{\ell}}^{w}\\
\vdots & \ddots & \vdots\\
\Pi_{UL,F_{\ell}1}^{w} & \cdots & \Pi_{UL,F_{\ell}F_{\ell}}^{w}
\end{array}\right]=\left[\begin{array}{ccc}
\Pi_{11}^{w} & \cdots & \Pi_{1F_{\ell}}^{w}\\
\vdots & \ddots & \vdots\\
\Pi_{F_{\ell}1}^{w} & \cdots & \Pi_{F_{\ell}F_{\ell}}^{w}
\end{array}\right],\\
\Pi_{UR}^{w}= & \left[\begin{array}{ccc}
\Pi_{UR,11}^{w} & \cdots & \Pi_{UR,1F_{k}}^{w}\\
\vdots & \ddots & \vdots\\
\Pi_{UR,F_{\ell}1}^{w} & \cdots & \Pi_{UR,F_{\ell}F_{k}}^{w}
\end{array}\right]=\left[\begin{array}{ccc}
\Pi_{1,F_{\ell}+1}^{w} & \cdots & \Pi_{1,F_{\ell}+F_{k}}^{w}\\
\vdots & \ddots & \vdots\\
\Pi_{F_{\ell},F_{\ell}+1}^{w} & \cdots & \Pi_{F_{\ell},F_{\ell}+F_{k}}^{w}
\end{array}\right],
\end{align*}
}where both $\Pi_{UL,fg}^{w}$ and $\Pi_{UR,fg}^{w}$ are $N\times N$
matrices. Equation \eqref{eq: D.WfnLfn/PY-Wfn} can then be rewritten
as
\begin{multline*}
d\ln\Lambda_{\ell_{f}n}=\sum_{g=1}^{F_{\ell}}\sum_{m=1}^{N}\Pi_{UL,fg\left(n,m\right)}^{w}d\ln\left(\frac{w_{gm}}{p}\right)+\sum_{g=1}^{F_{k}}\sum_{m=1}^{N}\Pi_{UR,fg\left(n,m\right)}^{w}d\ln\left(\frac{r_{g}}{p}\right)+\sum_{m=1}^{N}\Theta_{f\left(n,m\right)}^{w}d\ln A_{m}\\
\text{for }f=1,\ldots,F_{\ell}.
\end{multline*}
Substituting equation \eqref{eq: D.Rf/P} into this yields
\begin{multline}
d\ln\Lambda_{\ell_{f}n}=\sum_{g=1}^{F_{\ell}}\sum_{m=1}^{N}\Pi_{UL,fg\left(n,m\right)}^{w}d\ln\left(\frac{w_{gm}}{p}\right)-\sum_{g=1}^{F_{k}}\sum_{m=1}^{N}\Pi_{UR,fg\left(n,m\right)}^{w}d\ln q_{g}+\sum_{m=1}^{N}\Theta_{f\left(n,m\right)}^{w}d\ln A_{m}\\
\text{for }f=1,\ldots,F_{\ell}.\label{eq: D.WfnLfn/PY-Wfn=000026Qf}
\end{multline}
Substituting $\Lambda_{\ell_{f}n}=w_{fn}\ell_{fn}/py$ into this yields
\begin{multline*}
\sum_{g=1}^{F_{\ell}}\sum_{m=1}^{N}\left(\Pi_{UL,fg\left(n,m\right)}^{w}-I_{\left(f,g\right)}I_{\left(n,m\right)}\right)d\ln\left(\frac{w_{gm}}{p}\right)=\sum_{g=1}^{F_{\ell}}\sum_{m=1}^{N}I_{\left(f,g\right)}I_{\left(n,m\right)}d\ln\ell_{gm}\\
+\sum_{g=1}^{F_{k}}\sum_{m=1}^{N}\Pi_{UR,fg\left(n,m\right)}^{w}d\ln q_{g}-\sum_{m=1}^{N}\Theta_{f\left(n,m\right)}^{w}d\ln A_{m}-d\ln y\quad\text{for }f=1,\ldots,F_{\ell}.
\end{multline*}
Subtracting the equation for $(f,n)=(F_{\ell},N)$ from that for any
$f\in\{1,\ldots,F_{\ell}-1\}$ and $n\in\{1,\ldots,N\}$ or for any
$f=F_{\ell}$ and $n\in\{1,\ldots,N-1\}$ yields
\begin{multline}
\sum_{g=1}^{F_{\ell}}\sum_{m=1}^{N}\left[\Pi_{UL,fg\left(n,m\right)}^{w}-\Pi_{UL,F_{\ell}g\left(N,m\right)}^{w}-I_{\left(f,g\right)}I_{\left(n,m\right)}+I_{\left(F_{\ell},g\right)}I_{\left(N,m\right)}\right]d\ln\left(\frac{w_{gm}}{p}\right)\\
=\sum_{g=1}^{F_{\ell}}\sum_{m=1}^{N}\left(I_{\left(f,g\right)}I_{\left(n,m\right)}-I_{\left(F_{\ell},g\right)}I_{\left(N,m\right)}\right)d\ln\ell_{gm}+\sum_{g=1}^{F_{k}}\sum_{m=1}^{N}\left(\Pi_{UR,fg\left(n,m\right)}^{w}-\Pi_{UR,F_{\ell}g\left(N,m\right)}^{w}\right)d\ln q_{g}\\
-\sum_{m=1}^{N}\left(\Theta_{f\left(n,m\right)}^{w}-\Theta_{F_{\ell}\left(N,m\right)}^{w}\right)d\ln A_{m}.\label{eq: D.Wfn/P-Qf=000026Lfn}
\end{multline}
Solving the system of equations \eqref{eq: D.Wfn/P-Qf} and \eqref{eq: D.Wfn/P-Qf=000026Lfn}
for $d\ln(w_{fn}/p)$ and setting $d\ln\ell_{fn}=0$ yields
\begin{equation}
d\ln\left(\frac{w_{fn}}{p}\right)=\sum_{g=1}^{F_{k}}\Omega_{f\left(n,g\right)}^{q}d\ln q_{g}+\sum_{m=1}^{N}\Omega_{f\left(n,m\right)}^{A}d\ln A_{m},\label{eq: D.Wfn/P-Qf'}
\end{equation}
where $\Omega_{f(n,g)}^{q}$ is the $ng$-th element of the $N\times F_{k}$
submatrix $\Omega_{f}^{q}$ of the $F_{\ell}N\times F_{k}$ matrix{\footnotesize
\begin{align*}
\Omega^{q}= & \left[\begin{array}{ccc}
\left(\Omega_{1}^{q}\right)^{\prime} & \cdots & \left(\Omega_{F_{\ell}}^{q}\right)^{\prime}\end{array}\right]^{\prime}\\
= & \left[\begin{array}{ccccc}
-\epsilon_{\ell_{F_{\ell}N}\ell_{11}}^{\mathcal{C}} & \epsilon_{\ell_{11}\ell_{12}}^{\mathcal{C}}-\epsilon_{\ell_{F_{\ell}N}\ell_{12}}^{\mathcal{C}} & \cdots & \epsilon_{\ell_{11}\ell_{F_{\ell},N-1}}^{\mathcal{C}}-\epsilon_{\ell_{F_{\ell}N}\ell_{F_{\ell},N-1}}^{\mathcal{C}} & \epsilon_{\ell_{11}\ell_{F_{\ell}N}}^{\mathcal{C}}\\
\epsilon_{\ell_{12}\ell_{11}}^{\mathcal{C}}-\epsilon_{\ell_{F_{\ell}N}\ell_{11}}^{\mathcal{C}} & -\epsilon_{\ell_{F_{\ell}N}\ell_{12}}^{\mathcal{C}} & \cdots & \epsilon_{\ell_{12}\ell_{F_{\ell},N-1}}^{\mathcal{C}}-\epsilon_{\ell_{F_{\ell}N}\ell_{F_{\ell},N-1}}^{\mathcal{C}} & \epsilon_{\ell_{12}\ell_{F_{\ell}N}}^{\mathcal{C}}\\
\vdots & \vdots & \ddots & \vdots & \vdots\\
\epsilon_{\ell_{F_{\ell},N-1}\ell_{11}}^{\mathcal{C}}-\epsilon_{\ell_{F_{\ell}N}\ell_{11}}^{\mathcal{C}} & \epsilon_{\ell_{F_{\ell},N-1}\ell_{12}}^{\mathcal{C}}-\epsilon_{\ell_{F_{\ell}N}\ell_{12}}^{\mathcal{C}} & \cdots & -\epsilon_{\ell_{F_{\ell}N}\ell_{F_{\ell},N-1}}^{\mathcal{C}} & \epsilon_{\ell_{F_{\ell},N-1}\ell_{F_{\ell}N}}^{\mathcal{C}}\\
\Lambda_{\ell_{1}1} & \Lambda_{\ell_{1}2} & \cdots & \Lambda_{\ell_{F_{\ell}}N-1} & \Lambda_{\ell_{F_{\ell}}N}
\end{array}\right]^{-1}\\
 & \times\left[\begin{array}{ccc}
\epsilon_{\ell_{11}k_{1N}}^{\mathcal{C}}-\epsilon_{\ell_{F_{\ell}N}k_{1N}}^{\mathcal{C}} & \cdots & \epsilon_{\ell_{11}k_{F_{k}N}}^{\mathcal{C}}-\epsilon_{\ell_{F_{\ell}N}k_{F_{k}N}}^{\mathcal{C}}\\
\vdots & \vdots & \vdots\\
\epsilon_{\ell_{F_{\ell},N-1}k_{1N}}^{\mathcal{C}}-\epsilon_{\ell_{F_{\ell}N}k_{1N}}^{\mathcal{C}} & \cdots & \epsilon_{\ell_{F_{\ell},N-1}k_{F_{k}N}}^{\mathcal{C}}-\epsilon_{\ell_{F_{\ell}N}k_{F_{k}N}}^{\mathcal{C}}\\
\Lambda_{k_{1}} & \cdots & \Lambda_{k_{F_{k}}}
\end{array}\right],
\end{align*}
}and $\Omega_{f(n,m)}^{A}$ is the $nm$-th element of the $N\times N$
submatrix $\Omega_{f}^{A}$ of the $F_{\ell}N\times N$ matrix{\footnotesize
\begin{align*}
\Omega^{A}= & \left[\begin{array}{ccc}
\left(\Omega_{1}^{A}\right)^{\prime} & \cdots & \left(\Omega_{F_{\ell}}^{A}\right)^{\prime}\end{array}\right]^{\prime}\\
= & -\left[\begin{array}{ccccc}
-\epsilon_{\ell_{F_{\ell}N}\ell_{11}}^{\mathcal{C}} & \epsilon_{\ell_{11}\ell_{12}}^{\mathcal{C}}-\epsilon_{\ell_{F_{\ell}N}\ell_{12}}^{\mathcal{C}} & \cdots & \epsilon_{\ell_{11}\ell_{F_{\ell},N-1}}^{\mathcal{C}}-\epsilon_{\ell_{F_{\ell}N}\ell_{F_{\ell},N-1}}^{\mathcal{C}} & \epsilon_{\ell_{11}\ell_{F_{\ell}N}}^{\mathcal{C}}\\
\epsilon_{\ell_{12}\ell_{11}}^{\mathcal{C}}-\epsilon_{\ell_{F_{\ell}N}\ell_{11}}^{\mathcal{C}} & -\epsilon_{\ell_{F_{\ell}N}\ell_{12}}^{\mathcal{C}} & \cdots & \epsilon_{\ell_{12}\ell_{F_{\ell},N-1}}^{\mathcal{C}}-\epsilon_{\ell_{F_{\ell}N}\ell_{F_{\ell},N-1}}^{\mathcal{C}} & \epsilon_{\ell_{12}\ell_{F_{\ell}N}}^{\mathcal{C}}\\
\vdots & \vdots & \ddots & \vdots & \vdots\\
\epsilon_{\ell_{F_{\ell},N-1}\ell_{11}}^{\mathcal{C}}-\epsilon_{\ell_{F_{\ell}N}\ell_{11}}^{\mathcal{C}} & \epsilon_{\ell_{F_{\ell},N-1}\ell_{12}}^{\mathcal{C}}-\epsilon_{\ell_{F_{\ell}N}\ell_{12}}^{\mathcal{C}} & \cdots & -\epsilon_{\ell_{F_{\ell}N}\ell_{F_{\ell},N-1}}^{\mathcal{C}} & \epsilon_{\ell_{F_{\ell},N-1}\ell_{F_{\ell}N}}^{\mathcal{C}}\\
\Lambda_{\ell_{1}1} & \Lambda_{\ell_{1}2} & \cdots & \Lambda_{\ell_{F_{\ell}}N-1} & \Lambda_{\ell_{F_{\ell}}N}
\end{array}\right]^{-1}\\
 & \times\left[\begin{array}{ccc}
\Theta_{1\left(1,1\right)}^{w}-\Theta_{F_{\ell}\left(N,1\right)}^{w} & \cdots & \Theta_{1\left(1,N\right)}^{w}-\Theta_{F_{\ell}\left(N,N\right)}^{w}\\
\vdots & \vdots & \vdots\\
\Theta_{F_{\ell}\left(N-1,1\right)}^{w}-\Theta_{F_{\ell}\left(N,1\right)}^{w} & \cdots & \Theta_{F_{\ell}\left(N-1,N\right)}^{w}-\Theta_{F_{\ell}\left(N,N\right)}^{w}\\
\zeta_{1} & \cdots & \zeta_{N}
\end{array}\right].
\end{align*}
}The log-linearization of the average wages of labor $f$, $\overline{w}_{f}=\sum_{n=1}^{N}w_{fn}\ell_{fn}/\sum_{n=1}^{N}\ell_{fn}$,
yields
\[
d\ln\overline{w}_{f}=\sum_{n=1}^{N}\left(\frac{\Lambda_{\ell_{f}n}}{\Lambda_{\ell_{f}}}\right)d\ln w_{fn}+\sum_{n=1}^{N}\left(\frac{\Lambda_{\ell_{f}n}}{\Lambda_{\ell_{f}}}-\frac{\ell_{fn}}{\ell_{f}}\right)d\ln\ell_{fn}.
\]
The log change of the average wage of labor $f$ relative to labor
$g$ can be written as
\[
d\ln\left(\frac{\overline{w}_{f}}{\overline{w}_{g}}\right)=\sum_{n=1}^{N}\left(\frac{\Lambda_{\ell_{f}n}}{\Lambda_{\ell_{f}}}\right)d\ln w_{fn}-\sum_{n=1}^{N}\left(\frac{\Lambda_{\ell_{g}n}}{\Lambda_{\ell_{g}}}\right)d\ln w_{gn}+\sum_{n=1}^{N}\left(\frac{\Lambda_{\ell_{f}n}}{\Lambda_{\ell_{f}}}-\frac{\ell_{fn}}{\ell_{f}}\right)d\ln\ell_{fn}-\sum_{n=1}^{N}\left(\frac{\Lambda_{\ell_{g}n}}{\Lambda_{\ell_{g}}}-\frac{\ell_{gn}}{\ell_{g}}\right)d\ln\ell_{gn}.
\]
Substituting equation \eqref{eq: D.Wfn/P-Qf'} into this and setting
$d\ln\ell_{fn}=0$ yields equation \eqref{eq: D.Wf/Wg-Qf''}.

The log-linearization of the factor share, $\Lambda_{\ell_{f}}=\sum_{n=1}^{N}\Lambda_{\ell_{f}n}$,
yields
\[
d\ln\Lambda_{\ell_{f}}=\sum_{n=1}^{N}\frac{\Lambda_{\ell_{f}n}}{\Lambda_{\ell_{f}}}d\ln\Lambda_{\ell_{f}n}.
\]
Substituting equation \eqref{eq: D.WfnLfn/PY-Wfn=000026Qf} into this
yields
\begin{multline*}
d\ln\Lambda_{\ell_{f}}=\sum_{g=1}^{F_{\ell}}\sum_{m=1}^{N}\left(\sum_{n=1}^{N}\frac{\Lambda_{\ell_{f}n}}{\Lambda_{\ell_{f}}}\Pi_{UL,fg\left(n,m\right)}^{w}\right)d\ln\left(\frac{w_{gm}}{p}\right)\\
-\sum_{g=1}^{F_{k}}\sum_{m=1}^{N}\left(\sum_{n=1}^{N}\frac{\Lambda_{\ell_{f}n}}{\Lambda_{\ell_{f}}}\Pi_{UR,fg\left(n,m\right)}^{w}\right)d\ln q_{g}+\sum_{m=1}^{N}\left(\sum_{n=1}^{N}\frac{\Lambda_{\ell_{f}n}}{\Lambda_{\ell_{f}}}\Theta_{f\left(n,m\right)}^{w}\right)d\ln A_{m}.
\end{multline*}
Substituting equation \eqref{eq: D.Wfn/P-Qf'} into this yields equation
\eqref{eq: D.WfLf/PY-Qf''}, where
\begin{align*}
\Pi_{UL,ff\left(n,n\right)}^{w}= & \sum_{l\neq n}^{N}\Lambda_{\ell_{f}l}\left(1-\epsilon_{\ell_{fl}\ell_{fn}}^{\mathcal{C}}\right)+\sum_{l=1}^{N}\sum_{h\neq f}^{F}\Lambda_{\ell_{h}l}\left(1-\epsilon_{\ell_{hl}\ell_{fn}}^{\mathcal{C}}\right),\\
\Pi_{UL,ff\left(n,m\right)}^{w}= & \Lambda_{\ell_{f}m}\left(\epsilon_{\ell_{fn}\ell_{fm}}^{\mathcal{C}}-1\right)+\sum_{l\neq n,m}^{N}\Lambda_{\ell_{f}l}\left(\epsilon_{\ell_{fn}\ell_{fm}}^{\mathcal{C}}-\epsilon_{\ell_{fl}\ell_{fm}}^{\mathcal{C}}\right)+\sum_{l=1}^{N}\sum_{h\neq f}^{F}\Lambda_{\ell_{h}l}\left(\epsilon_{\ell_{fn}\ell_{fm}}^{\mathcal{C}}-\epsilon_{\ell_{hl}\ell_{fm}}^{\mathcal{C}}\right),\\
\Pi_{UL,fg\left(n,m\right)}^{w}= & \Lambda_{\ell_{g}m}\left(\epsilon_{\ell_{fn}\ell_{gm}}^{\mathcal{C}}-1\right)+\sum_{l\neq n}^{N}\Lambda_{\ell_{f}l}\left(\epsilon_{\ell_{fn}\ell_{gm}}^{\mathcal{C}}-\epsilon_{\ell_{fl}\ell_{gm}}^{\mathcal{C}}\right)+\sum_{l\neq m}^{N}\Lambda_{\ell_{g}l}\left(\epsilon_{\ell_{fn}\ell_{gm}}^{\mathcal{C}}-\epsilon_{g_{fl}\ell_{gm}}^{\mathcal{C}}\right)\\
 & +\sum_{l=1}^{N}\sum_{h\neq f,g}^{F}\Lambda_{\ell_{h}l}\left(\epsilon_{\ell_{fn}\ell_{gm}}^{\mathcal{C}}-\epsilon_{\ell_{hl}\ell_{gm}}^{\mathcal{C}}\right),
\end{align*}
\[
\sum_{m=1}^{N}\Pi_{UR,fg\left(n,m\right)}^{w}=\sum_{l=1}^{N}\Lambda_{k_{g}l}\left(\epsilon_{\ell_{fn}k_{gl}}^{\mathcal{C}}-1\right)+\sum_{l\neq n}^{N}\Lambda_{\ell_{f}l}\left(\epsilon_{\ell_{fn}k_{gl}}^{\mathcal{C}}-\epsilon_{\ell_{fl}k_{gl}}^{\mathcal{C}}\right)+\sum_{l=1}^{N}\sum_{h\neq f,g}^{F}\Lambda_{\ell_{h}l}\left(\epsilon_{\ell_{fn}k_{gl}}^{\mathcal{C}}-\epsilon_{\ell_{hl}k_{gl}}^{\mathcal{C}}\right).
\]

\section{Additional corollaries\label{sec: corolloaries}}
\begin{cor}
\label{cor: Oberfield-Raval}Suppose that production requires only
two factors. The aggregate elasticity of substitution \eqref{eq: morishima_y}
is identical to the aggregate elasticity of substitution \eqref{eq: morishima_c},
which can be expressed as a weighted average of the elasticities of
substitution in production and consumption:
\[
\epsilon_{\ell_{f}\ell_{g}}^{\mathcal{F}}=\epsilon_{\ell_{f}\ell_{g}}^{\mathcal{C}}=\sum_{n=1}^{N}\zeta_{n}\frac{\lambda_{\ell_{f}n}\lambda_{\ell_{g}n}}{\Lambda_{\ell_{f}}\Lambda_{\ell_{g}}}\epsilon_{\ell_{f}\ell_{g}n}^{\mathcal{C}}+\left(1-\sum_{n=1}^{N}\zeta_{n}\frac{\lambda_{\ell_{f}n}\lambda_{\ell_{g}n}}{\Lambda_{\ell_{f}}\Lambda_{\ell_{g}}}\right)\frac{1}{1-\eta}.
\]
\end{cor}
The first term represents the substitution between two factors within
sectors, and the second term represents the reallocation between sectors
with different factor intensities; the relative importance of the
two effects depends on the degree of factor intensity \citep{Oberfield_Raval_EM21}.
\begin{cor}
\label{cor: Hicks}Suppose that production requires only one type
of capital, $k$, and one type of labor, $\ell$. Let $r$ and $w$
denote their respective prices. Changes in the labor share are given
by
\[
\Delta\ln\Lambda_{\ell}=\frac{1-\epsilon}{\epsilon}\left(1-\Lambda_{\ell}\right)\Delta\ln\left(\frac{k}{\ell}\right)+\psi_{\left(\ell\right)}^{\ell}
\]
or
\[
\Delta\ln\Lambda_{\ell}=\left(1-\epsilon\right)\left(1-\Lambda_{\ell}\right)\Delta\ln\left(\frac{w}{r}\right)+\psi_{\left(\ell\right)}^{w},
\]
where $\epsilon$ is the aggregate elasticity of substitution between
capital and labor.
\end{cor}
In this case, the labor share can decline with a rise in the capital\textendash labor
ratio or the wage\textendash rental price ratio if $\epsilon>1$.
\citet{Karabarbounis_Neiman_QJE14} attribute the decline in the labor
share to capital-embodied technological change on the basis of findings
that the estimated elasticity of substitution exceeds one and that
the rental price of capital fell in many countries. \citet*{Elsby_Hobijn_Sahin_BPEA13}
attribute it to other factors on the basis of the observation that
the timing of decline in the labor share did not necessarily align
with the timing of capital deepening in the United States.
\begin{cor}
\label{cor: Katz_Murphy}Suppose that production requires one type
of capital ($k$) and two types of labor (skilled labor, $\ell_{h}$,
and unskilled labor, $\ell_{u}$), and the aggregate elasticities
of substitution satisfy $\epsilon_{\ell_{h}k}^{\mathcal{F}}=\epsilon_{\ell_{u}k}^{\mathcal{F}}=1$
and $\epsilon_{\ell_{h}\ell_{u}}^{\mathcal{F}}=\epsilon_{\ell_{u}\ell_{h}}^{\mathcal{F}}=\epsilon^{\mathcal{F}}$.
Equation \eqref{eq: D.Wf/Wg-Lf} becomes the \citet{Katz_Murphy_QJE92}
model:
\[
\Delta\ln\left(\frac{w_{h}}{w_{u}}\right)=-\frac{1}{\epsilon^{\mathcal{F}}}\Delta\ln\left(\frac{\ell_{h}}{\ell_{u}}\right)+\left(\psi_{\left(h\right)}^{\ell}-\psi_{\left(u\right)}^{\ell}\right).
\]
\end{cor}
This equation implies that the relative wages of skilled labor to
unskilled labor can increase only with disembodied technological change
biased towards skilled labor when the relative supply of skilled labor
to unskilled labor increases over time.
\begin{cor}
\label{cor: Krusell_etal}Suppose that production requires two types
of capital (capital equipment, $k_{e}$, and capital structures, $k_{s}$)
and two types of labor (skilled labor, $\ell_{h}$, and unskilled
labor, $\ell_{u}$), and the aggregate elasticities of substitution
satisfy $\epsilon_{\ell_{h}k_{s}}^{\mathcal{F}}=\epsilon_{\ell_{u}k_{s}}^{\mathcal{F}}=1$
and $1/\epsilon_{\ell_{u}\ell_{h}}^{\mathcal{F}}-1/\epsilon_{\ell_{h}\ell_{u}}^{\mathcal{F}}=1/\epsilon_{\ell_{h}k_{e}}^{\mathcal{F}}-1/\epsilon_{\ell_{u}k_{e}}^{\mathcal{F}}$.
Equation \eqref{eq: D.Wf/Wg-Lf} becomes the \citet*{Krusell_Ohanian_RiosRull_Violante_EM00}
model:
\[
\Delta\ln\left(\frac{w_{h}}{w_{u}}\right)=\left(\frac{\epsilon_{\ell_{u}k_{e}}^{\mathcal{F}}-\epsilon_{\ell_{h}k_{e}}^{\mathcal{F}}}{\epsilon_{\ell_{u}k_{e}}^{\mathcal{F}}\epsilon_{\ell_{h}k_{e}}^{\mathcal{F}}}\right)\Delta\ln\left(\frac{k_{e}}{\ell_{h}}\right)-\frac{1}{\epsilon_{\ell_{h}\ell_{u}}^{\mathcal{F}}}\Delta\ln\left(\frac{\ell_{h}}{\ell_{u}}\right)+\left(\psi_{\left(h\right)}^{\ell}-\psi_{\left(u\right)}^{\ell}\right).
\]
\end{cor}
This equation implies that the skill wage gap can increase with a
rise in capital equipment if capital equipment is more complementary
to skilled labor than to unskilled labor (i.e., $\epsilon_{\ell_{u}k_{e}}^{\mathcal{F}}>\epsilon_{\ell_{h}k_{e}}^{\mathcal{F}}$).
This effect is sufficiently large to account for a rise in the skill
premium in OECD countries when ICCT capital is distinguished from
non-ICCT capital \citep{Taniguchi_Yamada_LE22}.

\section{Data Description}

\subsection{Wages and hours worked\label{subsec: wages=000026hours}}

We adjust for changes in the age and education composition of the
labor force over time when we construct the data on wages and hours
worked. Each type of labor, $\ell_{f}$ for $f\in\left\{ mh,fh,mu,fu\right\} $,
can be divided into three age groups: young (aged between 15 and 29
years), middle (aged between 30 and 49 years), and old (aged 50 years
and older). In addition, unskilled labor consists of two education
groups: medium-skilled (entered college or completed high-school education)
and low-skilled (dropped out of high school or attended compulsory
education only), while skilled labor consists of one education group:
high-skilled (completed college). We assume that workers are perfect
substitutes within each type of labor.

Suppose that there is no need to make an adjustment to wages and hours
worked. Let checks denote unadjusted values. The wages for labor of
type $f$ in sector $n$, country $j$, and year $t$ could be calculated
as $\check{w}_{f,njt}=\sum_{a_{f}}\rho_{f,njt}^{a_{f}}\check{w}_{f,njt}^{a_{f}}$,
where $\rho_{f,njt}^{a_{f}}$ is the share of total hours worked by
group $a_{f}$ (i.e., $\rho_{f,njt}^{a_{f}}=\check{\ell}_{f,njt}^{a_{f}}/\sum_{a_{f}}\check{\ell}_{f,njt}^{a_{f}}$).
The hours worked by labor of type $f$ in sector $n$, country $j$,
and year $t$ could be calculated as $\check{\ell}_{f,njt}=\sum_{a_{f}}\check{\ell}_{f,njt}^{a_{f}}$.
We adjust for changes in the age and education composition of the
labor force by holding the share of each group constant when we calculate
wages and by using the time-invariant efficiency units as weights
when we calculate hours worked. Let $T_{j}$ denote the number of
years observed for country $j$. The composition-adjusted wages for
labor of type $f$ in sector $n$, country $j$, and year $t$ can
be calculated as $w_{f,njt}=\sum_{a_{f}}\overline{\rho}_{f,nj}^{a_{f}}\check{w}_{f,njt}^{a_{f}}$,
where $\overline{\rho}_{f,nj}^{a_{f}}$ is the sector- and country-specific
mean of $\rho_{f,njt}^{a_{f}}$ (i.e., $\overline{\rho}_{f,nj}^{a_{f}}=\sum_{t=1}^{T_{j}}\rho_{f,njt}^{a_{f}}/T_{j}$).
The composition-adjusted hours worked by labor of type $f$ in sector
$n$, country $j$, and year $t$ can be calculated as $\ell_{f,njt}=\sum_{a_{f}}(\overline{w}_{f,nj}^{a_{f}}/\overline{w}_{f,nj}^{a_{f^{\prime}}})\check{\ell}_{f,njt}^{a_{f}}$,
where the efficiency unit is measured by the sector- and country-specific
mean of $\check{w}_{f,njt}^{a_{f}}$ (i.e., $\overline{w}_{f,nj}^{a_{f}}=\sum_{t=1}^{T_{j}}\check{w}_{f,njt}^{a_{f}}/T_{j}$)
and normalized by $\overline{w}_{f,nj}^{a_{f^{\prime}}}$. We choose
middle-aged high-skilled labor as the base group for skilled labor
and middle-aged medium-skilled labor as the base group for unskilled
labor. Our results do not depend on the choice of the base group.

\subsection{Rental price of capital\label{subsec: rental_price}}

The rental price of capital ($r_{ft}$) relative to the output price
($p_{t}$) is determined by the relative investment price ($p_{ft}/p_{t}$),
depreciation rate ($\delta_{f}$), and interest rate ($\iota_{t}$)
for $f\in\left\{ i,o\right\} $. The investment price is calculated
by dividing the nominal value by the real value of investment. The
depreciation rate is the time average of those obtained from the capital
law of motion.

We calculate the rental price of capital in two ways. First, we adopt
the internal rate of return approach of \citet{OMahony_Timmer_EJ09},
who calculate the rental price of capital as
\begin{equation}
\frac{r_{ft}}{p_{t}}=\delta_{f}\left(\frac{p_{ft}}{p_{t}}\right)+\iota_{t}\left(\frac{p_{f,t-1}}{p_{t-1}}\right)-\left(\frac{p_{ft}}{p_{t}}-\frac{p_{f,t-1}}{p_{t-1}}\right),\label{eq: Euler}
\end{equation}
where the interest rate is the internal rate of return:
\[
\iota_{t}=\frac{\sum_{f}\left(\left.r_{ft}\right/p_{t}\right)k_{ft}-\sum_{f}\delta_{f}\left(\left.p_{ft}\right/p_{t}\right)k_{ft}+\sum_{f}\left(\left.p_{ft}\right/p_{t}-\left.p_{f,t-1}\right/p_{t-1}\right)k_{ft}}{\sum_{f}\left(\left.p_{f,t-1}\right/p_{t-1}\right)k_{ft}}.
\]
Second, we adopt the external rate of return approach of \citet{Niebel_Saam_RIW16},
who calculate the rental price of capital as
\[
\frac{r_{ft}}{p_{t}}=\delta_{f}\left(\frac{p_{ft}}{p_{t}}\right)+\iota_{t}\left(\frac{p_{f,t-1}}{p_{t-1}}\right)-\frac{1}{2}\left[\ln\left(\frac{p_{ft}}{p_{t}}\right)-\ln\left(\frac{p_{f,t-2}}{p_{t-2}}\right)\right]\left(\frac{p_{f,t-1}}{p_{t-1}}\right),
\]
and the interest rate as
\[
\iota_{t}=0.04+\frac{1}{5}\sum_{\tau=-2}^{2}\frac{cpi_{t-\tau}-cpi_{t-\tau-1}}{cpi_{t-\tau-1}},
\]
where $cpi$ is the consumer price index. The data can be obtained
from OECD.Stat.

\subsection{Non-ICCT equipment\label{subsec: nonICT}}

\begin{figure}[h]
\caption{ICCT equipment, non-ICCT equipment, and non-ICCT structures\label{fig: capital'}}

\begin{centering}
\subfloat[Investment prices\label{fig: capital_price'}]{
\centering{}\includegraphics[scale=0.6]{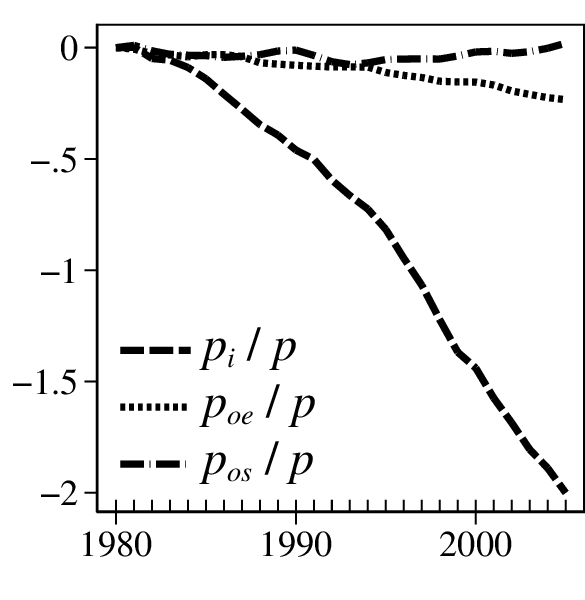}}\qquad{}\subfloat[Quantities\label{fig: capital_quantity'}]{
\centering{}\includegraphics[scale=0.6]{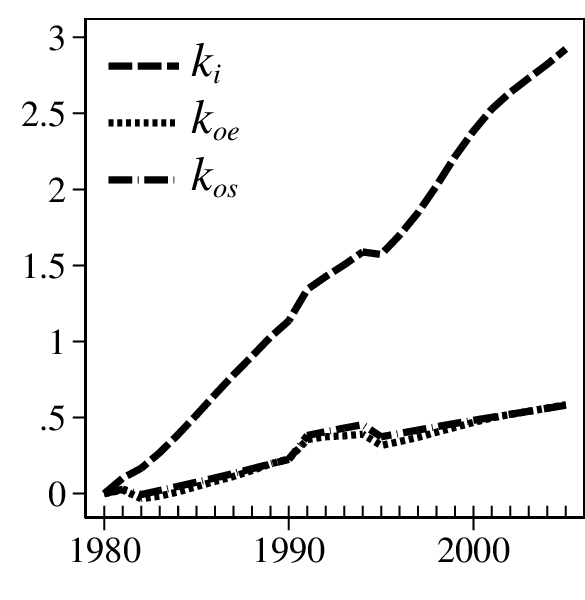}}
\par\end{centering}
\begin{singlespace}
\raggedright{}{\footnotesize\textit{Notes}}{\footnotesize : The investment
prices of ICCT equipment, non-ICCT equipment, and non-ICCT structures
are denoted by $p_{i}$, $p_{oe}$, and $p_{os}$, respectively, and
the output price is denoted by $p$. The quantities of ICCT equipment,
non-ICCT equipment, and non-ICCT structures are denoted by $k_{i}$,
$k_{oe}$, and $k_{os}$, respectively. All the series are logarithmically
transformed and normalized to zero in the year 1980.}{\footnotesize\par}
\end{singlespace}
\end{figure}

The trends in the rental prices of capital differ significantly between
ICCT equipment and non-ICCT structures but do not differ significantly
between non-ICCT equipment and non-ICCT structures (Figure \ref{fig: capital_price'}).
The rental price of ICCT equipment fell dramatically, but that of
non-ICCT equipment and structures remained almost unchanged. Meanwhile,
the quantities of ICCT equipment, non-ICCT equipment, and non-ICCT
structures increased. However, the rate of increase in ICCT equipment
is far greater than that in non-ICCT equipment and structures (Figure
\ref{fig: capital_quantity'}). The rate of increase in non-ICCT equipment
is almost the same as that in non-ICCT structures.

\section{Specification Issues}

This section is divided into four parts. First, we examine the consistency
of the relative magnitude of substitution parameters across different
specifications and discuss the empirical relevance of alternative
specifications. Second, we discuss the error terms in the estimating
equations. Third, we describe the weights on the substitution parameters
in the aggregate elasticities of substitution. Finally, we consider
the magnitude of the effects of technological change.

\subsection{Substitution parameters\label{subsec: specification1}}

As a starting point, we consider the use of a one-level CES function:
\begin{equation}
y=Ak_{o}^{\vartheta_{o}}\left[\left(1-\vartheta_{fh}-\vartheta_{mh}-\vartheta_{fu}-\vartheta_{mu}\right)k_{i}^{\varsigma}+\vartheta_{fh}\ell_{fh}^{\varsigma}+\vartheta_{mh}\ell_{mh}^{\varsigma}+\vartheta_{fu}\ell_{fu}^{\varsigma}+\vartheta_{mu}\ell_{mu}^{\varsigma}\right]^{\frac{1-\vartheta_{o}}{\varsigma}},\label{eq: ces1}
\end{equation}
where the degree of substitution between ICCT capital and the four
types of labor is governed by a single parameter $\varsigma$. We
suppress the sector subscript $n$ for notational simplicity but allow
all parameters to vary across sectors. Cost minimization implies that
the marginal rate of technical substitution of labor for capital equals
the relative price of labor to capital for all types of labor. This
condition leads to the following first-difference equations:
\begin{align}
\Delta\ln\left(\frac{w_{fh}}{r_{i}}\right) & =-\left(1-\varsigma_{fh}\right)\Delta\ln\left(\frac{\ell_{fh}}{k_{i}}\right)+\Delta\nu_{fh},\label{eq: D.Wfh/Ri_ces1}\\
\Delta\ln\left(\frac{w_{mh}}{r_{i}}\right) & =-\left(1-\varsigma_{mh}\right)\Delta\ln\left(\frac{\ell_{mh}}{k_{i}}\right)+\Delta\nu_{mh},\label{eq: D.Wmh/Ri_ces1}\\
\Delta\ln\left(\frac{w_{fu}}{r_{i}}\right) & =-\left(1-\varsigma_{fu}\right)\Delta\ln\left(\frac{\ell_{fu}}{k_{i}}\right)+\Delta\nu_{fu},\label{eq: D.Wfu/Ri_ces1}\\
\Delta\ln\left(\frac{w_{mu}}{r_{i}}\right) & =-\left(1-\varsigma_{mu}\right)\Delta\ln\left(\frac{\ell_{mu}}{k_{i}}\right)+\Delta\nu_{mu}.\label{eq: D.Wmu/Ri_ces1}
\end{align}
The substitution parameter $\varsigma$ can be estimated using any
one of equations \eqref{eq: D.Wfh/Ri_ces1}\textendash \eqref{eq: D.Wmu/Ri_ces1}.
If the production function \eqref{eq: ces1} is correctly specified,
the coefficients should be the same for all equations. Table \ref{tab: specification}
presents the results of parameter estimates and hypothesis testing
in different specifications. The substitution parameters are estimated
jointly by GMM using instruments described in Section \ref{subsec: identification}.
Panel A of Table \ref{tab: specification} indicates $\varsigma_{fh}<\varsigma_{mh}<\varsigma_{fu}<\varsigma_{mu}$
for each sector. The null hypothesis of $\varsigma_{fh}=\varsigma_{mh}=\varsigma_{fu}=\varsigma_{mu}$
is strongly rejected for manufacturing, utilities, and services sectors.
The specification \eqref{eq: ces1} is not valid for the three sectors,
though it is for construction sector.

Given the results above, we extend the production function \eqref{eq: ces1}
by incorporating a two-level CES function for the three sectors:
\begin{equation}
y=Ak_{o}^{\alpha}\left[\left(1-\vartheta_{mh}-\vartheta_{fu}-\vartheta_{mu}\right)D^{\varsigma^{\prime}}+\vartheta_{mh}\ell_{mh}^{\varsigma^{\prime}}+\vartheta_{fu}\ell_{fu}^{\varsigma^{\prime}}+\vartheta_{mu}\ell_{mu}^{\varsigma^{\prime}}\right]^{\frac{1-\alpha}{\varsigma^{\prime}}},\label{eq: ces2}
\end{equation}
where the degree of substitution between $k_{i}$ and the three types
of labor ($\ell_{mh}$, $\ell_{fu}$, and $\ell_{mu}$) is governed
by a parameter $\varsigma^{\prime}$. The degree of substitution between
$k_{i}$ and $\ell_{fh}$ is governed by the parameter $\sigma_{fh}$,
as in equation \eqref{eq: nestD}. Equations \eqref{eq: D.Wfh/Ri_ces1}\textendash \eqref{eq: D.Wmu/Ri_ces1}
are modified as follows:
\begin{align}
\Delta\ln\left(\frac{w_{fh}}{r_{i}}\right) & =-\left(1-\sigma_{fh}\right)\Delta\ln\left(\frac{\ell_{fh}}{k_{i}}\right)+\Delta\nu_{fh},\label{eq: D.Wfh/Ri_ces2}\\
\Delta\ln\left(\frac{w_{mh}}{w_{fh}}\right) & =-\left(1-\varsigma_{mh}^{\prime}\right)\Delta\ln\left(\frac{\ell_{mh}}{D}\right)+\left(1-\sigma_{fh}\right)\Delta\ln\left(\frac{\ell_{fh}}{D}\right)+\Delta\nu_{mh}^{\prime},\label{eq: D.Wmh/Wfh_ces2}\\
\Delta\ln\left(\frac{w_{fu}}{w_{fh}}\right) & =-\left(1-\varsigma_{fu}^{\prime}\right)\Delta\ln\left(\frac{\ell_{fu}}{D}\right)+\left(1-\sigma_{fh}\right)\Delta\ln\left(\frac{\ell_{fh}}{D}\right)+\Delta\nu_{fu}^{\prime},\label{eq: D.Wfu/Wfh_ces2}\\
\Delta\ln\left(\frac{w_{mu}}{w_{fh}}\right) & =-\left(1-\varsigma_{mu}^{\prime}\right)\Delta\ln\left(\frac{\ell_{mu}}{D}\right)+\left(1-\sigma_{fh}\right)\Delta\ln\left(\frac{\ell_{fh}}{D}\right)+\Delta\nu_{mu}^{\prime}.\label{eq: D.Wmu/Wfh_ces2}
\end{align}
The substitution parameter $\varsigma^{\prime}$ can be estimated
using any one of equations \eqref{eq: D.Wmh/Wfh_ces2}\textendash \eqref{eq: D.Wmu/Wfh_ces2}.
If the production function \eqref{eq: ces2} is correctly specified,
the coefficients on the first terms of the last three equations should
be the same. Panel B also indicates $\sigma_{fh}<\varsigma_{mh}^{\prime}<\varsigma_{fu}^{\prime}<\varsigma_{mu}^{\prime}$
for each sector. The null hypothesis of $\varsigma_{mh}^{\prime}=\varsigma_{fu}^{\prime}=\varsigma_{mu}^{\prime}$
is strongly rejected for manufacturing, utilities, and services sectors.
Specification \eqref{eq: ces2} is not valid for the three sectors.

Given the results above, we further extend the production function
\eqref{eq: ces2} by incorporating a three-level CES function for
the three sectors:
\begin{equation}
y=Ak_{o}^{\alpha}\left[\left(1-\vartheta_{fu}-\vartheta_{mu}\right)C^{\varsigma^{\prime\prime}}+\vartheta_{fu}\ell_{fu}^{\varsigma^{\prime\prime}}+\vartheta_{mu}\ell_{mu}^{\varsigma^{\prime\prime}}\right]^{\frac{1-\alpha}{\varsigma^{\prime\prime}}},\label{eq: ces3}
\end{equation}
where the degree of substitution between $k_{i}$ and the two types
of labor ($\ell_{fu}$ and $\ell_{mu}$) is governed by a parameter
$\varsigma^{\prime\prime}$. The degree of substitution between $k_{i}$
and $\ell_{mh}$ is governed by the parameter $\sigma_{mh}$ , as
in equation \eqref{eq: nestC}. Equations \eqref{eq: D.Wfh/Ri_ces2}\textendash \eqref{eq: D.Wmu/Wfh_ces2}
are modified as follows:
\begin{align}
\Delta\ln\left(\frac{w_{fh}}{r_{i}}\right) & =-\left(1-\sigma_{fh}\right)\Delta\ln\left(\frac{\ell_{fh}}{k_{i}}\right)+\Delta\nu_{fh},\label{eq: D.Wfh/Ri_ces3}\\
\Delta\ln\left(\frac{w_{mh}}{w_{fh}}\right) & =-\left(1-\sigma_{mh}\right)\Delta\ln\left(\frac{\ell_{mh}}{D}\right)+\left(1-\sigma_{fh}\right)\Delta\ln\left(\frac{\ell_{fh}}{D}\right)+\Delta\nu_{mh}^{\prime},\label{eq: D.Wmh/Wfh_ces3}\\
\Delta\ln\left(\frac{w_{fu}}{w_{fh}}\right) & =-\left(1-\varsigma_{fu}^{\prime\prime}\right)\Delta\ln\left(\frac{\ell_{fu}}{C}\right)+\left(1-\sigma_{fh}\right)\Delta\ln\left(\frac{\ell_{fh}}{D}\right)-\left(1-\sigma_{mh}\right)\Delta\ln\left(\frac{C}{D}\right)+\Delta\nu_{fu}^{\prime\prime},\label{eq: D.Wfu/Wfh_ces3}\\
\Delta\ln\left(\frac{w_{mu}}{w_{fh}}\right) & =-\left(1-\varsigma_{mu}^{\prime\prime}\right)\Delta\ln\left(\frac{\ell_{mu}}{C}\right)+\left(1-\sigma_{fh}\right)\Delta\ln\left(\frac{\ell_{fh}}{D}\right)-\left(1-\sigma_{mh}\right)\Delta\ln\left(\frac{C}{D}\right)+\Delta\nu_{mu}^{\prime\prime}.\label{eq: D.Wmu/Wfh_ces3}
\end{align}
The substitution parameter $\varsigma^{\prime\prime}$ can be estimated
using either equation \eqref{eq: D.Wfu/Wfh_ces3} or \eqref{eq: D.Wmu/Wfh_ces3}.
If the production function \eqref{eq: ces3} is correctly specified,
the coefficients on the first terms of the last two equations should
be the same. Panel C again indicates $\sigma_{fh}<\sigma_{mh}<\varsigma_{fu}^{\prime\prime}<\varsigma_{mu}^{\prime\prime}$
for each sector. The null hypothesis of $\varsigma_{fu}^{\prime\prime}=\varsigma_{mu}^{\prime\prime}$
is rejected for manufacturing, utilities, and services sectors. Specification
\eqref{eq: ces3} is not valid for the three sectors.

Finally, we extend the production function \eqref{eq: ces3} by incorporating
a four-level CES function, as in equation \eqref{eq: ces4}. Panel
D indicates $\sigma_{fh}<\sigma_{mh}<\sigma_{fu}<\sigma_{mu}$ for
each sector. The null hypotheses of $\sigma_{fh}=\sigma_{mh}$, $\sigma_{mh}=\sigma_{fu}$,
and $\sigma_{fu}=\sigma_{mu}$ are strongly rejected for manufacturing,
utilities, and services sectors. The results are consistent with specification
\eqref{eq: ces4}. The relative magnitude of the estimated substitution
parameters is consistent across specifications.

\begin{table}[ph]
\caption{Production function estimates in alternative specifications\label{tab: specification}}

\begin{centering}
\resizebox{1\textwidth}{!}{%
\begin{tabular}{lcccccccc}
\hline 
 & \multicolumn{4}{c}{Parameter estimates} &  & \multicolumn{3}{c}{Hypothesis testing}\tabularnewline
\cline{2-5}\cline{7-9}
 & \multicolumn{4}{c}{A. One-level CES} &  & \multicolumn{3}{c}{A. One-level CES}\tabularnewline
 & $\varsigma_{fh}$ & $\varsigma_{mh}$ & $\varsigma_{fu}$ & $\varsigma_{mu}$ &  & \multicolumn{3}{c}{$\varsigma_{fh}=\varsigma_{mh}=\varsigma_{fu}=\varsigma_{mu}$}\tabularnewline
\cline{2-5}\cline{7-9}
Manufacturing & \textendash 0.689 & 0.154 & 0.349 & 0.380 &  & \multicolumn{3}{c}{70.031}\tabularnewline
 & (0.319) & (0.057) & (0.055) & (0.048) &  & \multicolumn{3}{c}{{[}0.000{]}}\tabularnewline
Utilities & \textendash 1.120 & 0.081 & 0.221 & 0.339 &  & \multicolumn{3}{c}{31.199}\tabularnewline
 & (0.391) & (0.130) & (0.091) & (0.078) &  & \multicolumn{3}{c}{{[}0.000{]}}\tabularnewline
Construction & 0.096 & 0.453 & 0.450 & 0.507 &  & \multicolumn{3}{c}{3.229}\tabularnewline
 & (0.320) & (0.126) & (0.117) & (0.097) &  & \multicolumn{3}{c}{{[}0.358{]}}\tabularnewline
Services & \textendash 0.402 & 0.098 & 0.259 & 0.356 &  & \multicolumn{3}{c}{66.105}\tabularnewline
 & (0.128) & (0.063) & (0.051) & (0.054) &  & \multicolumn{3}{c}{{[}0.000{]}}\tabularnewline
\cline{2-5}\cline{7-9}
 & \multicolumn{4}{c}{B. Two-level CES} &  & \multicolumn{3}{c}{B. Two-level CES}\tabularnewline
 & $\sigma_{fh}$ & $\varsigma_{mh}^{\prime}$ & $\varsigma_{fu}^{\prime}$ & $\varsigma_{mu}^{\prime}$ &  & \multicolumn{3}{c}{$\varsigma_{mh}^{\prime}=\varsigma_{fu}^{\prime}=\varsigma_{mu}^{\prime}$}\tabularnewline
\cline{2-5}\cline{7-9}
Manufacturing & \textendash 0.689 & 0.287 & 0.467 & 0.506 &  & \multicolumn{3}{c}{105.444}\tabularnewline
 & (0.319) & (0.046) & (0.047) & (0.046) &  & \multicolumn{3}{c}{{[}0.000{]}}\tabularnewline
Utilities & \textendash 1.120 & 0.189 & 0.303 & 0.419 &  & \multicolumn{3}{c}{34.979}\tabularnewline
 & (0.391) & (0.105) & (0.071) & (0.064) &  & \multicolumn{3}{c}{{[}0.000{]}}\tabularnewline
Services & \textendash 0.402 & 0.343 & 0.499 & 0.607 &  & \multicolumn{3}{c}{56.943}\tabularnewline
 & (0.128) & (0.070) & (0.047) & (0.053) &  & \multicolumn{3}{c}{{[}0.000{]}}\tabularnewline
\cline{2-5}\cline{7-9}
 & \multicolumn{4}{c}{C. Three-level CES} &  & \multicolumn{3}{c}{C. Three-level CES}\tabularnewline
 & $\sigma_{fh}$ & $\sigma_{mh}$ & $\varsigma_{fu}^{\prime\prime}$ & $\varsigma_{mu}^{\prime\prime}$ &  & \multicolumn{3}{c}{$\varsigma_{fu}^{\prime\prime}=\varsigma_{mu}^{\prime\prime}$}\tabularnewline
\cline{2-5}\cline{7-9}
Manufacturing & \textendash 0.689 & 0.287 & 0.597 & 0.673 &  & \multicolumn{3}{c}{8.765}\tabularnewline
 & (0.319) & (0.046) & (0.047) & (0.051) &  & \multicolumn{3}{c}{{[}0.003{]}}\tabularnewline
Utilities & \textendash 1.120 & 0.189 & 0.384 & 0.560 &  & \multicolumn{3}{c}{38.617}\tabularnewline
 & (0.391) & (0.105) & (0.067) & (0.062) &  & \multicolumn{3}{c}{{[}0.000{]}}\tabularnewline
Services & \textendash 0.402 & 0.343 & 0.601 & 0.755 &  & \multicolumn{3}{c}{40.649}\tabularnewline
 & (0.128) & (0.070) & (0.038) & (0.049) &  & \multicolumn{3}{c}{{[}0.000{]}}\tabularnewline
\cline{2-5}\cline{7-9}
 & \multicolumn{4}{c}{D. Four-level CES} &  & \multicolumn{3}{c}{D. Four-level CES}\tabularnewline
 & $\sigma_{fh}$ & $\sigma_{mh}$ & $\sigma_{fu}$ & $\sigma_{mu}$ &  & $\sigma_{fh}=\sigma_{mh}$ & $\sigma_{mh}=\sigma_{fu}$ & $\sigma_{fu}=\sigma_{mu}$\tabularnewline
\cline{2-5}\cline{7-9}
Manufacturing & \textendash 0.689 & 0.288 & 0.600 & 0.776 &  & 10.540 & 108.465 & 14.488\tabularnewline
 & (0.319) & (0.046) & (0.043) & (0.054) &  & {[}0.001{]} & {[}0.000{]} & {[}0.000{]}\tabularnewline
Utilities & \textendash 0.962 & 0.184 & 0.404 & 0.631 &  & 15.300 & 6.140 & 23.543\tabularnewline
 & (0.362) & (0.104) & (0.063) & (0.058) &  & {[}0.000{]} & {[}0.013{]} & {[}0.000{]}\tabularnewline
Services & \textendash 0.413 & 0.340 & 0.598 & 0.861 &  & 25.515 & 22.783 & 61.220\tabularnewline
 & (0.134) & (0.067) & (0.034) & (0.048) &  & {[}0.000{]} & {[}0.000{]} & {[}0.000{]}\tabularnewline
\hline 
\end{tabular}}
\par\end{centering}
\raggedright{}{\footnotesize\textit{Notes}}{\footnotesize : The first
to fourth columns report the estimates of the substitution parameters
with standard errors clustered at the country level in parentheses.
The fifth to seventh columns report Wald statistics under the stated
hypotheses with }{\footnotesize\textit{p}}{\footnotesize -values in
square brackets.}{\footnotesize\par}
\end{table}

\subsection{Nesting structure\label{subsec: specification2}}

Our specification \eqref{eq: ces4} is a natural extension of the
one used in \citet{Fallon_Layard_JPE75} and \citet{Krusell_Ohanian_RiosRull_Violante_EM00}.
They estimate the production function of the form:
\begin{equation}
y=Ak_{o}^{\vartheta_{o}}\left[\left(1-\vartheta_{u}\right)\left[\left(1-\vartheta_{h}\right)k_{i}^{\varsigma_{h}}+\vartheta_{h}\ell_{h}^{\varsigma_{h}}\right]^{\frac{\varsigma_{u}}{\varsigma_{h}}}+\vartheta_{u}\ell_{u}^{\varsigma_{u}}\right]^{\frac{1-\vartheta_{o}}{\varsigma_{u}}},\label{eq: KORV}
\end{equation}
where the two substitution parameters ($\varsigma_{h}$, $\varsigma_{u}$)
are less than one, and the three share parameters ($\vartheta_{h}$,
$\vartheta_{u}$, and $\vartheta_{0}$) lie between zero and one.
There are three remarks regarding the specification of the production
function \eqref{eq: KORV}. First, new equipment is located inside
the CES function to allow for the possibility that capital-embodied
technological change may influence the skill wage gap. Second, new
equipment is placed in the lowest nest to allow for the possibility
that it may be more complementary to skilled labor than to unskilled
labor. Finally, the production function is confirmed to be concave
(i.e., the estimates of the substitution parameters are less than
one); however, not so if unskilled labor is placed in a nest lower
than that of skilled labor. This condition ensures that the relative
price of skills decreases with a rise in the relative supply of skills.

The production function \eqref{eq: KORV} proves to be useful in accounting
for changes in the skill premium in the United States and other advanced
countries \citep{Krusell_Ohanian_RiosRull_Violante_EM00,Taniguchi_Yamada_LE22}.
However, it is limited in that male and female labor are assumed to
be perfect substitutes. One simple extension is to incorporate the
following CES function into equation \eqref{eq: KORV}:
\begin{align*}
\ell_{h} & =\left[\left(1-\vartheta_{fh}\right)\ell_{mh}^{\varsigma_{fh}}+\vartheta_{fh}\ell_{fh}^{\varsigma_{fh}}\right]^{\frac{1}{\varsigma_{fh}}},\\
\ell_{u} & =\left[\left(1-\vartheta_{fu}\right)\ell_{mu}^{\varsigma_{fu}}+\vartheta_{fu}\ell_{fu}^{\varsigma_{fu}}\right]^{\frac{1}{\varsigma_{fu}}},
\end{align*}
where the two substitution parameters ($\varsigma_{fh}$, $\varsigma_{fu}$)
are less than one, and the two share parameters ($\vartheta_{fh}$,
$\vartheta_{fu}$) lie between zero and one. This specification relaxes
the assumption that male and female labor are perfect substitutes
yet still imposes the assumption that male and female labor are equally
substitutable with ICCT capital for each skill type. This assumption
implies that technological change embodied in new equipment has no
effect on the gender wage gap. Both the assumption and the implication
are inconsistent with the data, as shown above. Therefore, there are
only two alternatives. One is our preferred specification \eqref{eq: ces4}.
The other is the following specification:
\[
y=Ak_{o}^{\vartheta_{o}}\left[\left(1-\vartheta_{fu}\right)B^{\prime\varsigma_{fu}}+\vartheta_{fu}\ell_{fu}^{\varsigma_{fu}}\right]^{\frac{1-\vartheta_{o}}{\varsigma_{fu}}},
\]
where
\begin{align*}
B^{\prime} & =\left[\left(1-\vartheta_{mu}\right)C_{n}^{\prime\varsigma_{mu}}+\vartheta_{mu}\ell_{mu}^{\varsigma_{mu}}\right]^{\frac{1}{\varsigma_{mu}}},\\
C^{\prime} & =\left[\left(1-\vartheta_{fh}\right)D^{\prime\varsigma_{fh}}+\vartheta_{fh}\ell_{fh}^{\varsigma_{fh}}\right]^{\frac{1}{\varsigma_{fh}}},\\
D^{\prime} & =\left[\left(1-\vartheta_{mh}\right)k_{i}^{\varsigma_{mh}}+\vartheta_{mh}\ell_{mh}^{\varsigma_{mh}}\right]^{\frac{1}{\varsigma_{mh}}}.
\end{align*}
In this specification, male labor is located in a nest lower than
that of female labor for each skill type. The four substitution parameters
($\varsigma_{mh}$, $\varsigma_{fh}$, $\varsigma_{mu}$, $\varsigma_{fu}$)
are less than one, and the five share parameters ($\vartheta_{mh}$,
$\vartheta_{fh}$, $\vartheta_{mu}$, $\vartheta_{fu}$, $\alpha$)
lie between zero and one. The problem with this specification is that
the production function is not concave, as in the specification where
unskilled labor is located in a nest lower than that of skilled labor.
Specifically, the estimate of $\varsigma_{fh}$ exceeds one in all
sectors regardless of the use of instruments. This result contradicts
the economic principle that the relative wages of female skilled labor
should decrease with a rise in the relative supply of female skilled
labor. To summarize, there are three ways to extend the specification
of the production function \eqref{eq: KORV}. Female labor may be
located in a nest lower or higher than or in the same nest as that
of male labor. Our specification is consistent with the data, but
the other two are not.

\subsection{Error terms\label{subsec: error_terms}}

Error terms are added to equations \eqref{eq: D.Wfh/Ri}\textendash \eqref{eq: D.Wmh/Wmu}
for a couple of reasons. First, the gender wage gap and the skill
wage gap may be measured with errors. Second, the CES aggregates may
be approximated with errors. Finally, the share parameters may change
over time, possibly due to disembodied factor-biased technological
change. If the share parameters change over time, the first-order
Taylor expansion of equations \eqref{eq: nestB}\textendash \eqref{eq: nestD}
yields
\begin{align*}
\Delta\ln D_{n} & \simeq\Delta\ln\widetilde{D}_{n}+\Delta v_{D,n},\\
\Delta\ln C_{n} & \simeq\Delta\ln\widetilde{C}_{n}+\Delta v_{C,n},\\
\Delta\ln B_{n} & \simeq\Delta\ln\widetilde{B}_{n}+\Delta v_{B,n},
\end{align*}
where
\begin{align*}
\Delta v_{D,n}= & \frac{1}{\sigma_{fh,n}}\left[\left(1-\varphi_{fh,n}\right)\Delta\ln\left(1-\theta_{fh,n}\right)+\varphi_{fh,n}\Delta\ln\theta_{fh,n}\right],\\
\Delta v_{C,n}= & \left(1-\varphi_{mh,n}\right)\Delta v_{D,n}+\frac{1}{\sigma_{mh,n}}\left[\left(1-\varphi_{mh,n}\right)\Delta\ln\left(1-\theta_{mh,n}\right)+\varphi_{mh,n}\Delta\ln\theta_{mh,n}\right],\\
\Delta v_{B,n}= & \left(1-\varphi_{fu,n}\right)\Delta v_{C,n}+\frac{1}{\sigma_{fu,n}}\left[\left(1-\varphi_{fu,n}\right)\Delta\ln\left(1-\theta_{fu,n}\right)+\varphi_{fu,n}\Delta\ln\theta_{fu,n}\right].
\end{align*}
Substituting these equations into the marginal rate of technical substitution
equations yields a system of equations \eqref{eq: D.Wfh/Ri}\textendash \eqref{eq: D.Wmh/Wmu},
where the error terms are
\begin{align*}
\Delta v_{fh,n}= & \Delta\ln\theta_{fh,n}-\Delta\ln\left(1-\theta_{fh,n}\right).\\
\Delta v_{mh,n}= & -\left(\sigma_{mh,n}-\sigma_{fh,n}\right)\Delta v_{D,n}+\Delta\ln\theta_{mh,n}-\Delta\ln\left(1-\theta_{mh,n}\right)-\Delta\ln\theta_{fh,n},\\
\Delta v_{fu,n}= & \left(\sigma_{fu,n}-\sigma_{mh,n}\right)\Delta v_{C,n}+\left(\sigma_{mh,n}-\sigma_{fh,n}\right)\Delta v_{D,n}\\
 & +\Delta\ln\left(1-\theta_{fu,n}\right)-\Delta\ln\theta_{fu,n}+\Delta\ln\left(1-\theta_{mh,n}\right)+\Delta\ln\theta_{fh,n},\\
\Delta v_{mu1,n}= & -\left(\sigma_{mu,n}-\sigma_{fu,n}\right)\Delta v_{B,n}+\Delta\ln\theta_{mu,n}-\Delta\ln\left(1-\theta_{mu,n}\right)-\Delta\ln\theta_{fu,n},\\
\Delta v_{mu2,n}= & \left(\sigma_{mu,n}-\sigma_{fu,n}\right)\Delta v_{B,n}+\left(\sigma_{fu,n}-\sigma_{mh,n}\right)\Delta v_{C,n}\\
 & +\Delta\ln\left(1-\theta_{mu,n}\right)-\Delta\ln\theta_{mu,n}+\Delta\ln\left(1-\theta_{fu,n}\right)+\Delta\ln\theta_{mh,n},
\end{align*}

\subsection{Weights in the aggregate elasticities\label{subsec: weights}}

When the sectoral production function is of the nested CES form \eqref{eq: ces4},
the aggregate elasticity of substitution \eqref{eq: morishima_c}
can be rewritten as equation \eqref{eq: morishima_c''}, where each
substitution parameter is summed across sectors with the following
weights:
\begin{align*}
\pi_{\ell_{f}\ell_{g},n}^{fh} & =\begin{cases}
\left(\frac{\zeta_{n}\lambda_{\ell_{g}n}}{\Lambda_{\ell_{g}}}\right)\left(1-\frac{\lambda_{\ell_{g}n}}{\sum_{\ell_{h}\in\mathcal{L}_{1}}\lambda_{\ell_{h}n}}\right)+\left(\frac{\zeta_{n}\lambda_{\ell_{f}n}}{\Lambda_{\ell_{f}}}\right)\left(\frac{\lambda_{\ell_{g}n}}{\sum_{\ell_{h}\in\mathcal{L}_{1}}\lambda_{\ell_{h}n}}\right) & \text{if }\ell_{f},\ell_{g}\in\mathcal{L}_{1}\\
\left(\frac{\zeta_{n}\lambda_{\ell_{g}n}}{\Lambda_{\ell_{g}}}\right)\left(1-\frac{\lambda_{\ell_{g}n}}{\sum_{\ell_{h}\in\mathcal{L}_{1}}\lambda_{\ell_{h}n}}\right) & \text{if }\ell_{f}\notin\mathcal{L}_{1},\ell_{g}\in\mathcal{L}_{1}\\
0 & \text{if }\ell_{g}\notin\mathcal{L}_{1},
\end{cases}
\end{align*}
\begin{align*}
\pi_{\ell_{f}\ell_{g},n}^{mh} & =\begin{cases}
\left(\frac{\zeta_{n}\lambda_{\ell_{g}n}}{\Lambda_{\ell_{g}}}-\frac{\zeta_{n}\lambda_{\ell_{f}n}}{\Lambda_{\ell_{f}}}\right)\left(\frac{\lambda_{\ell_{g}n}}{\sum_{\ell_{h}\in\mathcal{L}_{1}}\lambda_{\ell_{h}n}}\right)\left(\frac{\lambda_{\ell_{mh}n}}{\sum_{\ell_{h}\in\mathcal{L}_{2}}\lambda_{\ell_{h}n}}\right)+\left(\frac{\zeta_{n}\lambda_{\ell_{f}n}}{\Lambda_{\ell_{f}}}\right)\left(\frac{\lambda_{\ell_{g}n}}{\sum_{\ell_{h}\in\mathcal{L}_{1}}\lambda_{\ell_{h}n}}\right) & \text{if }\ell_{f}=\ell_{mh},\ell_{g}\in\mathcal{L}_{1}\\
\left(\frac{\zeta_{n}\lambda_{\ell_{g}n}}{\Lambda_{\ell_{g}}}-\frac{\zeta_{n}\lambda_{\ell_{f}n}}{\Lambda_{\ell_{f}}}\right)\left(\frac{\lambda_{\ell_{g}n}}{\sum_{\ell_{h}\in\mathcal{L}_{1}}\lambda_{\ell_{h}n}}\right)\left(\frac{\lambda_{\ell_{mh}n}}{\sum_{\ell_{h}\in\mathcal{L}_{2}}\lambda_{\ell_{h}n}}\right) & \text{if }\ell_{f},\ell_{g}\in\mathcal{L}_{1}\\
\left(\frac{\zeta_{n}\lambda_{\ell_{g}n}}{\Lambda_{\ell_{g}}}\right)\left(\frac{\lambda_{\ell_{g}n}}{\sum_{\ell_{h}\in\mathcal{L}_{1}}\lambda_{\ell_{h}n}}\right)\left(\frac{\lambda_{\ell_{mh}n}}{\sum_{\ell_{h}\in\mathcal{L}_{2}}\lambda_{\ell_{h}n}}\right) & \text{if }\ell_{f}\notin\mathcal{L}_{2},\ell_{g}\in\mathcal{L}_{1}\\
\left(\frac{\zeta_{n}\lambda_{\ell_{g}n}}{\Lambda_{\ell_{g}}}\right)\left(1-\frac{\lambda_{\ell_{g}n}}{\sum_{\ell_{h}\in\mathcal{L}_{2}}\lambda_{\ell_{h}n}}\right)+\left(\frac{\zeta_{n}\lambda_{\ell_{f}n}}{\Lambda_{\ell_{f}}}\right)\left(\frac{\lambda_{\ell_{g}n}}{\sum_{\ell_{h}\in\mathcal{L}_{2}}\lambda_{\ell_{h}n}}\right) & \text{if }\ell_{f}\in\mathcal{L}_{1},\ell_{g}=\ell_{mh}\\
\left(\frac{\zeta_{n}\lambda_{\ell_{g}n}}{\Lambda_{\ell_{g}}}\right)\left(1-\frac{\lambda_{\ell_{g}n}}{\sum_{\ell_{h}\in\mathcal{L}_{2}}\lambda_{\ell_{h}n}}\right) & \text{if }\ell_{f}\notin\mathcal{L}_{1},\ell_{g}=\ell_{mh}\\
0 & \text{if }\ell_{g}\notin\mathcal{L}_{2},
\end{cases}
\end{align*}
\begin{align*}
\pi_{\ell_{f}\ell_{g},n}^{fu} & =\begin{cases}
\left(\frac{\zeta_{n}\lambda_{\ell_{g}n}}{\Lambda_{\ell_{g}}}-\frac{\zeta_{n}\lambda_{\ell_{f}n}}{\Lambda_{\ell_{f}}}\right)\left(\frac{\lambda_{\ell_{g}n}}{\sum_{\ell_{h}\in\mathcal{L}_{2}}\lambda_{\ell_{h}n}}\right)\left(\frac{\lambda_{\ell_{fu}n}}{\sum_{\ell_{h}\in\mathcal{L}_{3}}\lambda_{\ell_{h}n}}\right)+\left(\frac{\zeta_{n}\lambda_{\ell_{f}n}}{\Lambda_{\ell_{f}}}\right)\left(\frac{\lambda_{\ell_{g}n}}{\sum_{\ell_{h}\in\mathcal{L}_{2}}\lambda_{\ell_{h}n}}\right) & \text{if }\ell_{f}=\ell_{fu},\ell_{g}\in\mathcal{L}_{2}\\
\left(\frac{\zeta_{n}\lambda_{\ell_{g}n}}{\Lambda_{\ell_{g}}}-\frac{\zeta_{n}\lambda_{\ell_{f}n}}{\Lambda_{\ell_{f}}}\right)\left(\frac{\lambda_{\ell_{g}n}}{\sum_{\ell_{h}\in\mathcal{L}_{2}}\lambda_{\ell_{h}n}}\right)\left(\frac{\lambda_{\ell_{fu}n}}{\sum_{\ell_{h}\in\mathcal{L}_{3}}\lambda_{\ell_{h}n}}\right) & \text{if }\ell_{f},\ell_{g}\in\mathcal{L}_{2}\\
\left(\frac{\zeta_{n}\lambda_{\ell_{g}n}}{\Lambda_{\ell_{g}}}\right)\left(\frac{\lambda_{\ell_{g}n}}{\sum_{\ell_{h}\in\mathcal{L}_{2}}\lambda_{\ell_{h}n}}\right)\left(\frac{\lambda_{\ell_{fu}n}}{\sum_{\ell_{h}\in\mathcal{L}_{3}}\lambda_{\ell_{h}n}}\right) & \text{if }\ell_{f}\notin\mathcal{L}_{3},\ell_{g}\in\mathcal{L}_{2}\\
\left(\frac{\zeta_{n}\lambda_{\ell_{g}n}}{\Lambda_{\ell_{g}}}\right)\left(1-\frac{\lambda_{\ell_{g}n}}{\sum_{\ell_{h}\in\mathcal{L}_{3}}\lambda_{\ell_{h}n}}\right)+\left(\frac{\zeta_{n}\lambda_{\ell_{f}n}}{\Lambda_{\ell_{f}}}\right)\left(\frac{\lambda_{\ell_{g}n}}{\sum_{\ell_{h}\in\mathcal{L}_{3}}\lambda_{\ell_{h}n}}\right) & \text{if }\ell_{f}\in\mathcal{L}_{2},\ell_{g}=\ell_{fu}\\
\left(\frac{\zeta_{n}\lambda_{\ell_{g}n}}{\Lambda_{\ell_{g}}}\right)\left(1-\frac{\lambda_{\ell_{g}n}}{\sum_{\ell_{h}\in\mathcal{L}_{3}}\lambda_{\ell_{h}n}}\right) & \text{if }\ell_{f}\notin\mathcal{L}_{2},\ell_{g}=\ell_{fu}\\
0 & \text{if }\ell_{g}\notin\mathcal{L}_{3},
\end{cases}
\end{align*}
\begin{align*}
\pi_{\ell_{f}\ell_{g},n}^{mu} & =\begin{cases}
\left(\frac{\zeta_{n}\lambda_{\ell_{g}n}}{\Lambda_{\ell_{g}}}-\frac{\zeta_{n}\lambda_{\ell_{f}n}}{\Lambda_{\ell_{f}}}\right)\left(\frac{\lambda_{\ell_{g}n}}{\sum_{\ell_{h}\in\mathcal{L}_{3}}\lambda_{\ell_{h}n}}\right)\left(\frac{\lambda_{\ell_{mu}n}}{\sum_{\ell_{h}\in\mathcal{L}_{4}}\lambda_{\ell_{h}n}}\right)+\left(\frac{\zeta_{n}\lambda_{\ell_{f}n}}{\Lambda_{\ell_{f}}}\right)\left(\frac{\lambda_{\ell_{g}n}}{\sum_{\ell_{h}\in\mathcal{L}_{3}}\lambda_{\ell_{h}n}}\right) & \text{if }\ell_{f}=\ell_{mu},\ell_{g}\in\mathcal{L}_{3}\\
\left(\frac{\zeta_{n}\lambda_{\ell_{g}n}}{\Lambda_{\ell_{g}}}-\frac{\zeta_{n}\lambda_{\ell_{f}n}}{\Lambda_{\ell_{f}}}\right)\left(\frac{\lambda_{\ell_{g}n}}{\sum_{\ell_{h}\in\mathcal{L}_{3}}\lambda_{\ell_{h}n}}\right)\left(\frac{\lambda_{\ell_{mu}n}}{\sum_{\ell_{h}\in\mathcal{L}_{4}}\lambda_{\ell_{h}n}}\right) & \text{if }\ell_{f},\ell_{g}\in\mathcal{L}_{3}\\
\left(\frac{\zeta_{n}\lambda_{\ell_{g}n}}{\Lambda_{\ell_{g}}}\right)\left(\frac{\lambda_{\ell_{g}n}}{\sum_{\ell_{h}\in\mathcal{L}_{3}}\lambda_{\ell_{h}n}}\right)\left(\frac{\lambda_{\ell_{mu}n}}{\sum_{\ell_{h}\in\mathcal{L}_{4}}\lambda_{\ell_{h}n}}\right) & \text{if }\ell_{f}\notin\mathcal{L}_{4},\ell_{g}\in\mathcal{L}_{3}\\
\left(\frac{\zeta_{n}\lambda_{\ell_{g}n}}{\Lambda_{\ell_{g}}}\right)\left(1-\frac{\lambda_{\ell_{g}n}}{\sum_{\ell_{h}\in\mathcal{L}_{4}}\lambda_{\ell_{h}n}}\right)+\left(\frac{\zeta_{n}\lambda_{\ell_{f}n}}{\Lambda_{\ell_{f}}}\right)\left(\frac{\lambda_{\ell_{g}n}}{\sum_{\ell_{h}\in\mathcal{L}_{4}}\lambda_{\ell_{h}n}}\right) & \text{if }\ell_{f}\in\mathcal{L}_{3},\ell_{g}=\ell_{mu}\\
\left(\frac{\zeta_{n}\lambda_{\ell_{g}n}}{\Lambda_{\ell_{g}}}\right)\left(1-\frac{\lambda_{\ell_{g}n}}{\sum_{\ell_{h}\in\mathcal{L}_{4}}\lambda_{\ell_{h}n}}\right) & \text{if }\ell_{f}\notin\mathcal{L}_{3},\ell_{g}=\ell_{mu}\\
0 & \text{if }\ell_{g}\notin\mathcal{L}_{4},
\end{cases}
\end{align*}
\begin{align*}
\pi_{\ell_{f}\ell_{g},n}^{o} & =\begin{cases}
\left(\frac{\zeta_{n}\lambda_{\ell_{g}n}}{\Lambda_{\ell_{g}}}-\frac{\zeta_{n}\lambda_{\ell_{f}n}}{\Lambda_{\ell_{f}}}\right)\left(\frac{\lambda_{\ell_{g}n}}{\sum_{\ell_{h}\in\mathcal{L}_{4}}\lambda_{\ell_{h}n}}\right)\lambda_{k_{o}n} & \text{if }\ell_{f},\ell_{g}\in\mathcal{L}_{4}\\
\left(\frac{\zeta_{n}\lambda_{\ell_{g}n}}{\Lambda_{\ell_{g}}}\right)\left(\frac{\lambda_{\ell_{g}n}}{\sum_{\ell_{h}\in\mathcal{L}_{4}}\lambda_{\ell_{h}n}}\right)-\pi_{\ell_{f}\ell_{g},n}^{c} & \text{if }\ell_{f}=k_{o}\\
\frac{\zeta_{n}\lambda_{\ell_{g}n}}{\Lambda_{\ell_{g}}}-\pi_{\ell_{f}\ell_{g},n}^{c} & \text{if }\ell_{g}=k_{o},
\end{cases}
\end{align*}
and
\begin{align*}
\pi_{\ell_{f}\ell_{g},n}^{c} & =\begin{cases}
\left(\frac{\zeta_{n}\lambda_{\ell_{g}n}}{\Lambda_{\ell_{g}}}-\frac{\zeta_{n}\lambda_{\ell_{f}n}}{\Lambda_{\ell_{f}}}\right)\left(\frac{\lambda_{\ell_{g}n}}{\sum_{\ell_{h}\in\mathcal{L}_{4}}\lambda_{\ell_{h}n}}\right)\left(1-\lambda_{k_{o}n}\right) & \text{if }\ell_{g}\in\mathcal{L}_{4}\\
\left(\frac{\zeta_{n}\lambda_{\ell_{g}n}}{\Lambda_{\ell_{g}}}-\frac{\zeta_{n}\lambda_{\ell_{f}n}}{\Lambda_{\ell_{f}}}\right)\lambda_{\ell_{g}n} & \text{if }\ell_{g}=k_{o},
\end{cases}
\end{align*}
where the sets of factors are defined as $\mathcal{L}_{1}=\{k_{i},\ell_{fh}\}$,
$\mathcal{L}_{2}=\{k_{i},\ell_{fh},\ell_{mh}\}$, $\mathcal{L}_{3}=\{k_{i},\ell_{fh},\ell_{mh},\ell_{fu}\}$,
and $\mathcal{L}_{4}=\{k_{i},\ell_{fh},\ell_{mh},\ell_{fu},\ell_{mu}\}$.

\subsection{Magnitude of the effects of technological change\label{subsec: GE_effect_diff}}

We show that the effects of technological progress embodied in ICCT
capital tend to increase as the differences in the substitution parameters
in the production function increase and the income share of ICCT capital
relative to the income shares of other factors increases. For this
purpose, we assume here that the aggregate production function is
of the nested CES form \eqref{eq: ces4} (i.e., $N=1$) and that the
capital\textendash skill\textendash gender complementarity hypothesis
holds (i.e., $\sigma_{fh}<\sigma_{mh}<\sigma_{fu}<\sigma_{mu}$).
The effects of technological change embodied in ICCT capital on the
relative wages and income shares of the four types of labor can be
obtained by setting $d\ln q_{o}=d\ln A_{n}=0$ in equations \eqref{eq: D.Wf/Wg-Qf'}
and \eqref{eq: D.WfLf/PY-Qf'}, respectively.

We first consider the effects on the four income shares. The effect
on the income share of female skilled labor is
\begin{multline}
\frac{\partial\ln\Lambda_{\ell_{fh}}}{\partial\ln q_{i}}=\left(1-\epsilon_{k_{i}\ell_{fh}}^{\mathcal{C}}\right)\left(\frac{\Lambda_{k_{i}}}{\Lambda_{\ell_{fh}}}\right)\left(\frac{H_{fh}}{H}\right)+\left(\epsilon_{\ell_{mh}k_{i}}^{\mathcal{C}}-\epsilon_{\ell_{fh}k_{i}}^{\mathcal{C}}\right)\left(\frac{H_{mh}}{H}\right)\\
+\left(\epsilon_{\ell_{fu}k_{i}}^{\mathcal{C}}-\epsilon_{\ell_{fh}k_{i}}^{\mathcal{C}}\right)\left(\frac{H_{fu}}{H}\right)+\left(\epsilon_{\ell_{mu}k_{i}}^{\mathcal{C}}-\epsilon_{\ell_{fh}k_{i}}^{\mathcal{C}}\right)\left(\frac{H_{mu}}{H}\right),\label{eq: D.Lambda_Lfh/D.Qi}
\end{multline}
where
\[
H=H_{fh}+H_{mh}+H_{fu}+H_{mu},
\]
{\scriptsize
\begin{align*}
H_{fh}= & \left(\epsilon_{k_{i}\ell_{mu}}^{\mathcal{C}}+\frac{1}{\gamma}\right)\left(\epsilon_{k_{i}\ell_{fu}}^{\mathcal{C}}+\frac{1}{\gamma}\right)\left(\epsilon_{k_{i}\ell_{mh}}^{\mathcal{C}}+\frac{1}{\gamma}\right)\Lambda_{\ell_{fh}}+\left(\epsilon_{\ell_{mu}k_{i}}^{\mathcal{C}}-\epsilon_{\ell_{fu}k_{i}}^{\mathcal{C}}\right)\left(\epsilon_{\ell_{fu}k_{i}}^{\mathcal{C}}-\epsilon_{\ell_{mh}k_{i}}^{\mathcal{C}}\right)\left(\epsilon_{\ell_{mh}k_{i}}^{\mathcal{C}}-\epsilon_{\ell_{fh}k_{i}}^{\mathcal{C}}\right)\left(\frac{\Lambda_{\ell_{fh}}\Lambda_{\ell_{mh}}\Lambda_{\ell_{fu}}\Lambda_{\ell_{mu}}}{\Lambda_{k_{i}}\Lambda_{k_{i}}\Lambda_{k_{i}}}\right)\\
 & +\left(\epsilon_{k_{i}\ell_{fu}}^{\mathcal{C}}+\frac{1}{\gamma}\right)\left(\epsilon_{k_{i}\ell_{mu}}^{\mathcal{C}}+\frac{1}{\gamma}\right)\left(\epsilon_{\ell_{mh}k_{i}}^{\mathcal{C}}-\epsilon_{\ell_{fh}k_{i}}^{\mathcal{C}}\right)\left(\frac{\Lambda_{\ell_{fh}}\Lambda_{\ell_{mh}}}{\Lambda_{k_{i}}}\right)+\left(\epsilon_{k_{i}\ell_{mu}}^{\mathcal{C}}+\frac{1}{\gamma}\right)\left(\epsilon_{\ell_{fu}k_{i}}^{\mathcal{C}}-\epsilon_{\ell_{mh}k_{i}}^{\mathcal{C}}\right)\left(\epsilon_{\ell_{mh}k_{i}}^{\mathcal{C}}-\epsilon_{\ell_{fh}k_{i}}^{\mathcal{C}}\right)\left(\frac{\Lambda_{\ell_{fh}}\Lambda_{\ell_{mh}}\Lambda_{\ell_{fu}}}{\Lambda_{k_{i}}\Lambda_{k_{i}}}\right)\\
 & +\left(\epsilon_{k_{i}\ell_{mh}}^{\mathcal{C}}+\frac{1}{\gamma}\right)\left(\epsilon_{k_{i}\ell_{mu}}^{\mathcal{C}}+\frac{1}{\gamma}\right)\left(\epsilon_{\ell_{fu}k_{i}}^{\mathcal{C}}-\epsilon_{\ell_{fh}k_{i}}^{\mathcal{C}}\right)\left(\frac{\Lambda_{\ell_{fh}}\Lambda_{\ell_{fu}}}{\Lambda_{k_{i}}}\right)+\left(\epsilon_{k_{i}\ell_{fu}}^{\mathcal{C}}+\frac{1}{\gamma}\right)\left(\epsilon_{\ell_{mu}k_{i}}^{\mathcal{C}}-\epsilon_{\ell_{mh}k_{i}}^{\mathcal{C}}\right)\left(\epsilon_{\ell_{mh}k_{i}}^{\mathcal{C}}-\epsilon_{\ell_{fh}k_{i}}^{\mathcal{C}}\right)\left(\frac{\Lambda_{\ell_{fh}}\Lambda_{\ell_{mh}}\Lambda_{\ell_{mu}}}{\Lambda_{k_{i}}\Lambda_{k_{i}}}\right)\\
 & +\left(\epsilon_{k_{i}\ell_{mh}}^{\mathcal{C}}+\frac{1}{\gamma}\right)\left(\epsilon_{k_{i}\ell_{fu}}^{\mathcal{C}}+\frac{1}{\gamma}\right)\left(\epsilon_{\ell_{mu}k_{i}}^{\mathcal{C}}-\epsilon_{\ell_{fh}k_{i}}^{\mathcal{C}}\right)\left(\frac{\Lambda_{\ell_{fh}}\Lambda_{\ell_{mu}}}{\Lambda_{k_{i}}}\right)+\left(\epsilon_{k_{i}\ell_{mh}}^{\mathcal{C}}+\frac{1}{\gamma}\right)\left(\epsilon_{\ell_{mu}k_{i}}^{\mathcal{C}}-\epsilon_{\ell_{fu}k_{i}}^{\mathcal{C}}\right)\left(\epsilon_{\ell_{fu}k_{i}}^{\mathcal{C}}-\epsilon_{\ell_{fh}k_{i}}^{\mathcal{C}}\right)\left(\frac{\Lambda_{\ell_{fh}}\Lambda_{\ell_{fu}}\Lambda_{\ell_{mu}}}{\Lambda_{k_{i}}\Lambda_{k_{i}}}\right),
\end{align*}
\begin{align*}
H_{mh}= & \left(\epsilon_{k_{i}\ell_{fh}}^{\mathcal{C}}+\frac{1}{\gamma}\right)\left(\epsilon_{k_{i}\ell_{fu}}^{\mathcal{C}}+\frac{1}{\gamma}\right)\left(\epsilon_{k_{i}\ell_{mu}}^{\mathcal{C}}+\frac{1}{\gamma}\right)\Lambda_{\ell_{mh}}+\left(\epsilon_{k_{i}\ell_{fh}}^{\mathcal{C}}+\frac{1}{\gamma}\right)\left(\epsilon_{\ell_{mu}k_{i}}^{\mathcal{C}}-\epsilon_{\ell_{fu}k_{i}}^{\mathcal{C}}\right)\left(\epsilon_{\ell_{fu}k_{i}}^{\mathcal{C}}-\epsilon_{\ell_{mh}k_{i}}^{\mathcal{C}}\right)\left(\frac{\Lambda_{\ell_{mh}}\Lambda_{\ell_{fu}}\Lambda_{\ell_{mu}}}{\Lambda_{k_{i}}\Lambda_{k_{i}}}\right)\\
 & +\left(\epsilon_{k_{i}\ell_{fh}}^{\mathcal{C}}+\frac{1}{\gamma}\right)\left(\epsilon_{k_{i}\ell_{mu}}^{\mathcal{C}}+\frac{1}{\gamma}\right)\left(\epsilon_{\ell_{fu}k_{i}}^{\mathcal{C}}-\epsilon_{\ell_{mh}k_{i}}^{\mathcal{C}}\right)\left(\frac{\Lambda_{\ell_{mh}}\Lambda_{\ell_{fu}}}{\Lambda_{k_{i}}}\right)+\left(\epsilon_{k_{i}\ell_{fh}}^{\mathcal{C}}+\frac{1}{\gamma}\right)\left(\epsilon_{k_{i}\ell_{fu}}^{\mathcal{C}}+\frac{1}{\gamma}\right)\left(\epsilon_{\ell_{mu}k_{i}}^{\mathcal{C}}-\epsilon_{\ell_{mh}k_{i}}^{\mathcal{C}}\right)\left(\frac{\Lambda_{\ell_{mh}}\Lambda_{\ell_{mu}}}{\Lambda_{k_{i}}}\right),
\end{align*}
\[
H_{fu}=\left(\epsilon_{k_{i}\ell_{mu}}^{\mathcal{C}}+\frac{1}{\gamma}\right)\left(\epsilon_{k_{i}\ell_{mh}}^{\mathcal{C}}+\frac{1}{\gamma}\right)\left(\epsilon_{k_{i}\ell_{fh}}^{\mathcal{C}}+\frac{1}{\gamma}\right)\Lambda_{\ell_{fu}}+\left(\epsilon_{k_{i}\ell_{mh}}^{\mathcal{C}}+\frac{1}{\gamma}\right)\left(\epsilon_{k_{i}\ell_{fh}}^{\mathcal{C}}+\frac{1}{\gamma}\right)\left(\epsilon_{\ell_{mu}k_{i}}^{\mathcal{C}}-\epsilon_{\ell_{fu}k_{i}}^{\mathcal{C}}\right)\left(\frac{\Lambda_{\ell_{fu}}\Lambda_{\ell_{mu}}}{\Lambda_{k_{i}}}\right),
\]
\[
H_{mu}=\left(\epsilon_{k_{i}\ell_{fu}}^{\mathcal{C}}+\frac{1}{\gamma}\right)\left(\epsilon_{k_{i}\ell_{mh}}^{\mathcal{C}}+\frac{1}{\gamma}\right)\left(\epsilon_{k_{i}\ell_{fh}}^{\mathcal{C}}+\frac{1}{\gamma}\right)\Lambda_{\ell_{mu}}.
\]
}The effect on the income share of male skilled labor is
\begin{equation}
\frac{\partial\ln\Lambda_{\ell_{mh}}}{\partial\ln q_{i}}=\left(1-\epsilon_{k_{i}\ell_{mh}}^{\mathcal{C}}\right)\left(\frac{\Lambda_{k_{i}}}{\Lambda_{\ell_{mh}}}\right)\left(\frac{H_{mh}}{H}\right)+\left(\epsilon_{\ell_{fu}k_{i}}^{\mathcal{C}}-\epsilon_{\ell_{mh}k_{i}}^{\mathcal{C}}\right)\left(\frac{H_{fu}}{H}\right)+\left(\epsilon_{\ell_{mu}k_{i}}^{\mathcal{C}}-\epsilon_{\ell_{mh}k_{i}}^{\mathcal{C}}\right)\left(\frac{H_{mu}}{H}\right).\label{eq: D.Lambda_Lmh/D.Qi}
\end{equation}
The effect on the income share of female unskilled labor is
\begin{equation}
\frac{\partial\ln\Lambda_{\ell_{fu}}}{\partial\ln q_{i}}=\left(1-\epsilon_{k_{i}\ell_{fu}}^{\mathcal{C}}\right)\left(\frac{\Lambda_{k_{i}}}{\Lambda_{\ell_{fu}}}\right)\left(\frac{H_{fu}}{H}\right)+\left(\epsilon_{\ell_{mu}k_{i}}^{\mathcal{C}}-\epsilon_{\ell_{fu}k_{i}}^{\mathcal{C}}\right)\left(\frac{H_{mu}}{H}\right).\label{eq: D.Lambda_Lfu/D.Qi}
\end{equation}
The effect on the income share of male unskilled labor is
\begin{equation}
\frac{\partial\ln\Lambda_{\ell_{mu}}}{\partial\ln q_{i}}=\left(1-\epsilon_{k_{i}\ell_{mu}}^{\mathcal{C}}\right)\left(\frac{\Lambda_{k_{i}}}{\Lambda_{\ell_{mu}}}\right)\left(\frac{H_{mu}}{H}\right).\label{eq: D.Lambda_Lmu/D.Qi}
\end{equation}
The magnitude of the first terms in equations \eqref{eq: D.Lambda_Lfh/D.Qi}\textendash \eqref{eq: D.Lambda_Lmu/D.Qi}
depends on the substitution parameters in the production function
and the income share of ICCT capital relative to the income shares
of other factors.
\[
\left(1-\epsilon_{k_{i}\ell_{f}}^{\mathcal{C}}\right)\left(\frac{\Lambda_{k_{i}}}{\Lambda_{\ell_{f}}}\right)=\left(1-\frac{1}{1-\sigma_{f}}\right)\left(\frac{\Lambda_{k_{i}}}{\Lambda_{\ell_{f}}}\right)\qquad\text{for }f\in\left\{ mh,fh,mu,fu\right\} .
\]
The magnitude of the second to the last terms in equations \eqref{eq: D.Lambda_Lfh/D.Qi}\textendash \eqref{eq: D.Lambda_Lmu/D.Qi}
depends on the differences between the aggregate elasticities, which
in turn depend on the differences between the substitution parameters
in the production function and the income share of ICCT capital relative
to the income shares of other factors.
\begin{equation}
\epsilon_{\ell_{mh}k_{i}}^{\mathcal{C}}-\epsilon_{\ell_{fh}k_{i}}^{\mathcal{C}}=\left(\frac{1}{1-\sigma_{mh}}-\frac{1}{1-\sigma_{fh}}\right)\left(\frac{\Lambda_{k_{i}}}{\Lambda_{k_{i}}+\Lambda_{\ell_{fh}}}\right)>0,\label{eq: D.epsilon_mhki_fhki}
\end{equation}
\begin{equation}
\epsilon_{\ell_{fu}k_{i}}^{\mathcal{C}}-\epsilon_{\ell_{mh}k_{i}}^{\mathcal{C}}=\left(\frac{1}{1-\sigma_{fu}}-\frac{1}{1-\sigma_{mh}}\right)\left(\frac{\Lambda_{k_{i}}}{\Lambda_{k_{i}}+\Lambda_{\ell_{fh}}+\Lambda_{\ell_{mh}}}\right)>0,\label{eq: D.epsilon_fuki_mhki}
\end{equation}
\begin{equation}
\epsilon_{\ell_{mu}k_{i}}^{\mathcal{C}}-\epsilon_{\ell_{fu}k_{i}}^{\mathcal{C}}=\left(\frac{1}{1-\sigma_{mu}}-\frac{1}{1-\sigma_{fu}}\right)\left(\frac{\Lambda_{k_{i}}}{\Lambda_{k_{i}}+\Lambda_{\ell_{fh}}+\Lambda_{\ell_{mh}}+\Lambda_{\ell_{fu}}}\right)>0,\label{eq: D.epsilon_muki_fuki}
\end{equation}
\begin{equation}
\epsilon_{\ell_{mu}k_{i}}^{\mathcal{C}}-\epsilon_{\ell_{mh}k_{i}}^{\mathcal{C}}=\left(\epsilon_{\ell_{mu}k_{i}}^{\mathcal{C}}-\epsilon_{\ell_{fu}k_{i}}^{\mathcal{C}}\right)+\left(\epsilon_{\ell_{fu}k_{i}}^{\mathcal{C}}-\epsilon_{\ell_{mh}k_{i}}^{\mathcal{C}}\right)>0,\label{eq: D.epsilon_muki_mhki}
\end{equation}
\begin{equation}
\epsilon_{\ell_{fu}k_{i}}^{\mathcal{C}}-\epsilon_{\ell_{fh}k_{i}}^{\mathcal{C}}=\left(\epsilon_{\ell_{fu}k_{i}}^{\mathcal{C}}-\epsilon_{\ell_{mh}k_{i}}^{\mathcal{C}}\right)+\left(\epsilon_{\ell_{mh}k_{i}}^{\mathcal{C}}-\epsilon_{\ell_{fh}k_{i}}^{\mathcal{C}}\right)>0,\label{eq: D.epsilon_fuki_fhki}
\end{equation}
and
\begin{equation}
\epsilon_{\ell_{mu}k_{i}}^{\mathcal{C}}-\epsilon_{\ell_{fh}k_{i}}^{\mathcal{C}}=\left(\epsilon_{\ell_{mu}k_{i}}^{\mathcal{C}}-\epsilon_{\ell_{fu}k_{i}}^{\mathcal{C}}\right)+\left(\epsilon_{\ell_{fu}k_{i}}^{\mathcal{C}}-\epsilon_{\ell_{mh}k_{i}}^{\mathcal{C}}\right)+\left(\epsilon_{\ell_{mh}k_{i}}^{\mathcal{C}}-\epsilon_{\ell_{fh}k_{i}}^{\mathcal{C}}\right)>0.\label{eq: D.epsilon_muki_fhki}
\end{equation}
The effects of technological progress embodied in ICCT capital on
the income shares of male and female skilled labor are positive, while
those on the income shares of male and female unskilled labor are
negative, if $\sigma_{fh}<0$, $\sigma_{mh}\simeq0$, and $\sigma_{fu},\sigma_{mu}>0$.
Thus, the magnitude of the effects are all proportional to the differences
in the substitution parameters in the production function and the
income share of ICCT capital relative to the income shares of other
factors.

We next consider the effects on the four wage gaps. The effect on
the skilled gender wage gap is
\[
\frac{d\ln\left(\left.w_{mh}\right/w_{fh}\right)}{d\ln q_{i}}=-\frac{1}{H}\left(\epsilon_{k_{i}\ell_{mu}}^{\mathcal{C}}+\frac{1}{\gamma}\right)\left(\epsilon_{k_{i}\ell_{fu}}^{\mathcal{C}}+\frac{1}{\gamma}\right)\left(\epsilon_{\ell_{mh}k_{i}}^{\mathcal{C}}-\epsilon_{\ell_{fh}k_{i}}^{\mathcal{C}}\right)\left(1-\Lambda_{k_{o}}\right)<0.
\]
The magnitude of the effect is proportional to the difference between
the aggregate elasticities \eqref{eq: D.epsilon_mhki_fhki}. The effect
on the unskilled gender wage gap is
\[
\frac{d\ln\left(\left.w_{mu}\right/w_{fu}\right)}{d\ln q_{i}}=-\frac{1}{H}\left(\epsilon_{k_{i}\ell_{mh}}^{\mathcal{C}}+\frac{1}{\gamma}\right)\left(\epsilon_{k_{i}\ell_{fh}}^{\mathcal{C}}+\frac{1}{\gamma}\right)\left(\epsilon_{\ell_{mu}k_{i}}^{\mathcal{C}}-\epsilon_{\ell_{fu}k_{i}}^{\mathcal{C}}\right)\left(1-\Lambda_{k_{o}}\right)<0.
\]
The magnitude of the effect is proportional to the difference between
the aggregate elasticities \eqref{eq: D.epsilon_muki_fuki}. The effect
on the male skill wage gap is
\begin{multline*}
\frac{d\ln\left(\left.w_{mh}\right/w_{mu}\right)}{d\ln q_{i}}=\frac{1}{H}\left(\epsilon_{k_{i}\ell_{fh}}^{\mathcal{C}}+\frac{1}{\gamma}\right)\left(\epsilon_{k_{i}\ell_{fu}}^{\mathcal{C}}+\frac{1}{\gamma}\right)\left(\epsilon_{\ell_{mu}k_{i}}^{\mathcal{C}}-\epsilon_{\ell_{mh}k_{i}}^{\mathcal{C}}\right)\left(1-\Lambda_{k_{o}}\right)\\
+\frac{1}{H}\left(\epsilon_{k_{i}\ell_{fh}}^{\mathcal{C}}+\frac{1}{\gamma}\right)\left(\epsilon_{\ell_{mu}k_{i}}^{\mathcal{C}}-\epsilon_{\ell_{fu}k_{i}}^{\mathcal{C}}\right)\left(\epsilon_{\ell_{fu}k_{i}}^{\mathcal{C}}-\epsilon_{\ell_{mh}k_{i}}^{\mathcal{C}}\right)\left(1-\Lambda_{k_{o}}\right)\left(\frac{\Lambda_{\ell_{fu}}}{\Lambda_{k_{i}}}\right)>0.
\end{multline*}
The magnitude of the effect is proportional to the differences between
the aggregate elasticities \eqref{eq: D.epsilon_fuki_mhki}, \eqref{eq: D.epsilon_muki_fuki},
and \eqref{eq: D.epsilon_muki_mhki}. The effect on the female skill
wage gap is
\begin{multline*}
\frac{d\ln\left(\left.w_{fh}\right/w_{fu}\right)}{d\ln q_{i}}=\frac{1}{H}\left(\epsilon_{k_{i}\ell_{mu}}^{\mathcal{C}}+\frac{1}{\gamma}\right)\left(\epsilon_{k_{i}\ell_{mh}}^{\mathcal{C}}+\frac{1}{\gamma}\right)\left(\epsilon_{\ell_{fu}k_{i}}^{\mathcal{C}}-\epsilon_{\ell_{fh}k_{i}}^{\mathcal{C}}\right)\left(1-\Lambda_{k_{o}}\right)\\
+\frac{1}{H}\left(\epsilon_{k_{i}\ell_{mu}}^{\mathcal{C}}+\frac{1}{\gamma}\right)\left(\epsilon_{\ell_{fu}k_{i}}^{\mathcal{C}}-\epsilon_{\ell_{mh}k_{i}}^{\mathcal{C}}\right)\left(\epsilon_{\ell_{mh}k_{i}}^{\mathcal{C}}-\epsilon_{\ell_{fh}k_{i}}^{\mathcal{C}}\right)\left(1-\Lambda_{k_{o}}\right)\left(\frac{\Lambda_{\ell_{mh}}}{\Lambda_{k_{i}}}\right)>0.
\end{multline*}
The magnitude of the effect is proportional to the differences between
the aggregate elasticities \eqref{eq: D.epsilon_mhki_fhki}, \eqref{eq: D.epsilon_fuki_mhki},
and \eqref{eq: D.epsilon_fuki_fhki}. Thus, the magnitude of the effects
on the four wage gaps is proportional to the differences in the substitution
parameters in the production function and the income share of ICCT
capital relative to the income shares of other factors.

\section{Additional Results\label{sec: additional}}

Table \ref{tab: GE_effect_immobile} report the changes attributable
to the three types of technological change when labor is not perfectly
mobile across sectors. Table \ref{tab: GE_effect_by_country} reports
the effects of technological change embodied in ICCT capital on the
relative wages and income shares of the four types of labor by country
for the cases in which the supply of labor is perfectly inelastic
and unit elastic. Table \ref{tab: GE_effect_aggregate} compares the
magnitude of the changes in relative wages and income shares attributable
to ICCT equipment obtained by estimating and aggregating sectoral
production functions with that obtained by assuming and estimating
the aggregate production function.

\begin{table}[h]
\caption{Effects of technological change under imperfect labor mobility, 1980\textendash 2005\label{tab: GE_effect_immobile}}

\begin{centering}
\begin{tabular}{lr@{\extracolsep{0pt}.}lr@{\extracolsep{0pt}.}lr@{\extracolsep{0pt}.}l}
\hline 
 & \multicolumn{2}{c}{$\Delta\ln q_{i}$} & \multicolumn{2}{c}{$\Delta\ln q_{o}$} & \multicolumn{2}{c}{$\Delta\ln A_{n}$}\tabularnewline
\cline{2-7}
 & \multicolumn{6}{c}{Relative wages}\tabularnewline
\multirow{2}{*}{$\Delta\left(\frac{w_{mh}}{w_{fh}}\right)$} & \textendash 0&936 & \textendash 0&016 & 0&072\tabularnewline
 & (0&056) & (0&001) & (0&005)\tabularnewline
\multirow{2}{*}{$\Delta\left(\frac{w_{mu}}{w_{fu}}\right)$} & \textendash 0&062 & \textendash 0&001 & \textendash 0&004\tabularnewline
 & (0&006) & (0&000) & (0&001)\tabularnewline
\multirow{2}{*}{$\Delta\left(\frac{w_{mh}}{w_{mu}}\right)$} & 0&240 & 0&004 & \textendash 0&010\tabularnewline
 & (0&014) & (0&000) & (0&001)\tabularnewline
\multirow{2}{*}{$\Delta\left(\frac{w_{fh}}{w_{fu}}\right)$} & 1&337 & 0&022 & \textendash 0&104\tabularnewline
 & (0&058) & (0&001) & (0&005)\tabularnewline
\cline{2-7}
\multirow{2}{*}{$\Delta\left(\frac{\overline{w}_{M}}{\overline{w}_{F}}\right)$} & \textendash 0&251 & \textendash 0&004 & 0&014\tabularnewline
 & (0&015) & (0&000) & (0&001)\tabularnewline
\multirow{2}{*}{$\Delta\left(\frac{\overline{w}_{H}}{\overline{w}_{U}}\right)$} & 0&564 & 0&009 & \textendash 0&040\tabularnewline
 & (0&013) & (0&000) & (0&001)\tabularnewline
\cline{2-7}
 & \multicolumn{6}{c}{Labor shares}\tabularnewline
\multirow{2}{*}{$\Delta\left(\frac{w_{mh}\ell_{mh}}{py}\right)$} & 0&005 & 0&000 & 0&000\tabularnewline
 & (0&001) & (0&000) & (0&000)\tabularnewline
\multirow{2}{*}{$\Delta\left(\frac{w_{fh}\ell_{fh}}{py}\right)$} & 0&017 & 0&000 & \textendash 0&001\tabularnewline
 & (0&001) & (0&000) & (0&000)\tabularnewline
\multirow{2}{*}{$\Delta\left(\frac{w_{mu}\ell_{mu}}{py}\right)$} & \textendash 0&036 & \textendash 0&001 & 0&002\tabularnewline
 & (0&002) & (0&000) & (0&000)\tabularnewline
\multirow{2}{*}{$\Delta\left(\frac{w_{fu}\ell_{fu}}{py}\right)$} & \textendash 0&008 & 0&000 & 0&001\tabularnewline
 & (0&001) & (0&000) & (0&000)\tabularnewline
\cline{2-7}
\multirow{2}{*}{$\Delta\left(\frac{\sum_{f}w_{f}\ell_{f}}{py}\right)$} & 0&001 & 0&000 & 0&000\tabularnewline
 & (0&004) & (0&000) & (0&000)\tabularnewline
\hline 
\end{tabular}
\par\end{centering}
\begin{singlespace}
\raggedright{}{\footnotesize\textit{Notes}}{\footnotesize : The changes
in the levels of relative wages and labor shares attributable to technological
change are computed using equations \eqref{eq: D.Wf/Wg-Qf''} and
\eqref{eq: D.WfLf/PY-Qf''} of Proposition \ref{prop: D.Wf/Wg-Qf''}.
Standard errors in parentheses are computed using bootstrap with 500
replications.}{\footnotesize\par}
\end{singlespace}
\end{table}

\begin{sidewaystable}[ph]
\caption{Effects of technological change embodied in ICCT capital for each
country\label{tab: GE_effect_by_country}}

\begin{centering}
\resizebox{1\textwidth}{!}{\hspace{3.5cm}%
\begin{tabular}{lr@{\extracolsep{0pt}.}lr@{\extracolsep{0pt}.}lr@{\extracolsep{0pt}.}lr@{\extracolsep{0pt}.}lr@{\extracolsep{0pt}.}lr@{\extracolsep{0pt}.}lr@{\extracolsep{0pt}.}lr@{\extracolsep{0pt}.}lr@{\extracolsep{0pt}.}lr@{\extracolsep{0pt}.}lr@{\extracolsep{0pt}.}lr@{\extracolsep{0pt}.}l}
\hline 
 & \multicolumn{2}{c}{All} & \multicolumn{2}{c}{AUS} & \multicolumn{2}{c}{AUT} & \multicolumn{2}{c}{CZE} & \multicolumn{2}{c}{DEU} & \multicolumn{2}{c}{DNK} & \multicolumn{2}{c}{FIN} & \multicolumn{2}{c}{GBR} & \multicolumn{2}{c}{ITA} & \multicolumn{2}{c}{JPN} & \multicolumn{2}{c}{NLD} & \multicolumn{2}{c}{USA}\tabularnewline
\cline{2-25}
 & \multicolumn{24}{c}{Relative wages, $1/\gamma=0$}\tabularnewline
\multirow{2}{*}{$\Delta\left(\frac{w_{mh}}{w_{fh}}\right)$} & \textendash 0&886 & \textendash 1&351 & \textendash 0&565 & \textendash 0&770 & \textendash 0&879 & \textendash 3&325 & \textendash 0&264 & \textendash 0&938 & \textendash 0&552 & \textendash 1&318 & \textendash 1&301 & \textendash 0&770\tabularnewline
 & (0&052) & (0&082) & (0&033) & (0&048) & (0&053) & (0&239) & (0&013) & (0&057) & (0&036) & (0&099) & (0&092) & (0&043)\tabularnewline
\multirow{2}{*}{$\Delta\left(\frac{w_{mu}}{w_{fu}}\right)$} & \textendash 0&074 & \textendash 0&140 & \textendash 0&038 & \textendash 0&070 & \textendash 0&058 & \textendash 0&205 & \textendash 0&025 & \textendash 0&085 & \textendash 0&043 & \textendash 0&089 & \textendash 0&081 & \textendash 0&081\tabularnewline
 & (0&006) & (0&010) & (0&003) & (0&006) & (0&005) & (0&017) & (0&003) & (0&007) & (0&004) & (0&009) & (0&007) & (0&007)\tabularnewline
\multirow{2}{*}{$\Delta\left(\frac{w_{mh}}{w_{mu}}\right)$} & 0&262 & 0&412 & 0&226 & 0&306 & 0&216 & 1&105 & 0&102 & 0&225 & 0&223 & 0&188 & 0&419 & 0&176\tabularnewline
 & (0&014) & (0&022) & (0&014) & (0&016) & (0&011) & (0&075) & (0&005) & (0&012) & (0&015) & (0&008) & (0&022) & (0&008)\tabularnewline
\multirow{2}{*}{$\Delta\left(\frac{w_{fh}}{w_{fu}}\right)$} & 1&283 & 1&822 & 1&037 & 1&359 & 1&239 & 5&308 & 0&303 & 1&446 & 0&829 & 1&665 & 2&240 & 0&934\tabularnewline
 & (0&053) & (0&076) & (0&041) & (0&062) & (0&056) & (0&274) & (0&009) & (0&064) & (0&030) & (0&108) & (0&121) & (0&037)\tabularnewline
 & \multicolumn{24}{c}{Relative wages, $1/\gamma=1$}\tabularnewline
\multirow{2}{*}{$\Delta\left(\frac{w_{mh}}{w_{fh}}\right)$} & \textendash 0&486 & \textendash 0&737 & \textendash 0&315 & \textendash 0&420 & \textendash 0&470 & \textendash 1&599 & \textendash 0&155 & \textendash 0&503 & \textendash 0&299 & \textendash 0&660 & \textendash 0&655 & \textendash 0&431\tabularnewline
 & (0&036) & (0&045) & (0&019) & (0&027) & (0&028) & (0&110) & (0&006) & (0&030) & (0&017) & (0&047) & (0&044) & (0&024)\tabularnewline
\multirow{2}{*}{$\Delta\left(\frac{w_{mu}}{w_{fu}}\right)$} & \textendash 0&096 & \textendash 0&177 & \textendash 0&050 & \textendash 0&085 & \textendash 0&074 & \textendash 0&238 & \textendash 0&033 & \textendash 0&109 & \textendash 0&053 & \textendash 0&102 & \textendash 0&100 & \textendash 0&108\tabularnewline
 & (0&010) & (0&021) & (0&007) & (0&011) & (0&010) & (0&032) & (0&006) & (0&015) & (0&008) & (0&014) & (0&013) & (0&014)\tabularnewline
\multirow{2}{*}{$\Delta\left(\frac{w_{mh}}{w_{mu}}\right)$} & 0&264 & 0&422 & 0&230 & 0&303 & 0&212 & 0&992 & 0&105 & 0&229 & 0&206 & 0&177 & 0&400 & 0&189\tabularnewline
 & (0&013) & (0&020) & (0&012) & (0&017) & (0&011) & (0&058) & (0&009) & (0&012) & (0&011) & (0&012) & (0&022) & (0&011)\tabularnewline
\multirow{2}{*}{$\Delta\left(\frac{w_{fh}}{w_{fu}}\right)$} & 0&767 & 1&092 & 0&634 & 0&807 & 0&717 & 2&853 & 0&196 & 0&844 & 0&507 & 0&876 & 1&226 & 0&567\tabularnewline
 & (0&039) & (0&044) & (0&025) & (0&038) & (0&032) & (0&131) & (0&006) & (0&037) & (0&017) & (0&054) & (0&059) & (0&023)\tabularnewline
\cline{2-25}
 & \multicolumn{24}{c}{Labor shares, $1/\gamma=0$}\tabularnewline
\multirow{2}{*}{$\Delta\left(\frac{w_{mh}\ell_{mh}}{py}\right)$} & 0&006 & 0&005 & 0&004 & 0&004 & 0&005 & 0&015 & 0&002 & 0&004 & 0&004 & 0&005 & 0&009 & 0&002\tabularnewline
 & (0&001) & (0&001) & (0&001) & (0&001) & (0&001) & (0&002) & (0&000) & (0&001) & (0&000) & (0&001) & (0&001) & (0&001)\tabularnewline
\multirow{2}{*}{$\Delta\left(\frac{w_{fh}\ell_{fh}}{py}\right)$} & 0&017 & 0&016 & 0&016 & 0&022 & 0&015 & 0&022 & 0&013 & 0&008 & 0&010 & 0&006 & 0&007 & 0&021\tabularnewline
 & (0&001) & (0&001) & (0&001) & (0&001) & (0&001) & (0&001) & (0&000) & (0&000) & (0&000) & (0&000) & (0&000) & (0&001)\tabularnewline
\multirow{2}{*}{$\Delta\left(\frac{w_{mu}\ell_{mu}}{py}\right)$} & \textendash 0&039 & \textendash 0&074 & \textendash 0&025 & \textendash 0&027 & \textendash 0&024 & \textendash 0&107 & \textendash 0&013 & \textendash 0&037 & \textendash 0&032 & \textendash 0&032 & \textendash 0&050 & \textendash 0&036\tabularnewline
 & (0&002) & (0&004) & (0&002) & (0&002) & (0&002) & (0&007) & (0&001) & (0&002) & (0&002) & (0&002) & (0&003) & (0&002)\tabularnewline
\multirow{2}{*}{$\Delta\left(\frac{w_{fu}\ell_{fu}}{py}\right)$} & \textendash 0&008 & \textendash 0&011 & \textendash 0&006 & \textendash 0&007 & \textendash 0&005 & \textendash 0&027 & \textendash 0&005 & \textendash 0&006 & \textendash 0&005 & \textendash 0&004 & \textendash 0&004 & \textendash 0&008\tabularnewline
 & (0&001) & (0&001) & (0&001) & (0&001) & (0&001) & (0&003) & (0&000) & (0&001) & (0&001) & (0&000) & (0&001) & (0&001)\tabularnewline
 & \multicolumn{24}{c}{Labor shares, $1/\gamma=1$}\tabularnewline
\multirow{2}{*}{$\Delta\left(\frac{w_{mh}\ell_{mh}}{py}\right)$} & 0&014 & 0&014 & 0&010 & 0&010 & 0&011 & 0&028 & 0&004 & 0&009 & 0&008 & 0&012 & 0&019 & 0&010\tabularnewline
 & (0&001) & (0&001) & (0&001) & (0&001) & (0&001) & (0&002) & (0&000) & (0&000) & (0&000) & (0&001) & (0&001) & (0&001)\tabularnewline
\multirow{2}{*}{$\Delta\left(\frac{w_{fh}\ell_{fh}}{py}\right)$} & 0&020 & 0&020 & 0&020 & 0&026 & 0&017 & 0&024 & 0&016 & 0&009 & 0&012 & 0&006 & 0&008 & 0&026\tabularnewline
 & (0&001) & (0&001) & (0&001) & (0&001) & (0&001) & (0&001) & (0&001) & (0&000) & (0&000) & (0&000) & (0&000) & (0&001)\tabularnewline
\multirow{2}{*}{$\Delta\left(\frac{w_{mu}\ell_{mu}}{py}\right)$} & \textendash 0&070 & \textendash 0&128 & \textendash 0&046 & \textendash 0&046 & \textendash 0&042 & \textendash 0&173 & \textendash 0&024 & \textendash 0&066 & \textendash 0&052 & \textendash 0&051 & \textendash 0&083 & \textendash 0&067\tabularnewline
 & (0&005) & (0&009) & (0&004) & (0&004) & (0&004) & (0&016) & (0&003) & (0&006) & (0&005) & (0&005) & (0&008) & (0&006)\tabularnewline
\multirow{2}{*}{$\Delta\left(\frac{w_{fu}\ell_{fu}}{py}\right)$} & \textendash 0&007 & \textendash 0&009 & \textendash 0&005 & \textendash 0&008 & \textendash 0&004 & \textendash 0&019 & \textendash 0&007 & \textendash 0&004 & \textendash 0&003 & \textendash 0&003 & 0&001 & \textendash 0&010\tabularnewline
 & (0&001) & (0&002) & (0&001) & (0&001) & (0&001) & (0&005) & (0&001) & (0&001) & (0&001) & (0&001) & (0&001) & (0&001)\tabularnewline
\cline{2-25}
 & \multicolumn{24}{c}{Technological change embodied in ICCT capital}\tabularnewline
$\Delta\ln q_{i}$ & 2&001 & 2&552 & 2&005 & 1&333 & 1&379 & 3&365	 & 1&332 & 1&599 & 1&754 & 1&381 & 1&821 & 1&954\tabularnewline
\hline 
\end{tabular}\hspace{3.5cm}}
\par\end{centering}
\begin{singlespace}
\raggedright{}{\footnotesize\textit{Notes}}{\footnotesize : Country
names are abbreviated as follows: AUS, Australia; AUT, Austria; CZE,
the Czech Republic; DEU, Germany; DNK, Denmark; FIN, Finland; GBR,
the United Kingdom; ITA, Italy; JPN, Japan; NLD, the Netherlands;
and USA, the United States. The changes in the levels of relative
wages and labor shares attributable to technological change are computed
using equations \eqref{eq: D.Wf/Wg-Qf'} and \eqref{eq: D.WfLf/PY-Qf'}
of Proposition \ref{prop: D.Wf/Wg-Qf'}. The first column reproduces
the first and third columns of Table \ref{tab: GE_effect}. The second
to last columns report the effects evaluated at the sample means over
time for each country. Standard errors in parentheses are computed
using bootstrap with 500 replications.}{\footnotesize\par}
\end{singlespace}
\end{sidewaystable}

\begin{table}
\caption{Results from sectoral and aggregate production functions\label{tab: GE_effect_aggregate}}

\begin{centering}
\begin{tabular}{lr@{\extracolsep{0pt}.}lr@{\extracolsep{0pt}.}lr@{\extracolsep{0pt}.}lr@{\extracolsep{0pt}.}lr@{\extracolsep{0pt}.}lr@{\extracolsep{0pt}.}l}
\hline 
 & \multicolumn{4}{c}{$\Delta\ln k_{i}$} & \multicolumn{2}{c}{} & \multicolumn{2}{c}{} & \multicolumn{4}{c}{$\Delta\ln q_{i}$}\tabularnewline
 & \multicolumn{2}{c}{Sectoral} & \multicolumn{2}{c}{Aggregate} & \multicolumn{2}{c}{} & \multicolumn{2}{c}{} & \multicolumn{2}{c}{Sectoral} & \multicolumn{2}{c}{Aggregate}\tabularnewline
\cline{2-5}\cline{10-13}
 & \multicolumn{4}{c}{Relative wages} & \multicolumn{2}{c}{} & \multicolumn{2}{c}{} & \multicolumn{4}{c}{Relative wages}\tabularnewline
\multirow{2}{*}{$\Delta\ln\left(\frac{w_{mh}}{w_{fh}}\right)$} & \textendash 0&884 & \textendash 1&109 & \multicolumn{2}{c}{} & \multicolumn{2}{c}{$\Delta\left(\frac{w_{mh}}{w_{fh}}\right)$} & \textendash 0&886 & \textendash 1&015\tabularnewline
 & (0&070) & (0&097) & \multicolumn{2}{c}{} & \multicolumn{2}{c}{} & (0&052) & (0&061)\tabularnewline
\multirow{2}{*}{$\Delta\ln\left(\frac{w_{mu}}{w_{fu}}\right)$} & \textendash 0&070 & \textendash 0&075 & \multicolumn{2}{c}{} & \multicolumn{2}{c}{$\Delta\left(\frac{w_{mu}}{w_{fu}}\right)$} & \textendash 0&074 & \textendash 0&073\tabularnewline
 & (0&005) & (0&006) & \multicolumn{2}{c}{} & \multicolumn{2}{c}{} & (0&006) & (0&005)\tabularnewline
\multirow{2}{*}{$\Delta\ln\left(\frac{w_{mh}}{w_{mu}}\right)$} & 0&222 & 0&236 & \multicolumn{2}{c}{} & \multicolumn{2}{c}{$\Delta\left(\frac{w_{mh}}{w_{mu}}\right)$} & 0&262 & 0&254\tabularnewline
 & (0&014) & (0&017) & \multicolumn{2}{c}{} & \multicolumn{2}{c}{} & (0&014) & (0&015)\tabularnewline
\multirow{2}{*}{$\Delta\ln\left(\frac{w_{fh}}{w_{fu}}\right)$} & 1&036 & 1&270 & \multicolumn{2}{c}{} & \multicolumn{2}{c}{$\Delta\left(\frac{w_{fh}}{w_{fu}}\right)$} & 1&283 & 1&436\tabularnewline
 & (0&069) & (0&096) & \multicolumn{2}{c}{} & \multicolumn{2}{c}{} & (0&053) & (0&063)\tabularnewline
\multirow{2}{*}{$\Delta\ln\left(\frac{\overline{w}_{M}}{\overline{w}_{F}}\right)$} & \textendash 0&228 & \textendash 0&276 & \multicolumn{2}{c}{} & \multicolumn{2}{c}{$\Delta\left(\frac{\overline{w}_{M}}{\overline{w}_{F}}\right)$} & \textendash 0&249 & \textendash 0&276\tabularnewline
 & (0&017) & (0&022) & \multicolumn{2}{c}{} & \multicolumn{2}{c}{} & (0&014) & (0&015)\tabularnewline
\multirow{2}{*}{$\Delta\ln\left(\frac{\overline{w}_{H}}{\overline{w}_{U}}\right)$} & 0&454 & 0&531 & \multicolumn{2}{c}{} & \multicolumn{2}{c}{$\Delta\left(\frac{\overline{w}_{H}}{\overline{w}_{U}}\right)$} & 0&564 & 0&602\tabularnewline
 & (0&024) & (0&032) & \multicolumn{2}{c}{} & \multicolumn{2}{c}{} & (0&013) & (0&014)\tabularnewline
\cline{2-5}\cline{10-13}
 & \multicolumn{4}{c}{Labor shares} & \multicolumn{2}{c}{} & \multicolumn{2}{c}{} & \multicolumn{4}{c}{Labor shares}\tabularnewline
\multirow{2}{*}{$\Delta\ln\left(\frac{w_{mh}\ell_{mh}}{py}\right)$} & 0&092 & 0&101 & \multicolumn{2}{c}{} & \multicolumn{2}{c}{$\Delta\left(\frac{w_{mh}\ell_{mh}}{py}\right)$} & 0&006 & 0&006\tabularnewline
 & (0&015) & (0&018) & \multicolumn{2}{c}{} & \multicolumn{2}{c}{} & (0&001) & (0&001)\tabularnewline
\multirow{2}{*}{$\Delta\ln\left(\frac{w_{fh}\ell_{fh}}{py}\right)$} & 0&976 & 1&211 & \multicolumn{2}{c}{} & \multicolumn{2}{c}{$\Delta\left(\frac{w_{fh}\ell_{fh}}{py}\right)$} & 0&017 & 0&019\tabularnewline
 & (0&071) & (0&098) & \multicolumn{2}{c}{} & \multicolumn{2}{c}{} & (0&001) & (0&001)\tabularnewline
\multirow{2}{*}{$\Delta\ln\left(\frac{w_{mu}\ell_{mu}}{py}\right)$} & \textendash 0&130 & \textendash 0&135 & \multicolumn{2}{c}{} & \multicolumn{2}{c}{$\Delta\left(\frac{w_{mu}\ell_{mu}}{py}\right)$} & \textendash 0&039 & \textendash 0&037\tabularnewline
 & (0&004) & (0&003) & \multicolumn{2}{c}{} & \multicolumn{2}{c}{} & (0&002) & (0&002)\tabularnewline
\multirow{2}{*}{$\Delta\ln\left(\frac{w_{fu}\ell_{fu}}{py}\right)$} & \textendash 0&060 & \textendash 0&059 & \multicolumn{2}{c}{} & \multicolumn{2}{c}{$\Delta\left(\frac{w_{fu}\ell_{fu}}{py}\right)$} & \textendash 0&008 & \textendash 0&007\tabularnewline
 & (0&005) & (0&005) & \multicolumn{2}{c}{} & \multicolumn{2}{c}{} & (0&001) & (0&001)\tabularnewline
\multirow{2}{*}{$\Delta\ln\left(\frac{\sum_{f}w_{f}\ell_{f}}{py}\right)$} & \textendash 0&003 & 0&012 & \multicolumn{2}{c}{} & \multicolumn{2}{c}{$\Delta\left(\frac{\sum_{f}w_{f}\ell_{f}}{py}\right)$} & \textendash 0&001 & 0&006\tabularnewline
 & (0&007) & (0&009) & \multicolumn{2}{c}{} & \multicolumn{2}{c}{} & (0&004) & (0&004)\tabularnewline
\hline 
\end{tabular}
\par\end{centering}
\begin{singlespace}
\raggedright{}{\footnotesize\textit{Notes}}{\footnotesize : The figures
in the upper and lower halves of the first two columns are computed
using equation \eqref{eq: D.Wf/Wg-Lf} of Proposition \ref{prop: D.Wf/Wg-Lf}
and equation \eqref{eq: D.WfLf/PY-Lf} of Proposition \ref{prop: D.WfLf/PY-Lf},
while the figures in the upper and lower halves of the last two columns
are computed using equations \eqref{eq: D.Wf/Wg-Qf} and \eqref{eq: D.WfLf/PY-Qf}
of Proposition \ref{prop: D.Wf/Wg-Qf}. The figures in the first column
of the left panel are reproduced from the fifth columns of Tables
\ref{tab: relative_wages} and \ref{tab: labor_shares}, while those
in the first column of the right panel are reproduced from the first
column of Table \ref{tab: GE_effect}. The figures in the second columns
of the left and right panels are computed using the estimates of the
aggregate production function parameters reported in the last row
of Table \ref{tab: baseline}.}{\footnotesize\par}
\end{singlespace}
\end{table}

\end{document}